\documentclass[aip,jcp,superscriptaddress,reprint]{revtex4-2}
\usepackage[version=3]{mhchem} 
\usepackage{multirow}
\usepackage{graphicx} 
\usepackage{bm,amsmath,amsfonts,amssymb}
\usepackage[colorlinks=true,linkcolor=blue]{hyperref}
\usepackage{orcidlink}
\usepackage{comment}
\usepackage{hyperref}

\usepackage{float}
\makeatletter
\let\newfloat\newfloat@ltx
\makeatother
\usepackage{algorithm}
\usepackage{algpseudocode}
\makeatletter
\algrenewcommand\ALG@beginalgorithmic{\ttfamily}
\makeatother
\usepackage[frozencache]{minted} 
\usepackage{xr}
\makeatletter
\newcommand*{\addFileDependency}[1]{% argument=file name and extension
\typeout{(#1)}
\@addtofilelist{#1}
\IfFileExists{#1}{}{\typeout{No file #1.}}
}
\makeatother

\let\oldaddcontentsline\addcontentsline
\renewcommand{\addcontentsline}[3]{}

\begin{document}
\title{Multi-component Adaptive Sampling Configuration Interaction method in \texttt{openLOWDIN}: positronic and quantum nuclear systems}

\author{Jorge Charry\orcidlink{0000-0003-3069-2522}}
\email[Corresponding author:]{jorge.charry@researchers-hub.lu} 
\affiliation{HPC Platform, University of Luxembourg, L-4364 Esch-sur-Alzette, Luxembourg\looseness=-3}
\affiliation{Luxembourg Researchers Hub asbl, 223 rue de Luxembourg, L-4222, Esch-sur-Alzette, Luxembourg \looseness=-3}
\author{Matteo Barborini\orcidlink{0000-0001-7798-099X}}
\affiliation{HPC Platform, University of Luxembourg, L-4364 Esch-sur-Alzette, Luxembourg\looseness=-3}

%%%%%%%%%%%%%%%%%%%%%%%%%%%%%%%%%%%%%%%%%%%%%%%%%%%%%%%%%%%%%%%%%%%%%
%% Abstract
%%%%%%%%%%%%%%%%%%%%%%%%%%%%%%%%%%%%%%%%%%%%%%%%%%%%%%%%%%%%%%%%%%%%%
\begin{abstract}
We present a multi-component Selective Configuration Interaction (SCI) method based on the deterministic Adaptive Sampling CI (ASCI) algorithm for computing energies and wavefunction properties of molecular systems containing leptons of variable masses, or quantum nuclei. 
Near-FCI accuracy is achieved by incorporating second-order Epstein–Nesbet perturbation theory or by dressing the Hamiltonian via Davidson’s and co-workers shifted-Bk method. 
We validate the approach through benchmark positronic ($\ce{e^+LiH}$, $\ce{e^+OH^-}$, $\ce{e^+Mg}$) and quantum nuclear systems ($\ce{HCN}$), establishing a novel protocol for accurate convergence, based on the iterative adaptation of natural orbitals starting from those obtained by considering only cross-particle CISD excitations, to those obtained from SCI variational space, surpassing the accuracy of previous SCI studies.
Overall, we show that the present multi-component SCI framework offers a simple, systematically improvable, and cost-effective alternative to traditional CI approaches.
\end{abstract}

%%%%%%%%%%%%%%%%%%%%%%%%%%%%%%%%%%%%%%%%%%%%%%%%%%%%%%%%%%%%%%%%%%%%%
%% Title
%%%%%%%%%%%%%%%%%%%%%%%%%%%%%%%%%%%%%%%%%%%%%%%%%%%%%%%%%%%%%%%%%%%%%
\maketitle
%%%%%%%%%%%%%%%%%%%%%%%%%%%%%%%%%%%%%%%%%%%%%%%%%%%%%%%%%%%%%%%%%%%%%
%% Main text
%%%%%%%%%%%%%%%%%%%%%%%%%%%%%%%%%%%%%%%%%%%%%%%%%%%%%%%%%%%%%%%%%%%%%

\section{Introduction}

% General CI motivation
Since the foundations of quantum chemistry, the quest to solve the non-relativistic many-body Schrödinger equation for molecular systems has relied on single-particle approximations.
In this framework, the initial wavefunction is constructed as the Slater determinant of a finite set of single-particle orbitals and the electron–electron interactions are approximated by a mean-field potential, as done in the Hartree-Fock (HF) method.
To overcome the inherent limitations of such mean-field approach, the wavefunction is subsequently expanded as a linear combination of `configurations' (determiants), each representing different occupations of the single-particle orbitals, to systematically capture electron correlation. 
Referred to in early literature as the "superposition of configurations," this approach evolved into what is now universally known as the Configuration Interaction (CI) method, effectively transforming the continuous many-body partial differential equation into a solvable discrete matrix eigenvalue problem. \cite{sherrill_AQC_34_143_1999, cremer_WCMS_3_482_2013, helgaker_MET__523_2000}
% FCI
Incorporating the complete set of electron excitations within a finite basis set defines the Full Configuration Interaction (FCI) limit, which yields the exact variational energy within that single-particle space. 
However, as the system's size and the basis set size increase, the dimension of the Hilbert space of the determinants grows combinatorially, rendering the exact FCI calculations usually intractable. 
Over the passed decade, state-of-the-art implementations have pushed the boundaries of FCI, towards the successful handling of determinant spaces of billions ($10^{9}$) \cite{knowles_JCP_91_2396_1989, olsen_CPL_169_463_1990}, trillions ($10^{12}$) \cite{gao_JCTC_20_1185_2024} (for \ce{C_3H_8} with the STO-3G basis set), and more recently, quadrillion ($10^{15}$) of configurations (for the HBrTe system)\cite{shayit_NC_16_11016_2025}. 

% CI alternatives
To bypass this exponential bottleneck, the quantum chemistry community has historically pursued two primary alternative routes.

First, methods that truncate excitation ranks, as seen in truncated CI (e.g., CISD) or Coupled Cluster (CC) theory \cite{sherrill_AQC_34_143_1999, cremer_WCMS_3_482_2013, helgaker_MET__523_2000}. 
Today, Coupled Cluster with single, double, and perturbative triple excitations [CCSD(T)] stands prominently as the single-reference ``gold standard'' for van der Waals interacting molecules \cite{perera_CCCE__18_2024,nagy_CS_15_14556_2024}, although it often struggles with strongly correlated systems, bond-breaking pathways, or with transition metal complexes \cite{margraf_JCP_147_184101_2017, larsson_JACS_144_15932_2022}.

Second, Selective Configuration Interaction (SCI) methods, overcome the limitations of fixed-rank excitation truncations by selecting leading determinants within the whole FCI space, delivering near-FCI quality results while avoiding its prohibitive computational cost.
Rooted in the 1970s pioneering frameworks such as CIPSI (Configuration Interaction by Perturbation with multiconfigurational zeroth-order wavefunction Selected by Iterative process) \cite{huron_JCP_58_5745_1973}, the core idea behind the SCI methods is to dynamically expand the variational space by iteratively selecting and retaining only the most energetically significant configurations, estimating the remaining dynamical correlation through second-order perturbation theory,  exploiting the fact that the FCI Hilbert space is extraordinarily sparse.

Representative variants of SCI can be categorized in deterministic selective methods, that include modern formulations of CIPSI \cite{garniron_JCTC_15_3591_2019, chilkuri_JCTC_17_2868_2021}, heat-bath CI (HCI)\cite{holmes_JCTC_12_3674_2016}, adaptive sampling CI (ASCI) \cite{tubman_JCP_145_044112_2016,tubman_JCTC_16_2139_2020}, adaptive CI (ACI)\cite{schriber_JCP_144_161106_2016}, and iterative CI (iCI) \cite{liu_JCTC_12_1169_2016,zhang_JCTC_16_2296_2020},
and semi- or fully- stochastic variants such as FCI Quantum Monte Carlo (FCIQMC)\cite{booth_JCP_131_054106_2009,blunt_JCP_142_184107_2015},  Monte Carlo CI (MCCI) \cite{greer_JoCP_146_181_1998,prentice_JCP_151_164112_2019}, and semi-stochastic HCI (SHCI)\cite{sharma_JCTC_13_1595_2017}. 
For a comprehensive overview of SCI methods, we refer the reader to the blind test on benzene correlation energy\cite{eriksen_JPCL_11_8922_2020} and the perspective review by J. J. Eriksen\cite{eriksen_JPCL_12_418_2021}.

% Mmulticompoment SCI
Although SCI methods have proven remarkably effective for pure electronic structure calculations, their extension to multi-component systems including other types of quantum particles (quantum nuclei, positrons, or heavier Leptons ... etc.) remains significantly less explored.
Because the multi-component wavefunctions must capture both intra- and inter-species correlation effects, exact or near-exact treatments are particularly challenging. 
The few studies present in the literature to date include a couple of pioneering studies on quantum hydrogen nuclei using SHCI\cite{brorsen_JCTC_16_2379_2020,alaal_JCP_155_234107_2021}, while positronic systems have only been investigated within a single MCCI study \cite{coe_CPL_645_106_2016}. 
In both cases, it has been shown that it is possible to recover FCI-quality energies at a small, tractable computational cost, although there are remaining challenges in converging accurate protonic densities \cite{fajen_JCTC_17_965_2021,malbon_JCTC_21_3968_2025} and in predicting bound states for some weakly interacting positronic systems. 

To address this gap and expand the scope of selective methods for other quantum particle systems, in this work we present the implementation of the ASCI method for general multicomponent systems within the multi-component molecular approach integrated into the \texttt{openLowdin} software package \cite{flores-moreno_IJQC_114_50_2014,reyes_IJQC_119_1_2019}.
Additionally, we also tested different approaches for the starting molecular orbitals, among those the incorporation of the shifted-Bk (sBk) effective Hamiltonian \cite{gershgorn_IJoQC_2_751_1968, nitzsche_JCP_68_3103_1978, garniron_JCP_149_064103_2018} dressing method to account for external connected configurations omitted from the variational space during the diagonalization step, allowing us to obtain corrected wave functions and energies. 

This manuscript is organized as follows. 
First, we outline the theoretical background of the multi-component CI method for systems of distinguishable quantum particles, later summarizing the core principles of SCI. 
Next, we detail the ASCI algorithm, its perturbative corrections, and its implementation in \textit{openLowdin}. 
We then present and discuss results for molecular systems containing a positron and quantum hydrogen nuclei. 
Finally, we conclude with an assessment of multi-component SCI performance and outline future perspectives.
\section{Methods}
\subsection{Multi-component Hamiltonian}

Following the \texttt{openLOWDIN} code philosophy \cite{flores-moreno_IJQC_114_50_2014}, the non-relativistic Hamiltonian of a system comprised of N$_{sp}$ types of quantum species is written under the Any Particle Molecular Orbital (APMO)\cite{reyes_IJQC_119_1_2019} approach as: 
%----
\begin{multline}
\hat{H} = 
-\sum_{\alpha}^{N_{sp}}\sum_{i}^{N_{\alpha}}\frac{1}{2m_{\alpha}}\nabla_{i}^{2} 
+ \sum_{\alpha}^{N_{sp}} \sum_{i}^{N_{\alpha}} V_i^{\alpha}(\mathbf{r}^\alpha_i) + \\
+ \sum_{\alpha}^{N_{sp}} \sum_{i<j}^{N_{\alpha}} V_{ij}^{\alpha\alpha}(\mathbf{r}^\alpha_i,\mathbf{r}^\alpha_j) + \\
+ \sum_{\alpha < \beta}^{N_{sp}} \sum_{i}^{N_{\alpha}}\sum_{j}^{N_{\beta}} V_{ij}^{\alpha\beta}(\mathbf{r}^\alpha_i,\mathbf{r}^\beta_j)
+ V^{\text{ext}}
\label{eq:hamiltonian}.
\end{multline}
%----
Here, the indices $\{ \alpha, \beta, \ldots \}$ run through different types of quantum particles $N_{sp}$, each characterized by a mass $m_{\alpha}$, a charge $q_{\alpha}$, and satisfying a given spin statistics.
The indices $\{ i, j, \dots \}$ run through the $\{ N_\alpha, N_\beta, \ldots \}$ particles of a given type of quantum species, and the vectors $\mathbf{r}^{\alpha}_i$ are the associated three-dimensional Cartesian coordinates.
In the above equation, the first term is the kinetic energy, the second term is the one-body interaction between the quantum particles and all N$_{c}$ classical point charges in fixed positions $\mathbf{R}_J$ and charge $Z_J$, 
%----
\begin{equation}
 V_i^{\alpha}(\mathbf{r}_i) = \sum_{J}^{N_c} \frac{q_{\alpha} Z_J }{ |\mathbf{r}_i - \mathbf{R}_J| } .
\label{eq:one-body}
\end{equation}
%----
the third and fourth 
%----
\begin{equation}
V_{ij}^{\alpha\beta}(\mathbf{r}_i,\mathbf{r}_j) = \frac{q_{\alpha} q_{\beta}}{|\mathbf{r}_{i} - \mathbf{r}_{j}| } .
\label{eq:two-body}
\end{equation}
%----
are the intra-species and interspecies two-body potentials, respectively, and the last term is an external potential.

\subsection{Hartree-Fock wave function}

At the Hartree-Fock (HF) level, the ground-state wave function for a multi-species molecular system, $\Psi_0$, is typically represented as a product of single-determinant wave functions for each quantum species:
%----
\begin{multline}
\Psi_0(\{ \overline{\mathbf{r}}^{\alpha}, \alpha \in [1, N_{\text{sp}}] \}) \approx \Psi_{\text{HF}}(\{ \overline{\mathbf{r}}^{\alpha}, \alpha \in [1, N_{\text{sp}}] \}) =\\= \prod_{\alpha}^{N_{\text{sp}}} \Phi_{\alpha}(\overline{\mathbf{r}}^{\alpha}),
\label{eq:hf_wv}
\end{multline}
%----
where $\overline{\mathbf{r}}^{\alpha}$ denotes the complete set of $3N_{\alpha}$ coordinates associated with particle species $\alpha$. In the standard approach, each species wave function $\Phi_{\alpha}(\overline{\mathbf{r}}^{\alpha})$ is constructed as a Slater determinant of single-particle molecular orbitals $\phi_i(\mathbf{r}^{\alpha})$:
%----
\begin{equation}
\Phi_{\alpha}(\overline{\mathbf{r}}^{\alpha}) = \frac{1}{\sqrt{N_{\alpha}!}}
\begin{vmatrix}
\phi_1(\mathbf{r}_1^{\alpha}) & \phi_1(\mathbf{r}_2^{\alpha}) & \dots & \phi_1(\mathbf{r}_{N_{\alpha}}^{\alpha}) \\
\phi_2(\mathbf{r}_1^{\alpha}) & \phi_2(\mathbf{r}_2^{\alpha}) & \dots & \phi_2(\mathbf{r}_{N_{\alpha}}^{\alpha}) \\
\vdots & \vdots & \ddots & \vdots \\
\phi_{N_{\alpha}}(\mathbf{r}_1^{\alpha}) & \phi_{N_{\alpha}}(\mathbf{r}_2^{\alpha}) & \dots & \phi_{N_{\alpha}}(\mathbf{r}_{N_{\alpha}}^{\alpha})
\end{vmatrix},
\label{eq:det_a_long}
\end{equation}
%----
where $\mathbf{r}_k^{\alpha}$ represents the coordinates of the $k$-th particle of species $\alpha$. Finally, the molecular orbitals $\phi_i$ are expanded as linear combinations of atomic orbitals (LCAO) $\varphi$, determined by solving a set of coupled HF equations self-consistently across all quantum species.
\subsection{Second quantization}

To discuss the theory behind selected configuration interaction (SCI) methods, it is convenient to work within the second-quantization formalism. 
Within this framework, we introduce the Fock space, that is a linear vector space where each Slater determinant is represented in shorthand by an occupation number vector:
%------------------------
\begin{equation}
| \Phi \rangle_{\alpha} = | \phi_1, \phi_2, \dots, \phi_M\rangle_{\alpha}
\label{eq:det_a}
\end{equation}
%------------------------
where each occupation number $\phi_p$ is defined as
%------------------------
\begin{equation}
\phi_p = 
\left \{ 
\begin{array}{ll}
1 & \text{occupied}  \\
0 & \text{unoccupied}
\end{array} 
\right .
\end{equation}
%------------------------
The total wave function (Eq.~\ref{eq:hf_wv}) for the whole system can then be expressed as a tensor product of determinants for each quantum species:
%------------------------
\begin{equation}
|{\Psi}_0 \rangle= |\Phi_{\alpha}\rangle \otimes |\Phi_{\beta}\rangle \dots |\Phi_{N_{sp}}\rangle
= \bigotimes_{\alpha}^{N_{sp}} |\Phi_{\alpha}\rangle .
\label{eq:hf_wv_tensor}
\end{equation}
%------------------------
Under the second-quantization formalism, all operators and states are constructed using creation ($a^{\dagger}_p$) and annihilation ($a_p$) operators acting on a determinant state over a set of occupied or unoccupied orbitals $p, q, r, s, \ldots$:
%------------------------
\begin{multline}
a^{\dagger}_p |\Phi\rangle = a^{\dagger}_p | \phi_1, \phi_2, \dots 0_p \dots, \phi_M\rangle =\\ 
= | \phi_1, \phi_2, \dots 1_p, \dots, \phi_M \rangle
\label{eq:creation}
\end{multline}
%------------------------
and
%------------------------
\begin{multline}
a_p |\Phi\rangle = a_p | \phi_1, \phi_2, \dots 1_p \dots, \phi_M\rangle = \\
 = | \phi_1, \phi_2, \dots 0_p, \dots, \phi_M\rangle
\label{eq:annihilation}
\end{multline}
%------------------------
Thus, the Hamiltonian in Eq.~\ref{eq:hamiltonian} can be expressed in second-quantized form as
%------------------------
\begin{multline}
\hat{H} = \sum_{\alpha}^{N_{sp}} \sum_{p,q} h_{p,q} a^{\dagger}_p a_q +
\frac{1}{2}\sum_{\alpha}^{N_{sp}} \sum_{p,q,r,s} g_{p,q,r,s}^{\alpha} a^{\dagger}_p a^{\dagger}_r a_s a_q + \\
+ \sum_{\alpha <\beta}^{N_{sp}} \sum_{\substack{p,r \in \alpha \\ q,s \in \beta}} g_{p,q,r,s}^{\alpha\beta} a^{\dagger}_p a^{\dagger}_r a_s a_q 
+ V^{\text{ext}}
\label{eq:hamiltonian_sq}
\end{multline}
%------------------------
where $h_{p,q}$ are the one-body operators (kinetic and one-body integrals, Eq.~\ref{eq:one-body}), and $g^{\alpha}$ and $g^{\alpha\beta}$ are the intra- and inter-species two-body integrals over the molecular orbitals (Eq.~\ref{eq:two-body}):
%------------------------
\begin{multline}
h_{p,q} = ( p | \hat{h} | q ) =\\
= \int \phi^{*}_p (\mathbf{r}_1) \left( \frac{1}{2m_{\alpha}}\nabla_{1}^{2} 
+ V^{\alpha}_1(\mathbf{r}_1) \right) \phi_q(\mathbf{r}_1)  d\mathbf{r}_1
\label{eq:one_body_sq}
\end{multline}
%------------------------
%------------------------
\begin{multline}
g^{\alpha}_{p,q,r,s} = ( p q | r s )_{\alpha} =\\
= \int \phi^{*}_p (\mathbf{r}_1) \phi^{*}_r (\mathbf{r}_2)  V^{\alpha}_{12}(\mathbf{r}_1,\mathbf{r}_2) \phi_q (\mathbf{r}_1) \phi_s(\mathbf{r}_2)  d\mathbf{r}_1d\mathbf{r}_2
\label{eq:two_body_aa_sq}
\end{multline}
%------------------------
\begin{multline}
g^{\alpha\beta}_{p,q,r,s} = ( p q | r s )_{\alpha\beta} = \\
= \int \phi^{*}_p (\mathbf{r}_1) \phi^{*}_r (\mathbf{r}_2)  V^{\alpha\beta}_{12}(\mathbf{r}_1,\mathbf{r}_2) \phi_q (\mathbf{r}_1) \phi_s(\mathbf{r}_2)  d\mathbf{r}_1d\mathbf{r}_2
\label{eq:two_body_ab_sq}
\end{multline}
%------------------------
Subsequently, the HF wave function in Eq.~\ref{eq:hf_wv_tensor} can be expressed as a tensor product of consecutive creation operators acting on the vacuum state $|0\rangle$ (the empty state where all molecular orbitals are unoccupied):
%------------------------
\begin{multline}
|{\Psi}_{\text{HF}} \rangle = \bigotimes_{\alpha}^{N_{sp}} |\Phi_{\alpha}\rangle  =\\
= \bigotimes_{\alpha}^{N_{sp}} \left[ a^{\dagger}_1, \dots a^{\dagger}_p \dots a^{\dagger}_{N_{\alpha}} |\rangle \right] = \\ =\bigotimes_{\alpha}^{N_{sp}} \left[ \prod_p^{N_\alpha} a^{\dagger} |0\rangle \right] .
\label{eq:hf_wv_tensor_sq}
\end{multline}
%------------------------

\subsection{Configuration Interaction wave function}

The Configuration Interaction (CI) wave function is constructed as a linear combination of Slater determinants with different molecular orbitals' occupations:
\begin{equation}
| \Psi_{\text{CI}} \rangle = \sum_{I}^{\mathcal{N}_{\text{conf}}} c_I|\Psi_I\rangle = \sum_{I}^{\mathcal{N}_{\text{conf}}} c_I   
\bigotimes_{\alpha}^{N_{sp}} | \Phi_I \rangle_{\alpha} 
\label{eq:ci_wv}
\end{equation}

In standard CI approaches, these configurations are grouped according to their excitation level relative to the Hartree-Fock (HF) reference $|\Psi_{\text{HF}}\rangle$:
\begin{multline}
    | \Psi_{\text{CI}} \rangle = c_{\text{HF}}|\Psi_{\text{HF}}\rangle
    + \sum_{\alpha}\sum_{ia \in \alpha}^{\text{singles}} c_i^a|\Psi_i^a \rangle + \\
    +\sum_{\alpha,\beta}\sum_{\substack{ia\in\alpha\\jb\in\beta}}^{\text{doubles}} c_{ij}^{ab}|\Psi_{ij}^{ab} \rangle 
    +\sum_{\alpha,\beta.\gamma}\sum_{\substack{ia\in\alpha\\jb\in\beta\\kc\in\gamma}}^{\text{triples}} c_{ijk}^{abc}|\Psi_{ijk}^{abc} \rangle
    + \cdots
\label{eq:ci_wv_types}
\end{multline}

These terms correspond to single excitations from molecular orbital $i$ to $a$ ($|\Psi_i^a \rangle$), intraspecies double excitations ($|\Psi_{ij}^{ab} \rangle$ when $\{i,j,a,b\} \in \alpha$), interspecies double excitations ($|\Psi_{ij}^{ab} \rangle$ when $ia \in \alpha$ and $jb \in \beta$), as well as triples, quadrupoles, and higher excitations. 

The Full CI (FCI) limit is achieved when all possible excitations within a given basis set, covering all particle combinations across all quantum species, are included, yielding a total configuration space denoted as $\mathcal{N}_{\text{FCI}}$. Note, for example, that an interspecies triple excitation can be formed either as a product of three single excitations across three species or as a product of a single and a double excitation between two species.

The matrix elements of the Hamiltonian matrix $\mathbf{H}$ are constructed using the generalized Slater-Condon rules (see Supplementary Material). 
The CI coefficient vector $\mathbf{c}$ is then determined variationally by solving the CI eigenvalue problem $\textbf{Hc} = E\textbf{c}$.

Because the size of the Hamiltonian matrix $\mathbf{H}$ grows rapidly with the number of particles and the basis set size, storing the full matrix ($\mathcal{N}_{\text{conf}}^2$ scaling) becomes computationally prohibitive. 
Consequently, iterative diagonalization techniques for real symmetric matrices are necessary. 
The Jacobi-Davidson method \cite{bollhofer_CPC_177_951_2007} is widely used to target specific low-lying eigenvalues and eigenvectors. 
Within the \texttt{openLOWDIN} code, we have interfaced with the Jadamilu Fortran library \cite{bollhofer_CPC_177_951_2007} to compute selected eigenvalues of large, sparse symmetric matrices. 
This framework completely bypasses the need to store $\mathbf{H}$ explicitly by evaluating matrix-vector products of the form $\mathbf{Hc} = \boldsymbol{\sigma}$ on the fly. 
As a result, memory overhead is drastically minimized, requiring storage for only about $20\mathcal{N}_{\text{conf}}$ elements to house the eigenvectors, diagonal CI matrix elements, and temporary matrix-vector working vectors.
\subsection{Selective Configuration-Interaction methods}

At its core, Selective Configuration-Interaction (SCI) methods aim to achieve FCI-level accuracy by selecting only the most important configurations from the full expansion. 
This accelerates energy convergence while significantly reducing the number of configurations ($\mathcal{N}_{\text{SCI}} \ll \mathcal{N}_{\text{FCI}}$), thereby defining a computationally tractable subspace that exploits the inherent sparsity of the many-body wave function within the total Hilbert space \cite{huron_JCP_58_5745_1973}.

The SCI algorithms start by considering a reference wave function over a variational target space $\mathcal{T}^{(0)}$, typically composed of the single HF determinant or a small multi-configurational space, 
%----
\begin{equation}
| \Psi_{\text{SCI}} \rangle^{(0)} = \sum_{I \in \mathcal{T}^{(0)}} c_I \lvert \Psi_I \rangle
\label{eq:sci_wv_0}
\end{equation}
%----
and iteratively expands the target wave function space $\mathcal{T}^{(k+1)}$ by incorporating newly connected configurations $\mathcal{G}[\mathcal{T}^{(k)}]$ generated via single and double excitations from the current variational space $\mathcal{T}^{(k)}$; higher excitations yield vanishing matrix elements under the Slater-Condon rules:
%----
\begin{equation}
| \Psi_{\text{SCI}} \rangle^{(k+1)} = | \Psi_{\text{SCI}} \rangle^{(k)} + \sum_{I \in \mathcal{G}[\mathcal{T}^{(k)}] } c_I |\Psi_I \rangle .
\label{eq:sci_wv_k}
\end{equation}
%----
Numerous convergence and selection criteria have been proposed in the literature to construct the incremental space $\mathcal{G}[\mathcal{T}^{(k)}]$, giving rise to distinct flavors of SCI methods that are broadly classified as deterministic, stochastic, or semi-stochastic \cite{eriksen_JPCL_12_418_2021}.
At each step $k$, the variational energy $E^{(k)}$ and  coefficients $\textbf{c}^{(k)}$ of the linear expansion in the trial wavefunction $| \Psi_{\text{SCI}} \rangle^{(k)}$ are obtained by solving the eigenvalue problem 
\begin{equation}
\textbf{Hc}^{(k)} = E_{\text{SCI}}^{(k)}\textbf{c}^{(k)}
\label{eq:sci_eigen}
\end{equation}
within the reduced $\mathcal{T}^{(k)}$ space.
Finally, once the target space $\mathcal{T}^{(k)}$ reaches the desired final size $\mathcal{N}_{\text{SCI}}$, the variational energy $E_{\text{SCI}}^{(k)}$ can be corrected to account for all unselected ($M \notin \mathcal{T}^{(k)}$) yet connected ($M \in \mathcal{G}[\mathcal{T}^{(k)}]$) external determinants, via a second-order perturbative correction ($E_{\text{PT2}}$):
%----
\begin{equation}
E_{\text{SCI+PT2}} = E_{\text{SCI}}^{(k)} + \sum_{\substack{ M \notin \mathcal{T}^{(k)} \\ M \in \mathcal{G}[\mathcal{T}^{(k)}]} }^{\mathcal{N}_{\text{FCI}}} \frac{ \left| \langle \Psi_{\text{SCI}}^{(k)} | \hat{H} \rvert \Psi_M \rangle \right|^2}{E_{\text{SCI}}^{(k)} - \langle \Psi_M | \hat{H} | \Psi_M \rangle},   
\label{eq:pt2}
\end{equation}
%----
that yields a significantly more accurate total energy while avoiding the prohibitive computational cost of solving the full $\mathcal{N}_{\text{FCI}}$ eigenvalue problem.

However, a key limitation of this perturbative approach is that the wave function, and consequently any derived property, remain uncorrected for the external connected configurations outside the variational space $\mathcal{T}$. 
Within the framework of intermediate effective Hamiltonian theory, this limitation can be overcome by constructing and diagonalizing an effective Hamiltonian augmented with a dressing matrix. 
This strategy was originally introduced by Gershgorn and Shavitt in their Bk method \cite{gershgorn_IJoQC_2_751_1968} and later refined into the shifted-Bk (sBk) formalism by Davidson and co-workers \cite{nitzsche_JCP_68_3103_1978}. 
More recently, Garniron et al. \cite{garniron_JCP_149_064103_2018} extended the sBk approach to selective CI methods, demonstrating its effectiveness in systematically converging towards the FCI limit.

The foundational concept behind the sBk approach stems from L\"owdin's eigenvalue partitioning technique, expressed as
%-----------------------
\begin{equation}
\left(
\begin{array}{ccc}
\mathbf{H}^{(00)} & {^{\dagger}}\mathbf{H}^{(01)} & \mathbf{0} \\
\mathbf{H}^{(01)} & \mathbf{H}^{(11)}         & {^{\dagger}}\mathbf{H}^{(12)} \\
\mathbf{0}        & \mathbf{H}^{(12)}         & \mathbf{H}^{(22)} \\
\end{array}
\right)
\left(
\begin{array}{c}
\mathbf{c}^{(0)} \\ 
\mathbf{c}^{(1)} \\
\mathbf{c}^{(2)} \\
\end{array}
\right)
= 
E
\left(
\begin{array}{c}
\mathbf{c}^{(0)} \\ 
\mathbf{c}^{(1)} \\
\mathbf{c}^{(2)} \\
\end{array}
\right)
\label{eq:lowdin_hamiltonian}
\end{equation}
%-----------------------
where $\mathbf{H}^{(00)}$ is the Hamiltonian within the selected space, $\mathbf{H}^{(11)}$ is the Hamiltonian of the unselected but connected configurations, $\mathbf{H}^{(22)}$ is the Hamiltonian over configurations connected to $\mathbf{H}^{(11)}$, and $\mathbf{H}^{(\text{XY})}$ represent the coupling matrices between these partitioned subspaces. 

It has been demonstrated that this system of equations can be recast as an effective Schr\"odinger equation governed by the effective Hamiltonian
%-----------------------
\begin{equation}
\mathbf{H}^{\text{eff}} = \mathbf{H}^{(00)} + \mathbf{\Delta}
\label{eq:effective_hamiltonian}
\end{equation}
%-----------------------
 defined as the sum of the Hamiltonian in the selected space $\mathbf{H}^{(00)}$ and a dressing matrix
%-----------------------
\begin{multline}
\mathbf{\Delta} = {^{\dagger}}\mathbf{H}^{(01)} \left[ (E \mathbf{1} - \mathbf{H}^{(11)}) \right . -\\ - \left.  {^{\dagger}}\mathbf{H}^{(12)} ( E \mathbf{1} - \mathbf{H}^{(22)} )^{-1} \mathbf{H}^{(12)} \right]^{-1} \mathbf{H}^{(01)}.
\end{multline}
%-----------------------
This dressing matrix can be approximated by neglecting second-order coupling terms and retaining only the diagonal elements of $\mathbf{H}^{(11)}$, leading to
%-----------------------
 \begin{equation}
\mathbf{\Delta}^{\text{sBk}} = {^{\dagger}}\mathbf{H}^{(01)} \left[ (E \mathbf{1} - \mathbf{D}^{(11)}) \right]^{-1} \mathbf{H}^{(01)}
\end{equation}
%-----------------------
that in matrix element representation is expressed as
%-----------------------
\begin{equation}
\mathbf{\Delta}^{\text{sBk}}_{IJ} = \sum_{\substack{ M \notin \mathcal{T}^{(k)} \\ M \in \mathcal{G}[\mathcal{T}^{(k)}]} }^{\mathcal{N}_{\text{FCI}}} 
\frac{ \langle \Psi_{I}^{(k)} | \hat{H} \rvert \Psi_M \rangle \langle \Psi_{M} | \hat{H} \rvert \Psi_J^{(k)} \rangle }{E_{\text{SCI}}^{(k)} - \langle \Psi_M | \hat{H} | \Psi_M \rangle}.
\label{eq:sbk}
\end{equation}
%-----------------------
Notice that the dressing matrix does not depend directly on the coefficients of the linear combination of determinants in the SCI wave function. 

A Fortran format pseudo-code describing the above procedure is illustrated through Algorithm \ref{al:sci}.

%-----------------------
\begin{algorithm}
\caption{Selective Configuration Interaction (SCI)}
\label{al:sci}
\begin{minted}[escapeinside=||]{fortran}
real :: |$\mathbf{c}(\mathcal{N})$| 
type(configuration) :: |${\bf \vert \Psi \rangle} (\mathcal{N})$| 
! Initialize with |$\vert\Psi_{\text{HF}} \rangle$| 
|$E_{\text{SCI}}$| = |$E_{\text{HF}}$|; |$ \mathbf{c} (:)$| = 0; |$ \mathbf{c} (1)$| = 1 
while Convergence criteria is not met:
    |Select configurations| ( |$\mathbf{c}$|, |${\bf \vert \Psi \rangle}$| ) ! Eq. |\ref{eq:sci_wv_k} |
    |Solve| |$\textbf{Hc}^{(k)} = E_{\text{SCI}}^{(k)}\textbf{c}^{(k)}$|  ! Eq. |\ref{eq:sci_eigen} |
enddo while
|Compute PT2| ( |$\mathbf{c}$|, |${\bf \vert \Psi \rangle}$| ) ! Eq. |\ref{eq:pt2}| 
|Compute sBk| ( |$\mathbf{c}$|, |${\bf \vert \Psi \rangle}$| ) ! Eq. |\ref{eq:sbk}| 
\end{minted}
\end{algorithm}
%-----------------------

As discussed in the introduction, numerous SCI methods exist each with its own selection and convergence criteria, generally exhibiting comparable efficiency in exploring the FCI space \cite{eriksen_JPCL_11_8922_2020, eriksen_JPCL_12_418_2021}. 
However, during preliminary testing, we observed that determinant-driven selection criteria consistently yielded lower variational energies than integral-driven alternatives for an equivalent number of configurations, although this was not intended as a rigorous benchmark. 
Consequently, we adopted the Adaptive Sampling Configuration Interaction (ASCI) algorithm \cite{tubman_JCP_145_044112_2016} as the foundation for our first multi-component SCI implementation in \texttt{openLOWDIN}.
While the core methodology remains faithful to the original formulation, minor variations exist due to our specific code adaptations. 
In the following section, we detail our implementation, including memory management and parallelization strategies, with further algorithmic specifics provided in the Supplementary Material (see SM).

\subsection{Adaptive Sampling Configuration Interaction}

The Adaptive Sampling Configuration Interaction (ASCI) method can be divided in four steps: (i) search, (ii) selection and sorting, (iii) diagonalization and growth, and (iv) perturbative corrections\cite{tubman_JCP_145_044112_2016}.

\textbf{(i) Search.} A key feature of ASCI is the introduction of a core subspace $\mathcal{C}$---smaller than the target space ($\mathcal{C} \subset \mathcal{T}$)---which is dedicated to exploring and generating new configurations, thereby restricting $\mathcal{T}$ to the selection and diagonalization steps.
Consequently, the reference wave function for the search step reduces to
%----
\begin{equation}
| \Psi_{\mathcal{C}} \rangle^{(0)} = \sum_{I \in \mathcal{C}^{(0)}} c_I \lvert \Psi_I \rangle .
\label{eq:asci_wv_0}
\end{equation}
%----
These core configurations are used to estimate the amplitudes of the first-order perturbative wave function over all single and double excitations connected to $\mathcal{C}$, according to 
%----
\begin{equation}
A_I = \frac{ \sum_{ J \in \mathcal{C}^{(k)} } \hat{H}_{IJ} c_J^{(k)}
}{E_{\text{SCI}}^{(k)} - H_{II} } \quad I \in \mathcal{G}[\mathcal{C}^{(k)}]
\label{eq:asci_amplitude} .
\end{equation}
%----
These amplitudes serve as a metric to estimate the importance of the configurations connected to the core space, scaling approximately as $\mathcal{O}(\mathcal{N}_{\mathcal{C}} N_{\text{occ}}^2 N_{\text{vir}}^2)$ for each quantum species and every pair of species.

Ideally, one could compute the amplitudes for all connected configurations to the core $\mathcal{G}[\mathcal{C}^{(k)}]$, store them in memory, and then select the largest $N_{\mathcal{T}}$ configurations ranked by their absolute amplitude to construct the target space ${\mathcal{T}}$.
However, generating the complete set of unique (non-duplicate) configurations within $\mathcal{G}[\mathcal{C}^{(k)}]$ directly remains prohibitively expensive for practical applications due to their extremely large number.
To overcome this bottleneck, we introduce a temporary buffer space $\mathcal{B}$ to store candidate configurations $\mathbf{b}_{\text{conf}}$ and their corresponding amplitudes $\mathbf{b}$ that may enter the top $N_{\mathcal{T}}$ selection. 
Inherently, truncating via a buffer risks discarding configurations that receive small individual contributions to $A_I$ from many determinants $J$ (high connectivity) but accumulate a non-negligible total amplitude. 
To mitigate this effect, the buffer size is defined as a multiple of the target space, $N_\mathcal{B} = B_f N_\mathcal{T}$, where $B_f$ is a user-defined buffer factor that balances convergence accuracy against available memory.
Because this buffer is excluded from the computationally intensive and memory-demanding diagonalization step, $N_\mathcal{B}$ can be chosen to be substantially larger than the target space size $\mathcal{N}_{\mathcal{T}}$.

\begin{figure}[ht]
    \centering
    \includegraphics[width=1.0\linewidth]{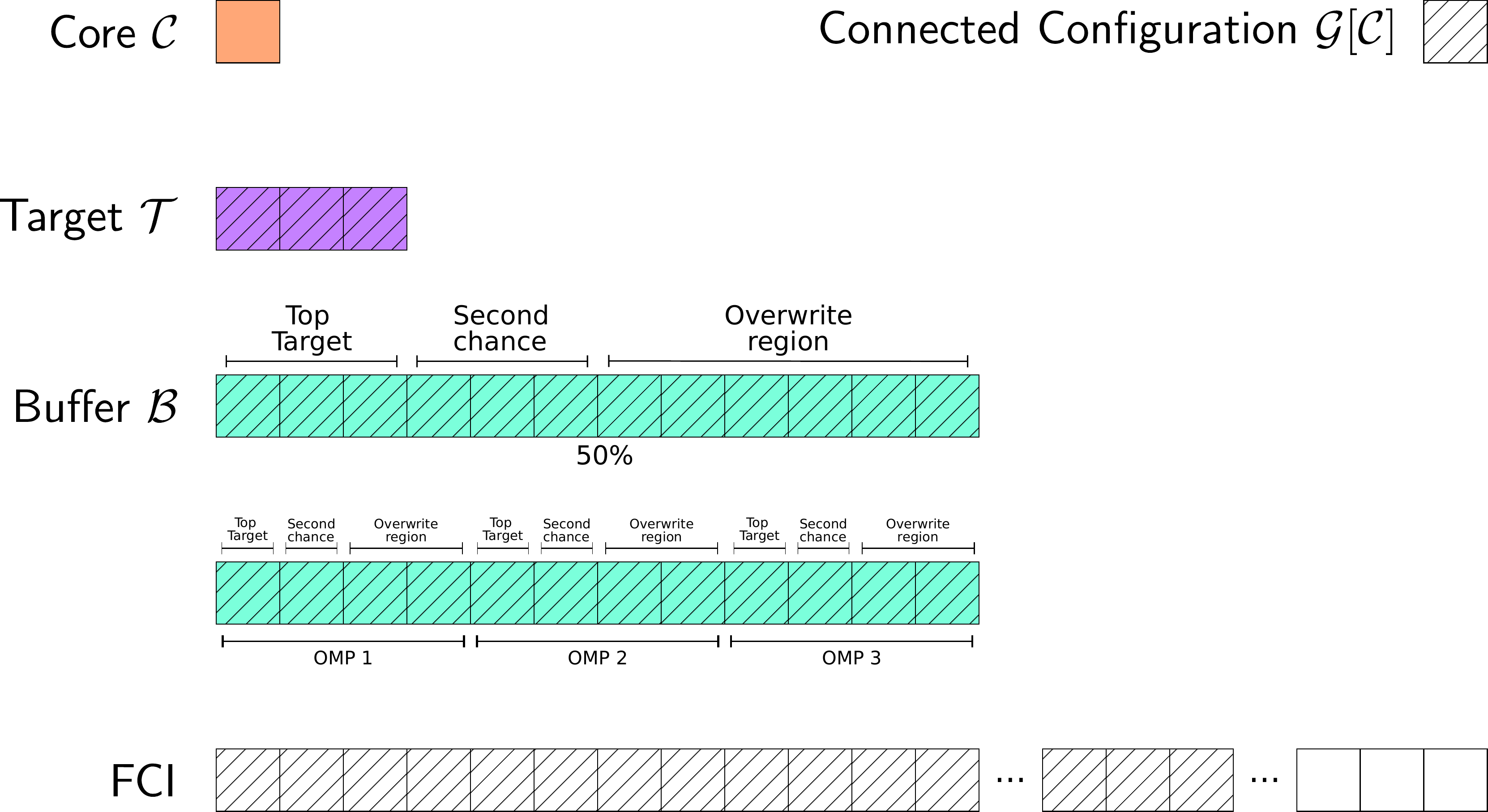}
    \caption{Graphical representation of the different CI subspaces defined within the ASCI implementation in openLOWDIN code}
    \label{fig:asci_spaces}
\end{figure}

Within this framework, Eq.~\ref{eq:asci_amplitude} is evaluated by looping independently over the core space index $J$, generating on-the-fly only the $I$ connected single, intra-species double, and inter-species double configurations. 
Their contributions to $A_{I}$ are computed and stored in a dedicated region of the buffer array assigned to each OpenMP thread, allowing the threads to operate independently within their own local memory without contention. 
This subspace partitioning scheme is illustrated in Fig.~\ref{fig:asci_spaces}.

\textbf{(ii) Selection and sorting}.
Once each OpenMP partition of $\mathbf{b}_{\text{conf}}$ reaches capacity, the configurations are sorted to identify and merge duplicates, summing their respective amplitudes. 
This step is necessary because configurations in our implementation are not indexed by a unique, canonical multi-component address.

Subsequently, the buffer arrays are sorted according to the absolute value of their amplitudes, $\text{sort}(\mathbf{b})$. 
At this point, the bottom 50$\%$ of the configurations are overwritten in subsequent iterations of the loop over Eq.~\ref{eq:asci_amplitude}, while the top $N_{\mathcal{T}}$ configurations $\text{top}_{\mathcal{N}_{\mathcal{T}}}(\text{sort}(\mathbf{b}))$ advance towards selection, and the remaining intermediate $0.5N_{\mathcal{B}}-N_{\mathcal{T}}$ configurations will retain the opportunity to accumulate amplitudes and survive till the end of the loop (see \ref{fig:asci_spaces}).
This procedure is repeated until all core configurations are processed, followed by a final global sort across the entire buffer space to gather the contributions from all OpenMP threads. 
The merge between the current core space and the top $\mathcal{N}_{\mathcal{T}}$ buffer amplitudes then defines the updated target space $ \mathcal{T}^{(k)} = \mathcal{C}^{(k)} \cup \text{top}_{\mathcal{N}_{\mathcal{T}}}(\mathcal{B}^{(k)})$.
Overall, using the quicksort algorithm, this operation scales as $\mathcal{O}(2\mathcal{N}_{\mathcal{B}} \log (\mathcal{N}_{\mathcal{B}}))$, and is executed whenever a buffer is completely full, and once more upon completion of the core space $\mathcal{C}$ search.

\textbf{(iii) Diagonalization and growth}.
Solving the eigenvalue problem of Eq.~\ref{eq:sci_eigen} within the target space yields the updated variational energy $E_{\text{SCI}}^{(k)}$ and the target space coefficients $\mathbf{t}$ from the corresponding eigenvectors. 
The next core space is subsequently defined by selecting the top $\mathcal{N}_{\mathcal{C}}$ elements of the sorted coefficient vector $\mathcal{C}^{(k+1)} = \text{top}_{\mathcal{N}_{\mathcal{C}}}( \text{sort} (\mathbf{t}))$.
At this point, three possible steps are available: (i) repeat from the initial `search' step  to refine the wave function within the current space size; (ii) expand the target space by a growth factor $T_f$ (e.g. $T_f = 2$) $\mathcal{N}_{\mathcal{T}} = \mathcal{N}_{\mathcal{T}} \cdot T_f$ and repeat from the initial `search' step; or (iii) terminate the iteration if energy convergence is achieved within a tolerance $\epsilon$ ($\epsilon = 10^{-5}$), \textit{i.e.} $|E_{\text{SCI}}^{(k)} -  E_{\text{SCI}}^{(k-1)}| < \epsilon$. 

The primary computational bottleneck in this step arises from evaluating the matrix elements for each matrix-vector product required by the iterative diagonalizer, which scales as $\mathcal{O}(\mathcal{N}_{\mathcal{T}}^2)$.

\textbf{(iv) Perturbative corrections}.
Once the convergence is reached, the second-order perturbative correction to the SCI energy is computed for the final variational SCI subspace using Eq. \ref{eq:pt2}.
However, this equation poses a significant computational challenge because the large number of connected configurations outside the SCI subspace ($I \notin \mathcal{T}^{(k)}$, $I \in \mathcal{G}[\mathcal{T}^{(k)}]$) are unknown \textit{a priori}. 
To address this, two primary approaches exist. 

The first approach recasts Eq.~\ref{eq:pt2} into a triple loop \cite{garniron_JCP_147_034101_2017,garniron_JCTC_15_3591_2019}
%------------------
\begin{equation}
E_{\text{PT2}} = \sum_{J}^{N_{\mathcal{T}}} \sum_{\substack{ M \notin \mathcal{T}^{(k)} \\ M \in \mathcal{G}[\mathcal{T}^{(k)}_J] \\ \\ M \notin \mathcal{G}[\mathcal{T}^{(k)}_{1:J-1}] } } \frac{ \left| \sum_{\substack{ I }}^{\mathcal{N}_{\mathcal{T}}} c_I  \langle \Psi_{I}^{(k)} |  \hat{H} \rvert \Psi_M \rangle \right|^2}{E_{\text{SCI}}^{(k)} - \langle \Psi_M | \hat{H} | \Psi_M \rangle} . 
\label{eq:asci_pt2_tripleloop}
\end{equation}
%------------------
Here, the outer and inner loops run over the target space $\mathcal{T}$, while the middle loop generates configurations from the outer index $\mathcal{G}[\mathcal{T}^{(k)}_J]$ restricting them to those not already present in the target space ($M \notin \mathcal{T}^{(k)}$) and not generated in any prior outer index ($M \notin \mathcal{G}[\mathcal{T}^{(k)}_{1:J-1}]$). This eliminates the need to store external configurations, though the operation count scales roughly as $\mathcal{O}(\mathcal{N}_{\mathcal{T}}^2 N_{\text{occ}}^2 N_{\text{vir}}^2)$.

The second approach partitions the external configurations $\mathcal{G}[\mathcal{T}^{(k)}]$ into $N_{\chi}$ non-overlapping batches according to a mapping condition $\chi$ that groups one or multiple configurations to prevent duplicates:
%------------------------
\begin{equation}
E_{\text{PT2}} = \sum_{\chi}^{N_{\chi}} \sum_{\substack{ M \notin \mathcal{T}^{(k)} \\ M \in \mathcal{G}[\mathcal{T}^{(k)}] \\ M \in \chi } }^{\mathcal{N}_{\mathcal{B}}} \frac{ \left| \sum_{\substack{ I }}^{\mathcal{N}_{\mathcal{T}}} c_I  \langle \Psi_{I}^{(k)} |  \hat{H} \rvert \Psi_M \rangle \right|^2}{E_{\text{SCI}}^{(k)} - \langle \Psi_M | \hat{H} | \Psi_M \rangle}   
\label{eq:asci_pt2_batches}
\end{equation}
%------------------------
Such a condition might involve considering all possible sets of excitations that include the occupation of the three highest unoccupied molecular orbitals for a specific quantum species (e.g. $\alpha$) \cite{tubman____2018}, allowing only configurations matching the current constraint $\chi$ to be generated simultaneously for evaluating partial sums in Eq.~\ref{eq:asci_pt2_batches}. 
Alternatively, ranges of 64-bit integer representations of the configurations can be utilized \cite{li_JCP_149_214110_2018}. 
Ideally, only external configurations satisfying these constraints are generated; when predictability is limited, the scaling becomes $\mathcal{O}(\mathcal{N}_{\chi} \mathcal{N}_{\mathcal{T}} N_{\text{occ}}^2 N_{\text{vir}}^2)$. 
Furthermore, both approaches can be combined with stochastic sampling of $\mathcal{G}[\mathcal{T}^{(k)}]$, as done in SHCI \cite{li_JCP_149_214110_2018}.

In our current implementation, we adopt the batch approach using the sum of the orbital indices as the mapping function. 
The range for each batch is defined by sampling a distribution previously computed in the buffer space and applying a transformation from a normal to a uniform distribution to achieve well-balanced batches. 
Although this approach requires generating all external configurations and checking whether they belong to the current branch, thereby introducing some computational overhead, it provides sufficient flexibility to control both memory and execution time by defining the number of batches. 
Consequently, we fix the $\text{top}_{\mathcal{N}_{\mathcal{T}}}$ configurations in the buffer with a large coefficient ($10^6$) so that any generated external duplicate will be merged there, while the remaining $N_{\mathcal{B}} - N_{\mathcal{T}}$ region stores the non-duplicated connected configurations that contribute the most to the target space in Eq.~\ref{eq:asci_pt2_batches}.

Alternatively to the above PT2 energy correction, it is possible to use the sBK dressing method to correct the wave function by solving the eigenvalue problem for the effective Hamiltonian in Eq. \ref{eq:effective_hamiltonian}. 
Similar to the diagonalization in step (iii),  the Davidson iterative method can be employed to avoid explicitly storing the Hamiltonian and dressing matrices by computing matrix-vector products $\mathbf{Hc} = \boldsymbol{\sigma}^{(0)}$, and $\mathbf{\Delta c} = \boldsymbol{\sigma}^{(1)}$. 
For maximum efficiency, the latter product can be computed by constructing all elements of a temporary vector $h_I(M) = \langle \Psi_{I}^{(k)} | \hat{H} \rvert \Psi_M \rangle $ for a given configuration $M$ connected to the target space, and followed by the summation over all products for $M$:
%----------------------
\begin{equation}
\mathbf{\sigma}_{I}^{(1)} = \sum_{\substack{ M \notin \mathcal{T}^{(k)} \\ M \in \mathcal{G}[\mathcal{T}^{(k)}]} }^{\mathcal{N}_{\mathcal{B}}} 
h_{I}(M) \cdot \left(
\sum_{\substack{ J \in \mathcal{T}^{(k)}}}^{\mathcal{N}_{\mathcal{T}}} 
\frac{ h_{J}(M) \cdot c_J  }{E_{\text{SCI}}^{(k)} - \langle \Psi_{M} | \hat{H} \rvert \Psi_M \rangle  }    
\right) .
\label{eq:sbk_davidson}
\end{equation}
%----------------------
Once again, the buffer space $\mathcal{B}$ is utilized, containing the non-duplicated largest configurations connected to the target space. 
In addition to the Hamiltonian shift for the matrix-vector product, the diagonal Hamiltonian vector must be updated to serve as the preconditioner required for iterative diagonalization:
%----------------------
\begin{equation}
\mathbf{D}_{I}^{\text{eff}}  = \mathbf{H}_{II}^{(00)} + \sum_{\substack{ M \notin \mathcal{T}^{(k)} \\ M \in \mathcal{G}[\mathcal{T}^{(k)}]} } 
\frac{ | h_{I}(M) |^2  }{E_{\text{SCI}}^{(k)} - \langle \Psi_{M} | \hat{H} \rvert \Psi_M \rangle  } .
\label{eq:sbk_davidson_diagonal}
\end{equation}
%----------------------
Note that evaluating the diagonal shift does not require storing any partial sums. 
In our current implementation, recomputing the buffer space per batch for each matrix-vector proved to be excessively expensive; therefore, we approximate Eq. \ref{eq:sbk_davidson} by using the buffer configurations generated from an initial PT2 estimation \ref{eq:asci_pt2_batches} on a single batch applied to every matrix-vector product call during the diagonalization. 

A Fortran pseudo-code summarizing these steps is presented in Algorithm~\ref{al:asci}, with further algorithmic specifics provided in the SM. 

\begin{algorithm}
\caption{ ASCI general procedure}
\label{al:asci}
\begin{minted}[escapeinside=||]{fortran}
real :: |$\mathbf{c}(\mathcal{N}_{\mathcal{C}})$|
real :: |$\mathbf{t}(\mathcal{N}_{\mathcal{T}})$|
real :: |$\mathbf{b}(\mathcal{N}_{\mathcal{B}})$|
type(configuration) :: |$\mathbf{c}_{conf}(\mathcal{N}_{\mathcal{C}})$|
type(configuration) :: |$\mathbf{t}_{conf}(\mathcal{N}_{\mathcal{T}})$|
type(configuration) :: |$\mathbf{b}_{conf}(\mathcal{N}_{\mathcal{B}})$|

! Initialize with |$\vert\Psi_{\text{HF}} \rangle$| 
k = 1; |$E_{\text{SCI}}$| = |$E_{\text{Hartree-Fock}}$|; |$ \mathbf{c} (:)$| = 0; |$ \mathbf{c}(1)$| = 1 
while do ( |$\Delta E$| <= 1E-5 ) ! Convergence criteria
    k = k + 1
    ! ASCI amplitudes Eq. |\ref{eq:asci_amplitude}|, Algorithm |\ref{al:asci_amplitudes}|
    call Compute_Amplitudes( |$\mathbf{c}$|, |$\mathbf{b}$|, |$\mathbf{c}_{conf}$|, |$\mathbf{b}_{conf}$| ) 
    ! |$\text{sort}(\mathcal{B}^{(k)})$| and collect OMP, Algorithm |\ref{al:asci_sort_buffer}|
    call Sort_Buffer ( |$\mathbf{b}$|, |$\mathbf{b}_{conf}$| ) 
    ! Form new Target, |$\mathcal{T}^{(k)} = \mathcal{C}^{(k)} \cup \text{top}_{\mathcal{N}_{\mathcal{T}}} (\text{sort}(\mathcal{B}^{(k)}))$| 
    |$\mathbf{t}(:)$| = |$\mathbf{b}(1:\mathcal{N}_{\mathcal{T}})$| ; |$\mathbf{t}_{conf}(:)$| = |$\mathbf{b}_{conf}(1:\mathcal{N}_{\mathcal{T}})$|  
    call Merge_Core_and_Target( |$\mathbf{c}$|, |$\mathbf{c}_{conf}$|, |$\mathbf{t}$|, |$\mathbf{t}_{conf}$| ) 
    ! Solve |$\textbf{Ht}^{(k)} = E_{\text{SCI}}^{(k)}\textbf{t}^{(k)}$|, Eq. |\ref{eq:sci_eigen}|
    call Eigen_Solver( |$\mathbf{t}$|, |$\mathbf{t}_{conf}$|, |$E_{SCI}$|) 
    ! Sort Target, |$\text{top}_{\mathcal{N}_{\mathcal{C}}}(\text{sort}(\mathcal{T}))$| 
    call Quick_Sort_Vector( |$\mathbf{t}$|, |$\mathbf{t}_{conf}$| ) 
    ! Form new Core, |$\mathcal{C}^{(k+1)}$| = |$\text{top}_{\mathcal{N}_{\mathcal{C}}}(\text{sort}(\mathcal{T}))$|
    |$\mathbf{c}(:)$| = |$\mathbf{t}(1:\mathcal{N}_{\mathcal{C}})$| ; |$\mathbf{c}_{conf}(:)$| = |$\mathbf{t}_{conf}(1:\mathcal{N}_{\mathcal{C}})$|  
    ! Increase Target size 
    |$\mathcal{N}_{\mathcal{T}}$| = |$\mathcal{N}_{\mathcal{T}}$| * |$T_{gf}$| 
enddo 
! Generate connected Target configurations
! Eq. |\ref{eq:asci_amplitude} |, Algorithm |\ref{al:asci_amplitudes}|  
call Compute_Amplitudes( |$\mathbf{t}$|, |$\mathbf{b}$|, |$\mathbf{t}_{conf}$|, |$\mathbf{b}_{conf}$| ) 
! |$\text{sort}(\mathcal{B})$| and collect OMP, Algorithm |\ref{al:asci_sort_buffer}|
call Sort_Buffer ( |$\mathbf{b}$|, |$\mathbf{b}_{conf}$| ) 
! PT2 energy correction, Eq. |\ref{eq:asci_pt2_batches}|, Algorithm | \ref{al:asci_pt2_batches}|
call Compute_PT2( |$E_{SCI}$|, |$E_{SCI+PT2}$|, |$\mathbf{t}_{conf}$|, |$\mathbf{b}_{conf}$| ) 
! sBk Hamiltonian correction Eq. |\ref{eq:sbk_davidson}|, Algorithm | \ref{al:asci_sBk}|
call Compute_sBK( |$E_{SCI}$|, |$\mathbf{t}_{conf}$|, |$\mathbf{b}_{conf}$| ) 
\end{minted}
\end{algorithm}

\subsection{Computational details}
Our SCI approach is implemented in the \texttt{openLOWDIN} code \cite{flores-moreno_IJQC_114_50_2014}, using the Jadamilu numerical library for the diagonalization of large sparse matrices\cite{bollhofer_CPC_177_951_2007}. 

For all calculations, the whole cartesian orbital space was employed except for the frozen core per atom ( 2 electrons for Li, C, N, O, and Mg )
Aug-cc-pVXZ electronic basis sets were taken from the Basis Set Exchange (BSE) online database \cite{pritchard_JCIM_59_4814_2019}. 
The positronic aug-cc-pVXZ basis sets used the same exponents and contractions as their corresponding atomic-center counterparts, except for magnesium, where we used the fully uncontracted version. 
For HCN, we employed the PB4-D basis set (4s3p2d) for the quantum hydrogen nuclei \cite{yu_JCP_152_244123_2020}.
Regarding the geometries, the Li-H internuclear distance was 3.015 and 3.348 bohr for LiH and e+LiH systems. 
The O-H internuclear distance was 1.8236 bohr for \ce{OH}, \ce{OH^-}, and \ce{e^+OH^-}.
And for HCN, the hydrogen atom was positioned at the origin, the carbon at 1.0801 \AA, and the nitrogen at 2.2484 \AA.

All calculations were performed on a single node of AION-ULHPC cluster, using 16 OMP cores of a 2x AMD Epyc ROME 7H12 @ 2.6 GHz [64c/280W] CPU, with a total available memory 256 GB DDR4 3200MT/s (8x16 GB DIMMs per socket, 16 DIMMs per node).

\begin{figure}[!ht]
    \centering
    \includegraphics[width=\linewidth]{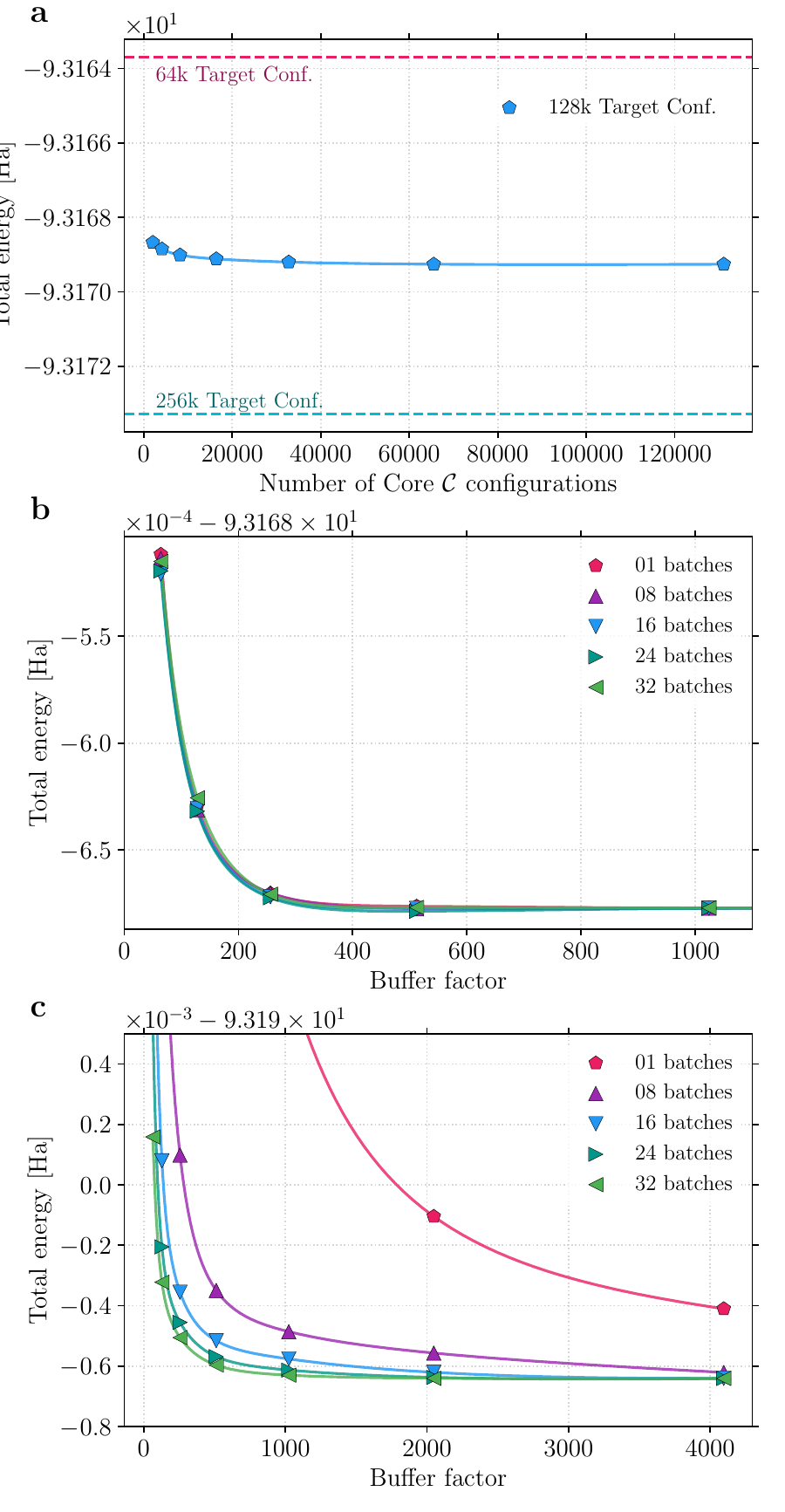}
    \caption{
    Convergence in the HCN molecule of the SCI energy with respect to the core (\textbf{a}) and buffer ((\textbf{b}) space sizes, as well as convergence of the SCI+PT2 energy (\textbf{c}) with buffer space factor $B_f$ and different number of batches for the PT2 partition, Eq. \ref{eq:asci_pt2_batches}. 
    Each SCI calculation considered 128k configurations for the target space. 
    Horizontal lines were added in panel (\textbf{a}) to show the relative energy changes with respect to the target variational space with half and double the number of configurations. 
    }
    \label{fig:core_and_buffer_convergence}
\end{figure}

\subsection{Convergence and algorithm scaling}

As discussed in the previous section, our implementation adopts the batch approach using the sum of the orbital indices as the mapping function, to construct the perturbative correction to the total energy in an efficient parallel way.

To determine the optimal sizes for the core and buffer space, we first performed a convergence analysis on the HCN system with one quantum proton, with a fixed number of target configurations, as depicted in Fig. \ref{fig:core_and_buffer_convergence}.
The results indicate that varying the sizes of the core (Fig. \ref{fig:core_and_buffer_convergence}a) and buffer space (Fig. \ref{fig:core_and_buffer_convergence}b) has a negligible impact on the SCI energy, with fluctuations over 100 times smaller than the energy shift between consecutive ASCI steps with a doubly growing target space. 
In contrast, the PT2 energies exhibit slower convergence with respect to the buffer space size (Fig.~\ref{fig:core_and_buffer_convergence}c); however, this can be accelerated by increasing the number of batches used to partition the PT2 summation in Eq.~\ref{eq:asci_pt2_batches}.

Based on these initial tests, for all SCI calculations in this work we employ a core space $\mathcal{C}$ of 2048 configurations, and a buffer space $\mathcal{B}$ 1024 times larger than the target variational space $\mathcal{T}$. 
The PT2 correction was evaluated in 16 batches according to Eq. \ref{eq:asci_pt2_batches}, while the sBk dressed Hamiltonian was computed on a single batch. 
The SCI variational calculations were divided into $log_2(\mathcal{N}_{\mathcal{T}}/2024)+1$ search, select, and diagonalization steps, starting with an initial $\mathcal{T}$ space of 2048 configurations and doubling it at each subsequent step.

\begin{table}[t]
\centering
\caption{
Computational time and maximum memory occupation for a single SCI-PT2 calculation for each system studied. 
SCI calculations were performed with a variational space of 131072 configurations.
}
\begin{tabular}{cccccccc}
\hline
\multirow{2}{*}{\textbf{System}} & \multirow{2}{*}{\textbf{N$_{e^-}^{orb}$}} & \multirow{2}{*}{\textbf{N$_{x^+}^{orb}$}} & \multirow{2}{*}{\textbf{N$_{e^-}^{act}$}} & \multicolumn{3}{c}{\textbf{Time(s)}}         & \multirow{2}{*}{\textbf{\begin{tabular}[c]{@{}c@{}}Max. \\ Mem. (GB)\end{tabular}}} \\ \cline{5-7}
                                 &                                           &                                           &                                           & \textbf{Total} & \textbf{SCI} & \textbf{PT2} &                                                                                     \\ \hline
e$^+$LiH                         & 158                                       & 154                                       & 2                                         & 8656           & 4099         & 2950         & 18                                                                                  \\
e$^+$OH$^-$                      & 47                                        & 48                                        & 8                                         & 3922           & 1477         & 1940         & 5                                                                                   \\
e$^+$Mg                          & 108                                       & 140                                       & 10                                        & 22967          & 6680         & 15083        & 16                                                                                  \\
HCN                              & 98                                        & 21                                        & 10                                        & 20394          & 4549         & 14945        & 31                                                                                  \\ \hline
\end{tabular}
\label{tab:res:cost}
\end{table}

\begin{figure*}[th]
    \centering
    \includegraphics[width=\linewidth]{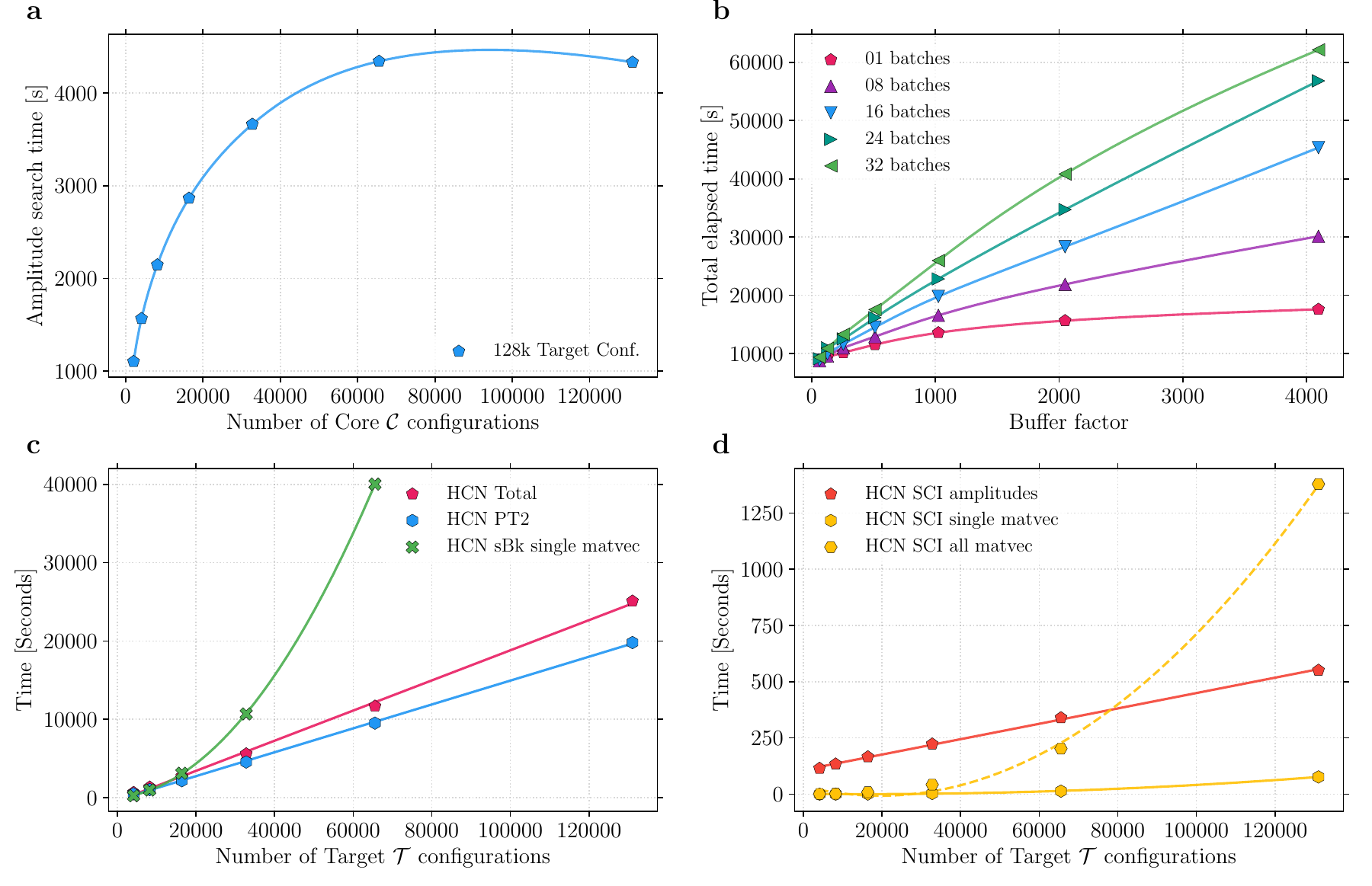}
    \caption{
    Computational time scaling on HCN as a function of the number of configurations.
    Elapsed time for the amplitude computation and selection Eq. \ref{eq:asci_amplitude}  with respect to the number of configurations in the core space (\textbf{a}). 
    Total elapsed time for a SCI+PT2 calculation with respect to the buffer factor space size $B_f$ (\textbf{b}) for different numbers of PT2 batches Eq. \ref{eq:asci_pt2_batches} with 128k target configurations.
    Total elapsed time for the variational and perturbative corrections calculations (\textbf{c}), as well as timing in major steps for the variational task (\textbf{d}) with respect to the number of target configurations. 
    }
    \label{fig:scaling}
\end{figure*}

To evaluate the computational efficiency of our multicomponent SCI implementations, we measured their execution time and memory consumption. 
Table \ref{tab:res:cost} summarizes the total time and peak memory footprint for the largest calculations across all systems using 128k configurations.
As expected, the maximum RAM of our SCI method scales linearly with the number of configurations mainly due to the size of the buffer space $\mathcal{B}$ configuration and amplitude vectors (see Fig. \ref{sm:fig:memory} in SM).
Furthermore, to better understand the effective algorithm scaling, we measured the execution times of the most critical steps in the SCI procedure, using the largest system (HCN) as a representative case (Fig. \ref{fig:scaling}).

For the SCI variational component, the main steps are the search and selection, which scale logarithmically $\mathcal{O}(log(\mathcal{N}_\mathcal{C}))$ with respect to the core space \ref{fig:scaling}a instead of linearly. 
This behavior arises from the exponential decay of the CI wave function coefficient distribution (See section Configurations distributions in SM), revealing one of the main advantages of ASCI approach \cite{tubman_JCP_145_044112_2016}.
It should be noted here that the PT2 step is independent of the core space \ref{eq:asci_pt2_batches}. 
On the other hand, both the variational and perturbative steps depend indirectly on the size of the buffer space $\mathcal{B}$, varying from a linear $\mathcal{O}(\mathcal{N}_\mathcal{B})$ to a logarithmic $\mathcal{O}(log(\mathcal{N}_\mathcal{B}))$ time scaling for high and low numbers of batches, respectively (See \ref{fig:scaling}b).
Thus, the trade-off between memory and time can be chosen by the user varying the buffer size. 

Finally, Fig. \ref{fig:scaling}c shows the scaling with respect to the target space $\mathcal{T}$, revealing that the SCI and SCI+PT2 computational time scales practically linearly, and most of the SCI+PT2 elapsed time is due to the PT2 step, in particular for the construction of the buffer amplitude vectors $\mathcal{O}(\mathcal{N}_\mathcal{T})$. 
On the other hand, the sBk dressed Hamiltonian method scales quadratically, hindering its applicability drastically. 
There, the bottleneck is the computation of the $\langle I | \hat{H} | M \rangle$ terms in Eq. \ref{eq:sbk_davidson}, leading to a total of $\mathcal{N}_\mathcal{B} \mathcal{N}_\mathcal{T}$ Hamiltonian elements per matrix-vector call required by the diagonalizer ( usually 15 calls ), leading to $\mathcal{O}(15\mathcal{N}_\mathcal{T}\mathcal{N}_\mathcal{B})$.
Although, by taking a closer look at the variational steps in Fig. \ref{fig:scaling}d, 
it's evident that in true the matrix diagonalization scales quadratically $\mathcal{O}(\mathcal{N}^2_\mathcal{T})$ per matrix-vector call, thus approximately $\mathcal{O}(15\mathcal{N}^2_\mathcal{T})$ for the whole run, being the major bottleneck for larger variational spaces.
In these cases, the cost of the matrix diagonalization was overshadowed mostly by the PT2 step and partially also by amplitude construction and selection, which scale linearly with the target space $\mathcal{O}(\mathcal{N}_\mathcal{T})$, Fig. \ref{fig:scaling}d. 

\section{Results}

\subsection{Electron-positron systems}

As prototype systems for electron-positron interaction we consider the three largest ones tackled in the previous MCCI investigation\cite{coe_CPL_645_106_2016}:
positronic lithium hydride \ce{e^+LiH}, where the HF reference is a bound state; positronic magnesium atom\ce{e^+Mg}, for which the HF reference is unbound; and positronic hydroxyl anion \ce{e^+OH^-}, also known as \ce{PsOH}, characterized for correctly binding a positron but failing to bind positronium to hydroxyl at HF level.  

%---------------------------------------------------------------------------
\subsubsection{Positronic lithium hydride}

The $e^+$LiH molecule is perhaps the most studied positronic molecule\cite{kurtz_JPBAMP_11_L479_1978, mella_CP_113_6154_2000, strasburger_JCP_114_615_2001, mella_JCP_116_2852_2002, bubin_JCP_120_6051_2004, buenker_JPCA_109_5956_2005, gianturco_PR-AMOP_73_1_2006, sirjoosingh_JCP_139_034103_2013, yamada_EPJD_68_1_2014, coe_CPL_645_106_2016, hofierka_N_606_688_2022, cassella_NC_15_5214_2024}.
With only two valence electrons and a single positron, it serves as an ideal benchmark for examining electron–electron and electron–positron correlation effects. 
Moreover, as a neutral diatomic molecule with a substantial dipole moment ($\sim 6$ Debye), it generates a strong electrostatic potential that is captured even by mean-field approximations such as HF. 
Nevertheless, the resulting HF binding energy is an order of magnitude smaller than that obtained from explicitly correlated methods. 

\begin{table}[th!]
\caption{
Total energies and positron affinities energies for  LiH and \ce{e^+LiH}.
MCCI and SCI results employed the aug-cc-pVQZ basis for electrons and positrons, with one electronic frozen orbital. 
For all methods, the Li-H internuclear distance was 3.015 and 3.348 bohr for LiH and \ce{e^+LiH} systems, respectively.
}\label{tab:resuls:e+lih}
\begin{tabular}{llll}
\hline
\textbf{Method} & \textbf{LiH E [Ha]} & \textbf{\ce{e^+ LiH} E [Ha]} & \textbf{PA [meV]} \\ \hline
HF              & -7.987250           & -7.991137                    & 105.8             \\
MCCI$^a$        & -8.024141           & -8.052732                    & 778.0             \\
HF/SCI[32k]     & -8.024226           & -8.052801                    & 777.6             \\
HF/SCI[128k]    & -8.024226           & -8.054224                    & 816.3             \\
HF/SCI[128k]+PT2& -8.024226           & -8.054231                    & 816.5             \\
FN-DMC$^b$      & -8.070045(38)       & -8.10718(11)                 & 1010(3)           \\
ECG$^c$         & -8.070538           & -8.107474                    & 1005.1            \\ \hline
\end{tabular}
\\ $^a$ Monte Carlo Configuration Interaction  \citenum{coe_CPL_645_106_2016} with 35093 configurations. 
$^b$ Diffusion Monte Carlo \citenum{kita_JCP_131_134310_2009} extrapolated to zero time-step.
$^c$ Explicitly Correlated Gaussians\citenum{strasburger_JCP_114_615_2001}.
\end{table}

\begin{figure}[tbh]
    \centering
    \includegraphics[width=0.995\linewidth]{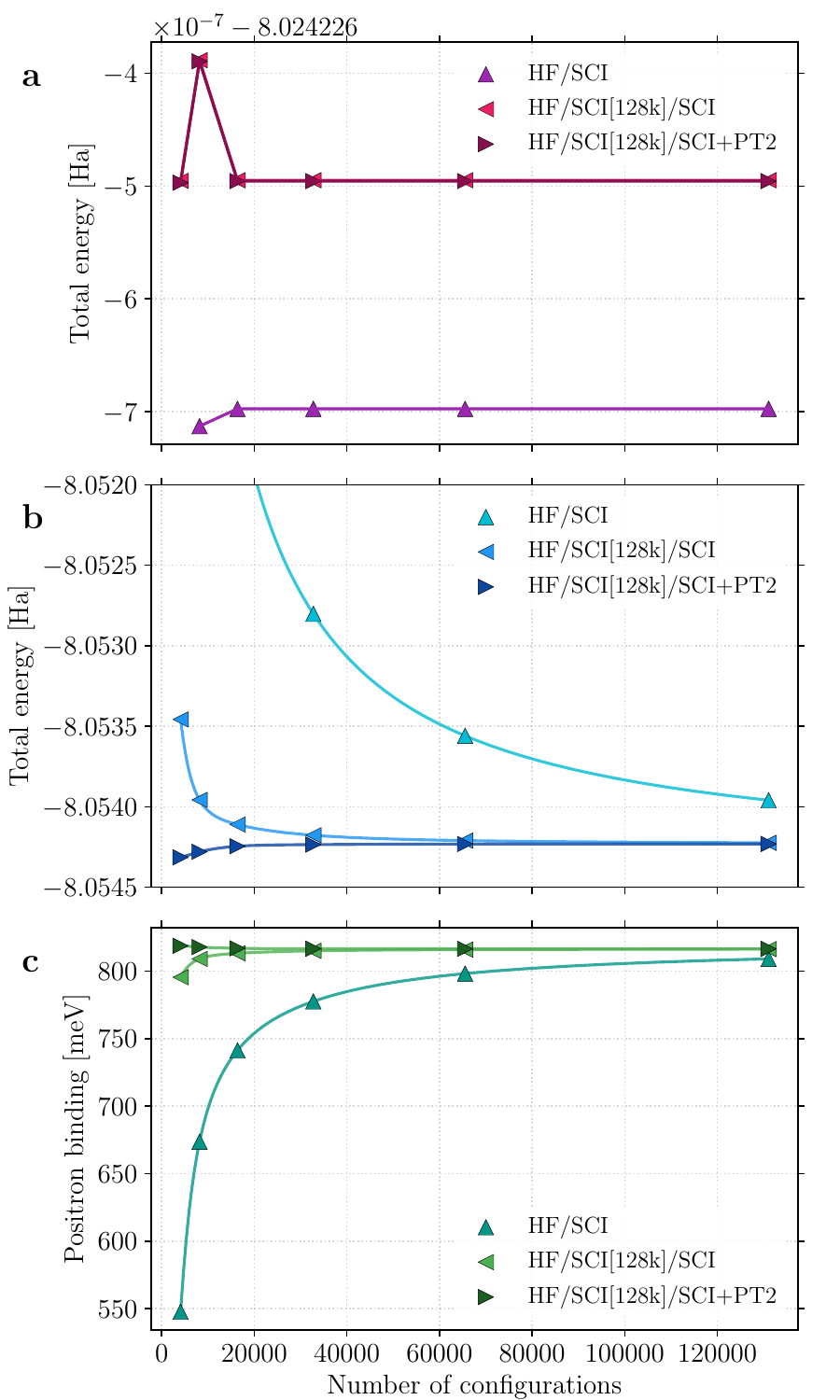}
    \caption{
    Convergence in positronic lithium hydride \ce{e^+LiH} for the SCI and SCI+PT2 energies as a function of the size of the variational space.
    Total energies for the electronic system (\textbf{a}) and the electron-positron (\textbf{b}), as well as positron binding energies (\textbf{c}) computed as $E_{\ce{A}} - E_{\ce{e^+A}}$.
    The slash symbol ($/$) is used to indicate the natural orbital updated between two calculations. 
    For restarts from a previous SCI calculation, the number inside square brackets indicates the size of the variational space.
    }
    \label{fig:results:e+lih}
\end{figure}

Rather than exhaustively revisiting this system, the present work aims to compare our SCI implementation against the MCCI approach \cite{coe_CPL_645_106_2016}, by employing the same molecular geometry and basis set.
Table \ref{tab:resuls:e+lih} compares these two approaches alongside reference values from Diffusion Monte Carlo (DMC) \cite{kita_JCP_131_134310_2009} and explicitly correlated Gaussian (ECG) \cite{strasburger_JCP_114_615_2001} calculations evaluated at the same internuclear distance. 
For a comparable variational space of roughly $32\text{k}$ configurations, our binding energy is only $0.4\text{ meV}$ lower than the MCCI result — a minor discrepancy probably attributable to slight differences in basis set exponents, selection convergence criteria, or the use of Cartesian orbitals in the \texttt{OpenLOWDIN} code rather than spherical harmonics. 
In both cases, selected configuration interaction reduces the binding energy error from 900 meV ($90\%$) down to 230 meV ($23\%$).
We further examine the convergence of the SCI variational energy and its second-order perturbative correction (PT2) with respect to the variational space size in Fig. \ref{fig:results:e+lih}. 
Because neutral LiH is strongly single-reference in character, its total energy converges rapidly (Fig. \ref{fig:results:e+lih}a). 
Conversely, \ce{e^+LiH} converges markedly more slowly, showing incomplete convergence even at 32k configurations (Fig. \ref{fig:results:e+lih}b). 
To accelerate convergence, we perform a subsequent SCI calculation in the basis of natural orbitals (see Supplemental Material) obtained from an initial SCI wavefunction, which is a well-established strategy in SCI methodologies\cite{illas_JCP_89_6376_1988}. 
Using natural orbitals yields rapid convergence even with a modest variational space of 4k configurations, recovering a binding energy of 795 meV, and reaches 816.3 meV at 128k configurations, which is within 0.2 meV of the extrapolated SCI+PT2 estimate (Fig. \ref{fig:results:e+lih}c). 
As expected, the PT2-corrected energies converge even faster toward the same full-CI limit, approaching it from below due to the non-variational nature of perturbation theory.

%-------------------------------------
\begin{table*}[tbh]
\centering
\caption{
Total energies and positron affinities for \ce{OH}, \ce{OH^-} and \ce{e^+OH^-}.
SCI results employed the D-aug-cc-pVDZ basis for electrons and positrons, with one electronic frozen orbital. 
The \ce{O-H} internuclear distance was 1.8236 bohr for all systems.
}\label{tab:results:e+oh-}
\begin{tabular}{lllllll}
\hline
\multicolumn{1}{c}{}                & \multicolumn{3}{c}{\textbf{Energy [Ha]}}                                                                                   & \multicolumn{1}{c}{}                    & \multicolumn{1}{c}{}                    & \multicolumn{1}{c}{}                  \\ \cline{2-4}
\multicolumn{1}{c}{\textbf{Method}} & \multicolumn{1}{c}{\textbf{\ce{OH}}} & \multicolumn{1}{c}{\textbf{\ce{OH^-}}} & \multicolumn{1}{c}{\textbf{\ce{e^+ OH^-}}} & \multicolumn{1}{c}{\textbf{PoBE [eV]}} & \multicolumn{1}{c}{\textbf{PsBE [eV]}} & \multicolumn{1}{c}{\textbf{EA [eV]}} \\ \hline
HF                                  & -75.404653                           & -75.396542                             & -75.571774                                 & 4.768                                   & -2.255                                  & -0.221                                \\
MCCI$^a$                            & -75.580576                           & -75.637845                             & -75.826953                                 & 5.146                                   & -0.099                                  & 1.558                                 \\
MCCI$^b$                            & -75.433741                           & -75.502055                             & -75.701751                                 & 5.434                                   & 0.490                                   & 1.859                                 \\
HF/SCI                              & -75.585863                           & -75.641111                             & -75.836829                                 & 5.326                                   & 0.026                                   & 1.503                                 \\
HF/SCI+PT2                          & -75.588184                           & -75.649002                             & -75.851357                                 & 5.506                                   & 0.358                                   & 1.655                                 \\
SCI[128k]/SCI                                 & -75.588368                           & -75.648642                             & -75.848082                                 & 5.427                                   & 0.264                                   & 1.640                                 \\
SCI[128k]/SCI+PT2                             & -75.588533                           & -75.650087                             & -75.854126                                 & 5.552                                   & 0.424                                   & 1.675                                 \\
FN-DMC$^c$                          & -75.6286(37)                         & -75.6973(38)                           & -75.9024(42)                               & 5.58(15)                                & 0.65(15)                                & 1.87(14)                              \\
FN-DMC$^d$                          & -75.7213(8)                          & -75.7874(4)                            & -75.9815(9)                                & 5.58(3)                                 & 0.28(3)                                 & 1.80(2)                               \\ \hline
\end{tabular} \\
 $^{a,b}$ Monte Carlo Configuration Interaction \citenum{coe_CPL_645_106_2016}. $^a$ One doubly occupied frozen electronic orbital, D-aug-cc-pVDZ basis for electrons and positrons. $^b$ Three doubly occupied frozen electronic orbital, aug-cc-pVDZ+Ryd basis for electrons and positrons.
 $^c$ Diffusion Monte Carlo \citenum{yoshida_PRA_54_964_1996} with a time-step of 0.0025 ha$^{-1}$ using an \ce{O-H} distance of 0.970 \AA and 0.964 \AA. for both \ce{OH^-} and \ce{e^+OH^-}. 
 $^d$ Diffusion Monte Carlo \citenum{bressanini_JCP_109_5931_1998} with a time-step of 0.0005 ha$^{-1}$ using an \ce{O-H} distance of 1.833 a.u for \ce{OH}, \ce{OH^-} and \ce{e^+OH^-}. 

\end{table*}
%-------------------------------------

%-------------------------------------
\begin{figure}[tbh]
    \centering
    \includegraphics[width=0.995\linewidth]{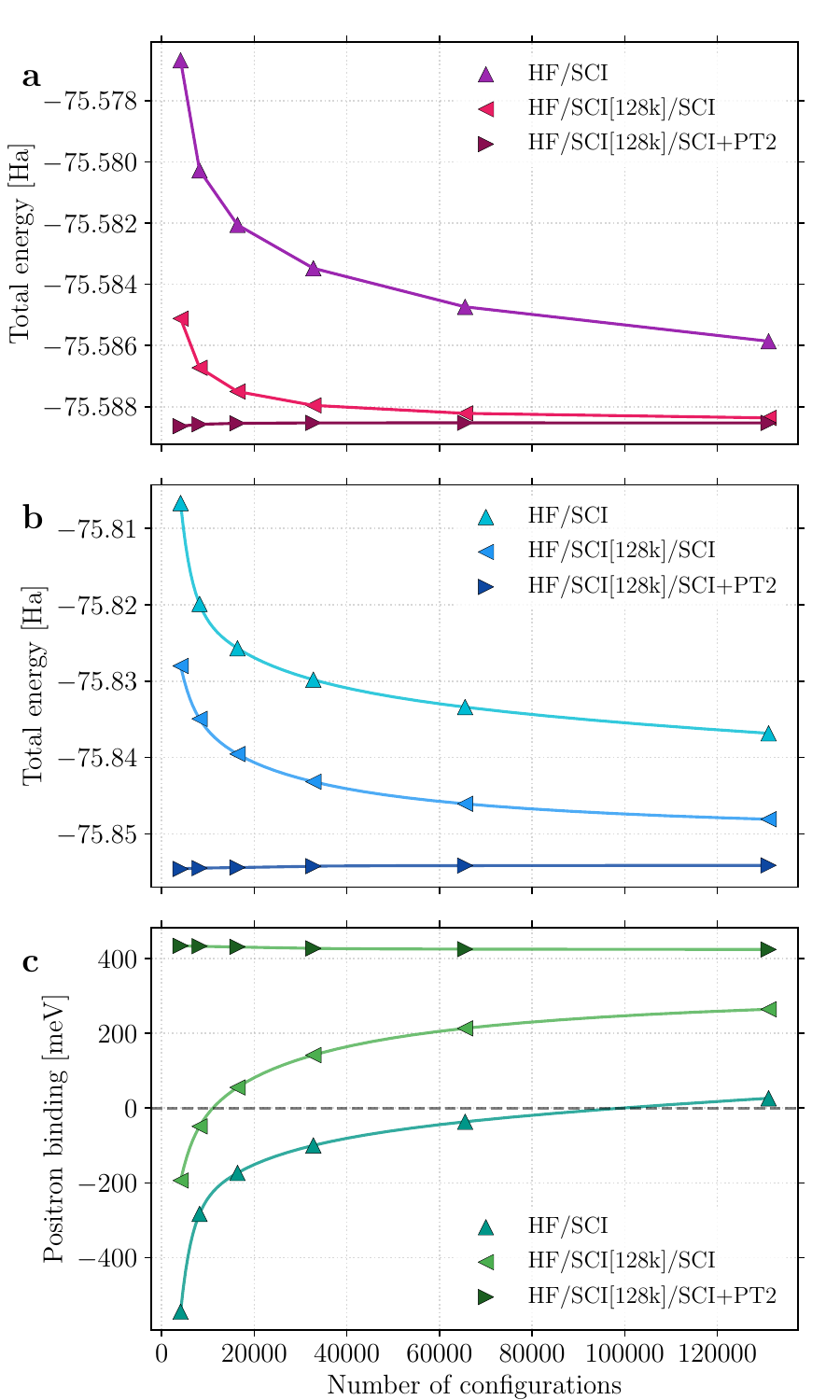}
    \caption{
    Convergence in positronic hydroxyl anion \ce{e^+OH^-} for the SCI and SCI+PT2 energies as a function of the size of the variational space.
    Total energies for the electronic system OH (\textbf{a}) and the electron-positron \ce{e^+OH^-} (\textbf{b}), as well as positronium binding energies (\textbf{c}) computed as $E_{\ce{A}} +E_{\ce{Ps}}- E_{\ce{e^+A^-}}$.
    The slash symbol ($/$) is used to indicate the natural orbital passing between two calculations. 
    For restarts from a previous SCI calculation, the number inside square brackets indicates the size of the variational space.
    }
    \label{fig:results:e+oh-}
\end{figure}
%-------------------------------------
 
%---------------------------------------------------------------------------
\subsubsection{Positronic hydroxyl anion}

The next system under study is \ce{e+OH^-}, in which the positron strongly binds to the charged \ce{OH^-} fragment due to the intense electrostatic attraction to the negatively charged oxygen atom. 
However, the lowest energy dissociation channel of \ce{e+OH^-}, rather than that of the detachment of the positron from the hydroxide ion (\ce{OH^-}), corresponds to the electron capture that forms a positronium `atom' (\ce{e^-e^+}) and a neutral hydroxyl radical (\ce{OH}). 
Consequently, Table \ref{tab:results:e+oh-} reports both the positron binding energy (PoBE) and the positronium binding energy (PsBE), defined respectively as $E_{\ce{A^-}} - E_{\ce{e^+A^-}}$ and $E_{\ce{A}} + E_{\ce{Ps}} - E_{\ce{e^+A^-}}$, alongside the total energies of all participating species.
As expected, HF qualitatively captures positron binding energy (PoBE$>0$)but erroneously predicts the system to be unbound with respect to positronium detachment (PsBE $<0$). 
Similarly, MCCI calculations \cite{coe_CPL_645_106_2016} yield an unbound state when using 113781 configurations. 
To remedy this, Coe and Paterson \cite{coe_CPL_645_106_2016} performed an additional calculation with 49633 configurations by freezing three orbitals (correlating $6$ of the $10$ electrons rather than $8$) and augmenting the basis set with oxygen-centered Rydberg functions to improve the electron affinity, thereby recovering a positive PsBE. 
In contrast, using $128\text{k}$ configurations, our SCI approach achieves a positive PsBE of 0.026 eV while freezing only a single core orbital and without modifying the basis set. 
Restarting the SCI calculation in the basis of natural orbitals generated from this initial wave function further improves the PsBE to 0.264 eV.
As shown in Fig. \ref{fig:results:e+oh-}, the energy convergence for this system is markedly slower than what seen for \ce{e^+LiH}. 
Specifically, the SCI total energy of neutral \ce{OH} reaches the SCI+PT2 limit only after the natural orbital restart, whereas \ce{e^+OH^-} remains unconverged, indicating that an even larger variational space is required. 
Once again, the PT2 correction effectively bridges this remaining correlation gap relative to DMC benchmarks, yielding a nearly constant extrapolated energy across all variational space sizes within the constraints of the modest basis set used here.

%---------------------------------------------------------------------------
\subsubsection{Positronic magnesium atom}

The final positronic system examined in this work is positron binding to a magnesium atom, \ce{e^+Mg}. 
This case is substantially more challenging than the preceding systems because the HF reference state is completely unbound; in the self-consistent field (SCF) solution, the positronic orbital delocalizes as much as the basis set permits to minimize nuclear repulsion and positron kinetic energy. 
Consequently, post-HF methods converge extremely slowly for such non-polar systems with respect to basis set size, CI expansion, and perturbation order\cite{romero_JCP_141_114103_2014, coe_CPL_645_106_2016, upadhyay_JCTC_20_9879_2024, cassella_NC_15_5214_2024}. 
%---------------------------------------------------------------------------
\begin{table*}[tbh]
\centering
\caption{
Total energies and positron affinities for \ce{Mg}, and \ce{e^+Mg}.
SCI results employed the aug-cc-pVQZ basis for electrons and positrons, with one electronic frozen orbital. 
}
\begin{tabular}{lllr}
\hline
\multicolumn{1}{c}{}                & \multicolumn{2}{c}{\textbf{Energy [Ha]}}                                        & \multicolumn{1}{c}{}                    \\
\multicolumn{1}{c}{\textbf{Method}} & \multicolumn{1}{c}{\textbf{Mg}} & \multicolumn{1}{c}{\textbf{\ce{e^+ Mg}}} & \multicolumn{1}{c}{\textbf{PoBE [meV]}} \\ \hline
HF                                  & -199.614244                          & -199.599145                              & -410.9                                  \\
MCCI$^a$                            & -199.6777                            & -199.6594                                & -498                                    \\
CISD$^{ep}$                         & -199.614244                          & -199.629397                              & 412.3                                   \\
CISD$^{ep}\ ^b$                      &                                      &                                          & 386.0                                   \\
CISDT$^{eep}\ ^b$                    &                                      &                                          & 167.8                                   \\
HF/SCI                              & -199.768297                          & -199.755326                              & -353.0                                  \\
HF/SCI[128k]SCI                     & -199.770976                          & -199.768885                              & -56.9                                   \\
CISD$^{ep}$/SCI                     & -199.768297                          & -199.769149                              & 23.2                                    \\
CISD$^{ep}$/SCI[128k]/SCI           & -199.770976                          & -199.773419                              & 66.5                                    \\
CISD$^{ep}$/SCI[128k]/SCI+PT2       & -199.771383                          & -199.777417                              & 164.2                                   \\
FCI$^c$                             &                                      &                                          & 463.7                                   \\
FCSVM$^c$                           &                                      &                                          & 460.7                                   \\ 
DMC(SD)$^b$                         &                                      &                                          & 90(27)                                   \\ 
DMC(NO CISD$^{ep}$)$^b$             &                                      &                                          & 471(14)                                   \\ \hline
\end{tabular}\\
$^a$ Monte Carlo Configuration Interaction with aug-cc-pVQZ basis, one frozen core electronic orbitals from \citenum{coe_CPL_645_106_2016}. 
$^b$ CI and Diffusion Monte Carlo results from \citenum{upadhyay_JCTC_20_9879_2024}, the superscript $ep$ indicates the double excitations are restricted up to electron-positron and $eep$ to triples electron-electron-positron, and NO denotes the use of natural orbitals as the DMC trial wavefunction. 
$^c$ Extrapolated FCI and fixed-core stochastic variational method (FCSVM) from reference \citenum{bromley_PRA_73_032507_2006} employed a two-valence-electron and one-positron Hamiltonian model with a polarization potential. 
\label{tab:results:e+mg}
\end{table*}
\begin{figure}[thb]
    \centering
    \includegraphics[width=0.995\linewidth]{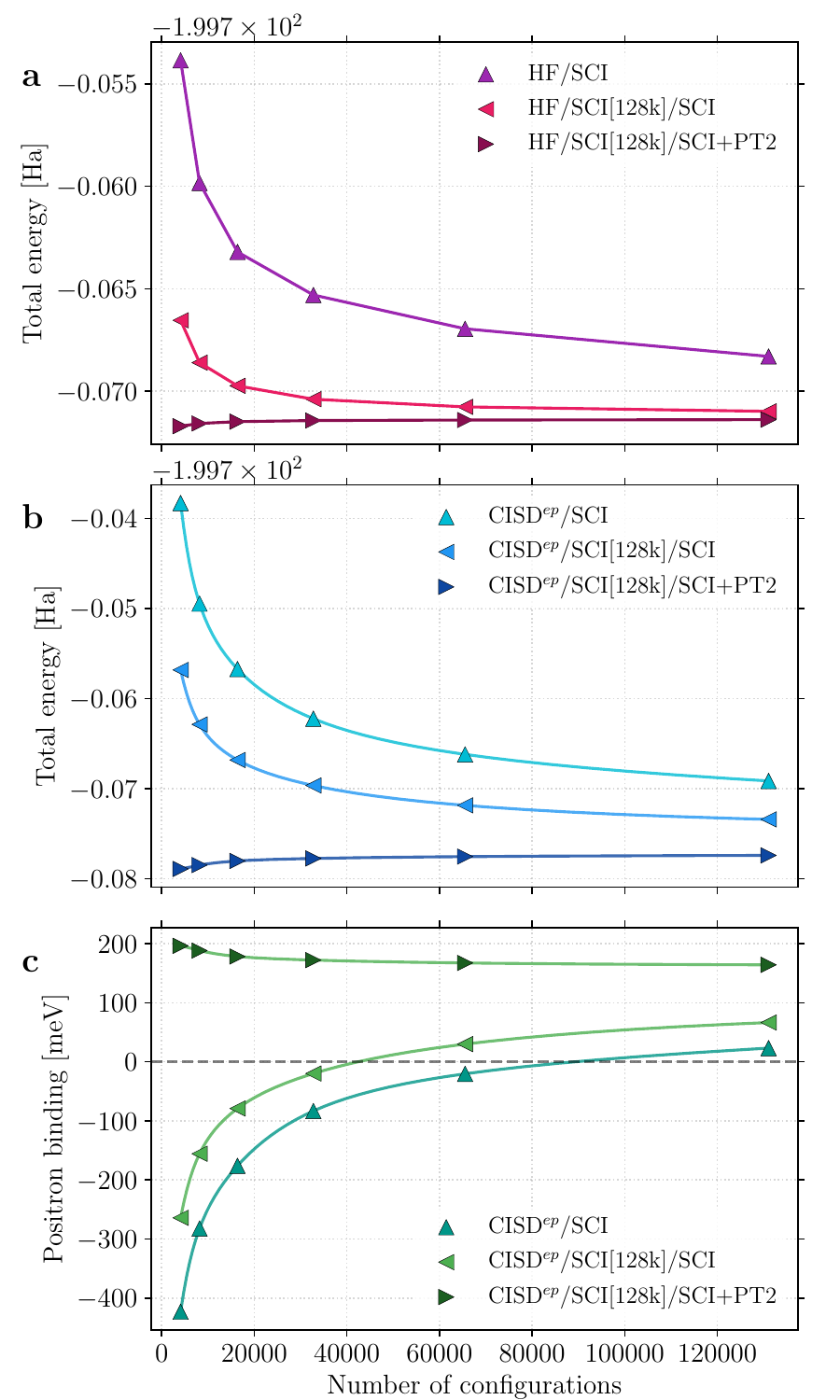}
    \caption{Convergence in positronic magnesium atom \ce{e^+Mg} for the SCI and SCI+PT2 energies as a function of the size of the variational space.
    Total energies for the electronic system (\textbf{a}) and the electron-positron (\textbf{b}), as well as positron binding energies (\textbf{c}) computed as $E_{\ce{A}} - E_{\ce{e^+A}}$.
    The slash symbol ($/$) is used to indicate the natural orbital passing between two calculations. 
    For restarts from a previous SCI calculation, the number inside square brackets indicates the size of the variational space.
    }
    \label{fig:results:e+mg}
\end{figure}
%---------------------------------------------------------------------------
As a result, both the MCCI calculations of Coe and Paterson\cite{coe_CPL_645_106_2016} and our initial SCI calculations, whether initialized from HF orbitals or standard SCI natural orbitals, yield negative PoBEs, as summarized in Table \ref{tab:results:e+mg}.

However, prior studies have demonstrated that CI singles and doubles restricted to interspecies electron–positron excitations, denoted here as $\text{CISD}^{ep}$, can yield remarkably accurate PoBEs by neglecting purely electronic correlation, often agreeing closely with benchmark theoretical and experimental values \cite{upadhyay_JCTC_20_9879_2024, yoshida_EPJD_78_10_2024}. 

Specifically, for \ce{e^+Mg}, Upadhyay et al. \cite{upadhyay_JCTC_20_9879_2024} reported a PoBE of 386 meV using $\text{CISD}^{ep}$, a result we reproduce with our implementation (see Table \ref{tab:results:e+mg}). 
Although reintroducing electronic correlation via $\text{CISDT}^{eep}$ (including up to electron–electron–positron triple excitations) reduces the binding energy to 168 meV, a $\text{CISD}^{ep}$ natural orbital wave function serves as an effective trial state for DMC, yielding a PoBE of 471(14) meV and a substantially lower variational energy than single-determinant DMC (90(27) meV). 
For comparison, 460 meV has long served as the benchmark PoBE based on FCI and fixed-core stochastic variational method (FCSVM) calculations using a two-valence-electron, one-positron model Hamiltonian with a core-polarization potential \cite{bromley_PRA_73_032507_2006}, a value recently also confirmed by Upadhyay et al. \cite{upadhyay_JCTC_20_9879_2024} via CI-QMC.

Motivated by these findings, we initiate our SCI procedure from a $\text{CISD}^{ep}$ wave function to mitigate the slow convergence of the positronic state. 
This $\text{CISD}^{ep}$/SCI approach recovers a positive PoBE of 23 meV, which increases to 66 meV upon a second natural orbital restart and reaches 164 meV with the addition of PT2 correction (Fig. \ref{fig:results:e+mg}).
The remaining discrepancy relative to the 460 meV reference is primarily attributable to basis set incompleteness and the lack of exponents optimized specifically for \ce{e^+Mg}.
It is also possible that previously reported positron bindings underestimated the energy in Mg atom, thus increasing the PoBE energy.

The origin of this slow convergence is further clarified by comparing the summed squared CI weights across excitation classes for all three positronic systems (see Fig. S1 in the Supplemental Material). 
Specifically, in \ce{e^+Mg}, mixed electron–positron excitations carry a larger cumulative weight than their purely electronic counterparts, whereas the opposite trend holds for \ce{e^+LiH} and \ce{e^+OH^-}. 
Consequently, the SCI expansion must incorporate a much larger number of positronic single excitations to adequately describe the strongly multi-determinantal positronic wave function, which introduces an imbalance between the descriptions of the neutral atom and the positron–atom complex.

Nevertheless, it should be noted that converging the magnesium PoBE to the complete basis set limit lies beyond the scope of the present methodological comparison, which focuses on benchmarking our SCI implementation and identifying the remaining bottlenecks in SCI approaches.

\subsection{Quantum Nuclei}

\subsubsection{Hydrogen cyanide}

For quantum nuclear systems, we selected hydrogen cyanide (\ce{HCN}) with a quantum proton, as it is among the most extensively studied benchmarks for both the multicomponent SHCI method \cite{yu_JCP_152_244123_2020, brorsen_JCTC_16_2379_2020, alaal_JCP_155_234107_2021, fajen_JCTC_17_965_2021, malbon_JCTC_21_3968_2025} and orbital-based multicomponent approaches in general. 
Here, the primary challenge is to recover physically accurate ground-state protonic densities. 
For quantum protons, it is well established that mean-field HF yields overly localized protonic densities \cite{reyes_IJQC_119_1_2019, pavosevic_JCTC_15_338_2019, schiebel_JCP_165_114302_2026} relative to benchmark explicitly correlated Gaussian (ECG) methods \cite{hoshino_JCP_135_24111_2011, matyus_JCP_137_0_2012, sirjoosingh_JCP_139_034102_2013, bubin_JCP_152_204102_2020}. 
Among \textit{ab initio} multicomponent wave-function methods, coupled cluster with singles and doubles (CCSD) has proven to be the most robust for predicting proton densities, whereas standard CISD offers little improvement over HF \cite{pavosevic_JCTC_15_338_2019, yu_JCP_152_244123_2020}, although recent $\text{MCSCF}/\text{CISD}^{ep}$ applied to smaller active spaces \cite{malbon_JCTC_21_3968_2025} still seem to qualitatively preserve accurate protonic densities.
On the other hand, the SHCI method \cite{li_JCP_149_214110_2018} has been successfully extended to treat \ce{HCN}. 
Brorsen et al. \cite{brorsen_JCTC_16_2379_2020} demonstrated that SHCI can approach Fourier grid Hamiltonian (FGH) densities, a numerically exact benchmark for a proton moving on an electronic potential energy surface, comparable in accuracy to ECG, using roughly 17 million configurations with the CI space truncated at quadruple excitations (CISDTQ) combined this with CASSCF for additional orbital relaxation via SCI(CISDTQ)/CASSCF \cite{fajen_JCTC_17_965_2021}. 
Subsequent studies have explored truncations up to triple electronic excitations coupled with single protonic excitations \cite{alaal_JCP_155_234107_2021}. 

%------------------------
\begin{figure}[t]
    \centering
    \includegraphics[width=0.995\linewidth]{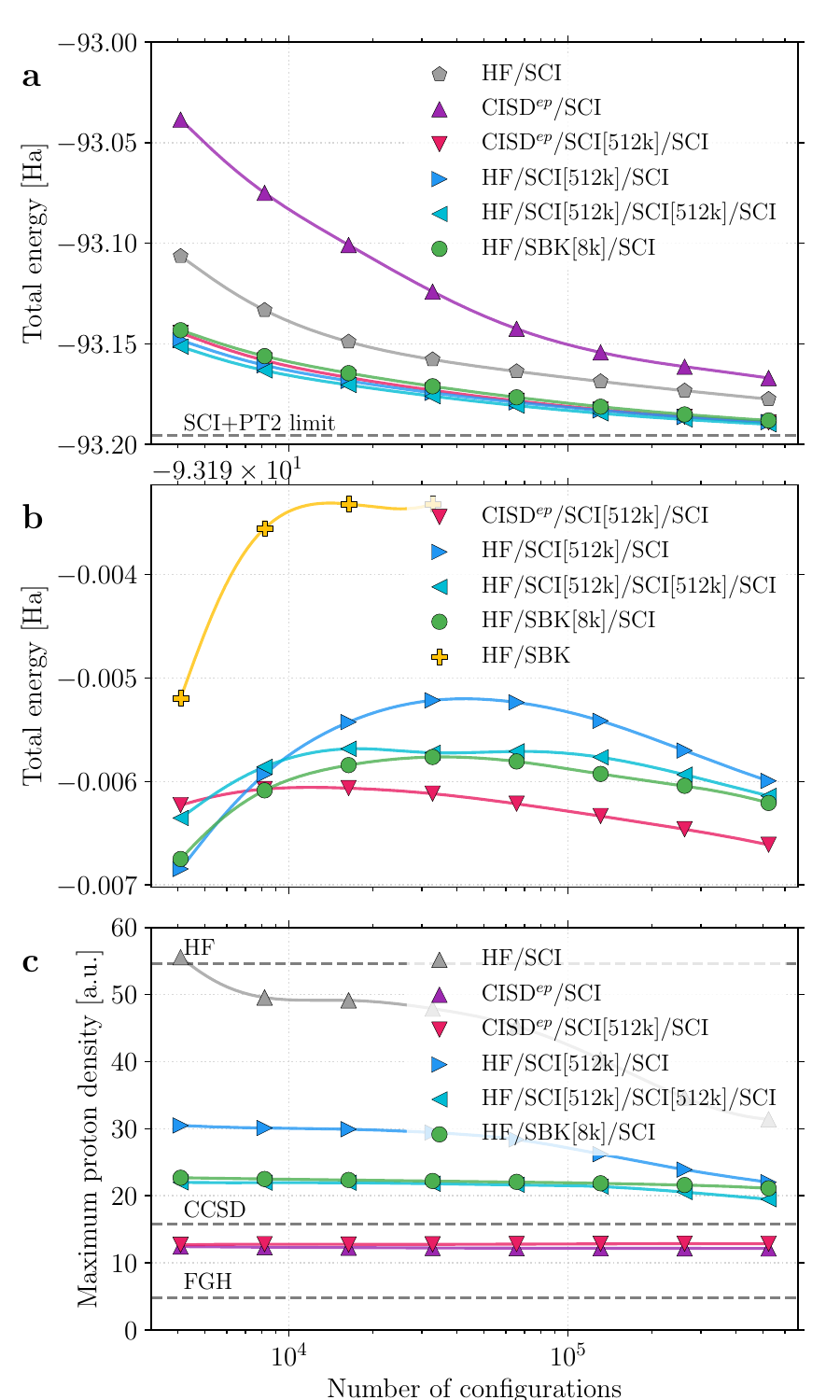}
    \caption{Convergence in \ce{HCN} with quantum hydrogen nuclei for the SCI (\textbf{a}) and SCI+PT2 (\textbf{b}) energies, as well as maximum proton density (\textbf{c}) as a function of the size of the variational space.
    The slash symbol ($/$) is used to indicate the natural orbital passing between two calculations. 
    For restarts from a previous SCI calculation, the number inside square brackets indicates the size of the variational space.
    (Data for CCSD\cite{yu_JCP_152_244123_2020}, FGH\cite{yu_JCP_152_244123_2020} and HCI\cite{brorsen_JCTC_16_2379_2020} using the WebPlotDigitizer online tool \cite{WebPlotDigitizer})
    }
    \label{fig:results:hcn}
\end{figure}
%------------------------

%------------------------
\begin{figure}[t]
    \centering
    \includegraphics[width=1.047\linewidth]{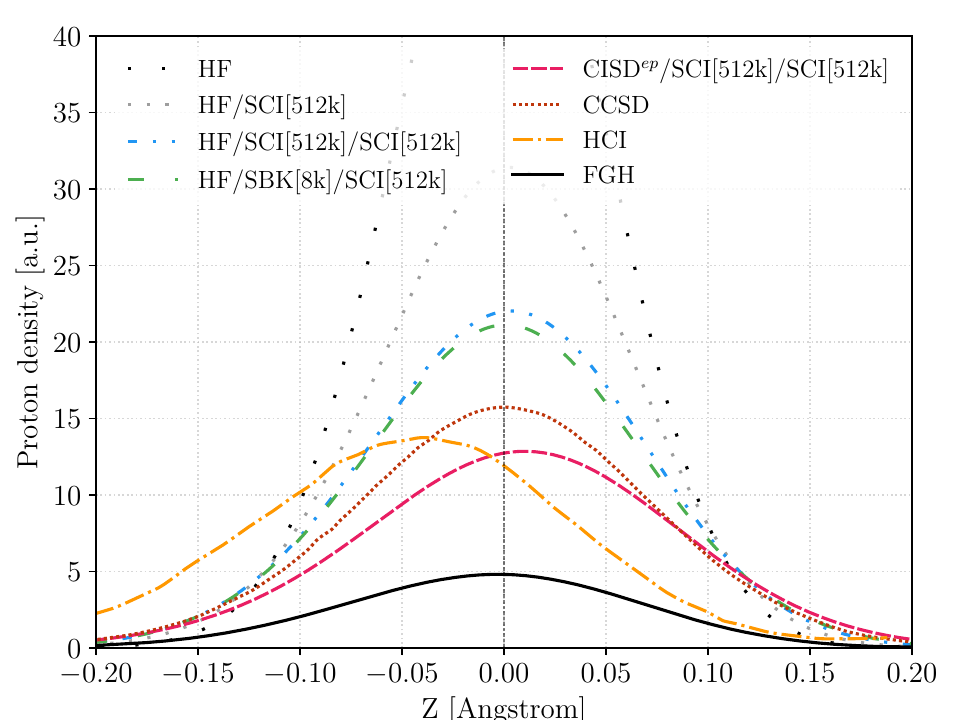}
    \caption{
    On-axis one-dimensional quantum proton density for HCN molecule computed with different methods. 
    The slash symbol ($/$) is used to indicate the natural orbital passing between two calculations. 
    For restarts from a previous SCI calculation, the number inside square brackets indicates the size of the variational space.
    (Data for CCSD\cite{yu_JCP_152_244123_2020}, FGH\cite{yu_JCP_152_244123_2020} and HCI\cite{brorsen_JCTC_16_2379_2020} using the WebPlotDigitizer online tool \cite{WebPlotDigitizer})
    }
    \label{fig:results:hcn_densities}
\end{figure}
%------------------------

In contrast, the present work evaluates our multicomponent SCI implementation without imposing any predefined excitation-rank truncation on the CI expansion.

Accordingly, we examined the effect of three different sets of starting orbitals on the convergence of the SCI variational energy (Fig. \ref{fig:results:hcn}a), the perturbative energy corrections (Fig. \ref{fig:results:hcn}b), and the maximum height of the protonic density (Fig. \ref{fig:results:hcn}c), which serves as a simple proxy for proton delocalization.

In the first approach, the calculation is initialized from Hartree–Fock orbitals (denoted HF/SCI). 
While this scheme exhibits the second-fastest energy convergence, it suffers from the slowest convergence of the maximum protonic density. 
Performing a subsequent natural orbital restart (HF/SCI/SCI) shifts both the energy and the density closer to the reference benchmark using a variational space four times smaller; however, the convergence rate plateaus at small variational space sizes before accelerating again at larger configuration counts. 
A second natural orbital restart yields no appreciable gain in variational energy, although the density continues to move toward the reference value just as in the preceding step. 
Across all of these restarts, the perturbatively corrected SCI+PT2 energy remains nearly constant relative to the variation observed in the unperturbed SCI energies.

The second approach initializes the calculation using natural orbitals generated from an initial SCI run dressed with the diagonal shift correction of Eq. \ref{eq:sbk_davidson} (denoted HF/sBk/SCI). 
With this protocol, the HF/sBk/SCI and HF/sBk/SCI+PT2 energies are comparable to their respective HF/SCI/SCI and HF/SCI/SCI+PT2 counterparts, while the resulting protonic densities lie noticeably closer to the FGH reference even in compact variational spaces. 
Nevertheless, this strategy presents a clear drawback: when the variational space is overly small (e.g., 4k configurations), the dressed sBk densities appear unphysical, whereas at very large variational spaces they revert closely to the HF/SCI results, diminishing the advantage. 
Note also that the eigenvalue of the effective Hamiltonian in Eq. \ref{eq:sbk} (the HF/sBk eigenvalue) lies slightly above the SCI+PT2 energies due to approximations in the buffer space and batch count (Fig. \ref{fig:results:hcn}b). 
Furthermore, constructing the sBk wave function becomes computationally prohibitive for large target spaces, as discussed later in this manuscript. 
Even so, restarting from an 8k-configuration sBk wave function reproduces the energies and densities of a double-restarted HF/SCI calculation with 512k configurations at a substantially lower computational cost.

The third and final approach involves performing an initial $\text{CISD}^{ep}$ calculation to generate natural orbitals for a subsequent SCI run ($\text{CISD}^{ep}$/SCI) without restricting the excitation rank. 
Although this scheme yields the highest variational and perturbative energies, which only slowly converge toward the same energy limit for large configuration counts, the resulting protonic density begins much closer to the FGH reference, even for compact variational spaces.

Having established the energy and peak-height convergence, we examine the full one-dimensional protonic densities obtained from representative methods in Fig. \ref{fig:results:hcn_densities}. 
Specifically, we compare our densities obtained from HF, HF-initialized SCI (HF/SCI), natural-orbital-restarted SCI ($\text{SCI}[512\text{k}]/\text{SCI}$), the dressed-Hamiltonian scheme (HF/sBk/SCI), and $\text{CISD}^{ep}$-initialized SCI ($\text{CISD}^{ep}$/SCI) -- each evaluated with the largest variational space of 512k configurations -- against reference CCSD and FGH results \citenum{yu_JCP_152_244123_2020} as well as multicomponent HCI benchmarks \citenum{brorsen_JCTC_16_2379_2020}. 
As anticipated, HF produces the most overly localized density; HF/sBk/SCI and SCI/SCI reduce the peak height roughly halfway toward the reference, yet remain noticeably more localized than CCSD. 
Ultimately, $\text{CISD}^{ep}$/SCI emerges as our best-performing protocol, achieving accuracy comparable to CCSD and HCI while avoiding any predefined truncation of the CI expansion and requiring only 512k configurations, which are one order of magnitude fewer than the $\sim$ 17 million configurations reported for HCI (exact determinant count unspecified) \cite{brorsen_JCTC_16_2379_2020}. 
Currently, the multireference MRSD$^{ep}$ approach \cite{malbon_JCTC_21_3968_2025} remains the closest match to the FGH benchmark. 
Note also that in several cases the proton density profiles are slightly shifted around the origin. 
This artifact stems from the limited ability of fixed-center Gaussian basis functions to describe nuclear translation, forcing the CI expansion to inefficiently incorporate higher excitations merely to adjust the mean proton position.

\section{Conclusions}

In this work, we have presented the implementation of a multicomponent SCI methodology based on the adaptive sampling configuration interaction (ASCI) algorithm \cite{tubman_JCP_145_044112_2016} within the any-particle molecular orbital (APMO) framework. 
Overall, we demonstrated that multicomponent SCI provides a conceptually straightforward and systematically improvable alternative to conventional multi-component CI, which can offer superior efficiency and scalability on modern high-performance computing (HPC) architectures. 

For positronic systems, our results agree closely with prior MCCI benchmarks while successfully capturing bound states in challenging regimes where earlier approaches failed, without restricting the active electron count, particularly for positron binding to \ce{Mg} and positronium binding to \ce{OH^-}.

In the case of quantum nuclear effects in \ce{HCN}, our total energies converge rapidly and even faster than previously published approaches upon incorporating second-order perturbative (PT2) corrections, mirroring the behavior observed in the convergence of the positronic systems. 

Although spatial protonic densities converge slowly when initialized from standard Hartree–Fock orbitals, employing natural orbitals generated from a tailored CISD reference (excluding purely electronic double excitations) markedly accelerates convergence toward the Fourier grid Hamiltonian (FGH) benchmark. 
Consequently, our approach achieves an accuracy comparable to multicomponent CCSD and semistochastic heat-bath CI (SHCI), yet requires substantially smaller variational spaces, of about 512k configurations instead of 17 million previously reported, without imposing any truncation on the CI excitation rank. 
Similarly, the dressed effective Hamiltonian based on the sBk scheme \cite{garniron_JCTC_15_3591_2019} shifts the protonic density toward the reference benchmark, although our current implementation exhibits numerical instabilities and becomes computationally prohibitive at large configuration counts.

Through this work, we have established that SCI approaches within the APMO framework provide a versatile, high-accuracy tool for generating multideterminantal trial wave functions, benchmarking and parameterizing lower-level effective models, and optimizing multicomponent basis sets.

Building upon these developments, this framework opens several promising avenues for future research: (i) algorithmic and low-level code optimizations, such as bit-packed integer representations of configurations; (ii) distributed-memory scaling via Message Passing Interface (MPI) parallelization; (iii) orbital localization schemes to exploit the short-range nature of positron–electron and nucleus–electron correlations in larger systems \cite{moncada_JCP_158_134101_2023a}; (iv) deploying SCI expansions as multideterminantal trial wave functions for diffusion quantum Monte Carlo (QMC) \cite{giner_CJC_91_879_2013,caffarel__1234_15_2016, upadhyay_JCTC_20_9879_2024, charrymartinez_JCTC_18_2267_2022, barborini_JCP_164_062501_2026}, auxiliary-field quantum Monte Carlo (AFQMC) \cite{mahajan_JCP_156_174111_2022}, and transcorrelated (TC) methods \cite{haupt_JCP_163_144113_2025}; and (v) extending the methodology to exotic quantum systems, including heavier leptons, quantum harmonic oscillators, and polaritonic systems. Together, these advances will establish multicomponent SCI as a scalable, benchmark-grade engine for resolving complex correlation effects across physics and chemistry.
%%%%%%%%%%%%%%%%%%%%%%%%%%%%%%%%%%%%%%%%%%%%%%%%%%%%%%%%%%%%%%%%%%%%%
%% Acknoledgements
%%%%%%%%%%%%%%%%%%%%%%%%%%%%%%%%%%%%%%%%%%%%%%%%%%%%%%%%%%%%%%%%%%%%%
\begin{acknowledgments}
The calculations presented in this paper were carried out using the HPC facilities of the University of Luxembourg~\cite{VBCG_HPCS14} {\small (see \href{http://hpc.uni.lu}{hpc.uni.lu})}. 
The authors gratefully acknowledge the ULHPC team for their expert support.
\end{acknowledgments}

\section*{Author contributions:}
JC: Conceptualization (lead), Data curation, Formal Analysis, Investigation (lead), Methodology, Software, Supervision (lead), Validation, Visualization, Writing – original draft, Writing – review \& editing (equal).
MB: Conceptualization (supporting), Resources, Investigation (supporting),  Supervision (supporting), Writing – review \& editing(equal).

\section*{Competing interests:}
There are no competing interests to declare.

\section*{Code and data availability}
Data for this study, including raw results, inputs, geometry, basis sets, and outputs, as well as post-processing scripts, are available in the Zenodo archive \url{https://doi.org/10.5281/zenodo.24283604}.

\section*{Supporting Material}
Additional equations for the generalized Slater-Condon rules, reduced density matrix, natural orbitals, and MCSCF.
Algorithms S3 to S9. 
Core and buffer space convergence analysis. 
Excitation populations and configuration distributions per system.

%%%%%%%%%%%%%%%%%%%%%%%%%%%%%%%%%%%%%%%%%%%%%%%%%%%%%%%%%%%%%%%%%%%%%
%% Appendices
%%%%%%%%%%%%%%%%%%%%%%%%%%%%%%%%%%%%%%%%%%%%%%%%%%%%%%%%%%%%%%%%%%%%%
%\appendix

%%%%%%%%%%%%%%%%%%%%%%%%%%%%%%%%%%%%%%%%%%%%%%%%%%%%%%%%%%%%%%%%%%%%%
%% Bibliography
%%%%%%%%%%%%%%%%%%%%%%%%%%%%%%%%%%%%%%%%%%%%%%%%%%%%%%%%%%%%%%%%%%%%%
\bibliography{references.bib}
\bibliographystyle{achemso}

\begin{figure*}
    \centering
    \includegraphics[width=0.4\textwidth]{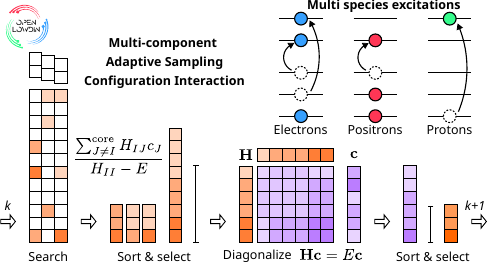}
    \caption{Graphical abstract}
    \label{fig:TOC}
\end{figure*}

\clearpage

% 1. Switch layout if needed (e.g., from two-column to one-column)
\onecolumngrid

% 2. Reset counters
\setcounter{section}{0}
\setcounter{figure}{0}
\setcounter{table}{0}
\setcounter{equation}{0}
\setcounter{page}{1}

\renewcommand*{\figurename}{\textbf{Figure}}
\newcommand*\mycirc[1]{%
  \begin{tikzpicture}
    \node[draw, circle, fill=black, inner sep=0.6pt]{#1};
  \end{tikzpicture}
}
\renewcommand{\thefigure}{\textbf{S\arabic{figure}}}
\newcommand*\mycirctwo[1]{%
  \begin{tikzpicture}
    \node[ circle, fill=jctcgreen, inner sep=1pt]{#1};
  \end{tikzpicture}
}

\renewcommand*{\tablename}{\textbf{Table}}
\renewcommand{\thetable}{\textbf{S\arabic{table}}}

\setcounter{algorithm}{2} 
\renewcommand{\thealgorithm}{\textbf{S\arabic{algorithm}}}

\renewcommand{\theequation}{S\arabic{equation}} % Prefix with 'S.'

% --- SI Title Page ---
\begin{center}
  { \Large \bf \sffamily Supplementary Material \\ ``Multi-component Adaptive Sampling Configuration Interaction method in openLOWDIN: positronic and quantum nuclear systems''
}\\[6mm]
{\large Jorge Charry$^{a,b}$, Matteo Barborini$^{a}$
}\\[6mm]
{\normalsize 
$^{a}$HPC Platform, University of Luxembourg, L-4364 Esch-sur-Alzette, Luxembourg\\[2mm]
$^{b}$Luxembourg Researchers Hub asbl, 223 rue de Luxembourg, L-4222, Esch-sur-Alzette, Luxembourg \\[6mm]

Date: \today \\[6mm]
}
\end{center}

\vspace{1.5em}

%Restore \addcontentsline so only SI sections get recorded
\let\addcontentsline\oldaddcontentsline

% Customize the TOC heading and render it
\tableofcontents

\section{Slater-Condon rules}
Given the CI wave function, the corresponding CI Hamiltonian matrix elements are computed following Slater’s rules, taking into account that the multi-species potential operator is still a two-body operator at most.
Thus, we can distinguish five different cases:

Identical determinants,
\begin{multline}
\langle \Phi_I | \hat{H} | \Phi_I \rangle =
\sum_{\alpha}^{N_s} \sum_i^{N_{\alpha}} (i|\hat{h}|i)  
+\sum_{\alpha}^{N_s} \sum_i^{N_{\alpha}}\sum_{j>i}^{N_{\alpha}} \left[ (ii|jj) - (ij|ji) \right] 
+\sum_{\alpha}^{N_s}\sum_{\beta > \alpha}^{N_s} \sum_{\substack{i \in \alpha \\ j \in \beta}}^{N_{\alpha}N_{\beta}} (ii|jj)
\label{eq:sc_rules_0}
\end{multline}
One orbital different,
\begin{gather}
| \Phi_I \rangle = | \cdots \phi_i \cdots \rangle_{\alpha} \otimes \bigotimes_{\beta \ne \alpha}^{N_s} | \Phi_I\rangle_{\beta}, \quad\quad
| \Phi_J \rangle = | \cdots \phi_j \cdots \rangle_{\alpha} \otimes \bigotimes_{\beta \ne \alpha}^{N_s} | \Phi_J\rangle_{\beta} \notag \\ 
\langle \Phi_I | \hat{H} | \Phi_J \rangle = (i|\hat{h}|i) +  \sum_k^{N_{\alpha}} [(ij|kk]-[ik|kj)] 
+\sum_{\beta \ne \alpha}^{N_s} \sum_{k \in\beta}^{N_{\beta}}(ij|kk), \ \ i,j \in \alpha
\label{eq:sc_rules_1}
\end{gather}
Two different orbitals, same species,
\begin{gather}
| \Phi_I \rangle = | \cdots \phi_i \cdots \phi_j \cdots \rangle_{\alpha} \otimes \bigotimes_{\beta \ne \alpha}^{N_s} | \Phi_I\rangle_{\beta}, \quad\quad
| \Phi_J \rangle = | \cdots \phi_k \cdots \phi_l \cdots \rangle \otimes \bigotimes_{\beta \ne \alpha}^{N_s} | \Phi_J\rangle_{\beta} \notag \\
\langle \Phi_I | \hat{H} | \Phi_J \rangle = (ik|jl)-(il|jk), \ \ i,j,k,l \in \alpha
\label{eq:sc_rules_2aa}
\end{gather}
One orbital difference for two different quantum species,
\begin{gather}
| \Phi_I \rangle = | \cdots \phi_i \cdots \rangle_{\alpha} \otimes
| \cdots \phi_k \cdots \rangle_{\beta} \otimes 
\bigotimes_{\substack{\gamma \ne \alpha.\beta}}^{N_s} | \Phi_I\rangle_{\gamma}, \quad\quad
| \Phi_J \rangle = | \cdots \phi_j \cdots \rangle_{\alpha} \otimes
| \cdots \phi_l \cdots \rangle_{\beta} \otimes 
\bigotimes_{\substack{\gamma \ne \alpha,\beta}}^{N_s} | \Phi_J \rangle_{\gamma} \notag \\
\langle \Phi_I |  \hat{H} |  \Phi_J \rangle =  (ij| kl), \ \ i,j \in \alpha ,k,l \in \beta
\label{eq:sc_rules_2ab}
\end{gather}
And finally, for more than two spin orbitals different within the same species or more than two different species, differing in one spin orbital, the matrix element is
\begin{equation}
\langle \Phi_I | \hat{H} | \Phi_J \rangle = 0
\label{eq:sc_rules_3}
\end{equation}

\section{Natural Orbitals}

The one-particle reduced density matrix (1-RDM) is 
\begin{align}
D_{p,q}^{\alpha} =& \sum_{IJ} c_{I} c_{J} \langle \Psi_I | a_{p}^{\dagger}a_q | \Psi_J \rangle \notag \\
=&\sum_{I} c_{I} c_{I} \langle \Psi_I | a_{p}^{\dagger}a_q | \Psi_I \rangle
+\sum_{I<J} (1 + \mathcal{P}_{IJ}) c_{I} c_{J} \langle \Psi_I | a_{p}^{\dagger}a_q | \Psi_J \rangle 
\quad \left\{ p,q \in \alpha \right\}
\label{eq:2rdm_a}
\end{align}
where $\mathcal{P}_{IJ}$ is a permutation operator for switching configurations index $I$ and $J$. 
The contributions to each matrix element from the pair of determinants $I$ and $J$ are given by the number of different orbitals between $I$ and $J$ due to their orthogonality, allowing us to keep only the equal configurations on the bra and ket after applying the annihilation and creation operators $a_{p}^{\dagger}a_q$.

%-------------------------------------------------------
{\bf (i) identical case $I=J$.}
Here, $a_q$ can only annihilate an orbital $q$ if it's occupied in the configuration $J$, and in order to restore the configuration $I$ in the ket, $a_{p}^{\dagger}$ can only create a particle in the same orbital $q$.
\begin{align}
^{(0)}D_{p,q}^{\alpha} =& \sum_{I} c_{I}  c_{I} \delta_{pq} \to {^{(0)}}D_{p,p}^{\alpha}
\label{eq:2rdm_a_0} \\
&\left\{ q \in |\Psi_I \rangle \right\} \notag
\end{align}
%-------------------------------------------------------
{\bf (ii) One-orbital difference ($j \to i$).} 
The only non-zero term is the one that restores the configuration $J$ to $I$, or vice-versa
\begin{align}
\sum_{I<J} c_{I} c_{J} \cdot \text{phase(I,J)} (\delta_{pi} \delta_{qj} + \delta_{pj} \delta_{qi} ) 
\label{eq:2rdm_a_1} \to {^{(1)}}D_{i,j}^{\alpha} + {^{(1)}}D_{j,i}^{\alpha}  \\
\left\{ i \in |\Psi_I \rangle,  i \notin |\Psi_J \rangle, j \in |\Psi_J \rangle, j \notin |\Psi_I \rangle \right \} \notag
\end{align}
where $\text{phase} (I,J) = (-1)^{N_{perm}}$ changes the sign according to the $N_{perm}$ minimum number of swaps required to sort the orbital indices for both configurations into their maximum coincidence.  
\begin{equation}
D_{p,q}^{\alpha} = {^{(0)}}D_{p,q}^{\alpha} + {^{(1)}}D_{p,q}^{\alpha}
\label{eq:2rdm_a_01}
\end{equation}

Once the 1-RDMs have been computed for each quantum species, their matrix diagonalization will provide the natural orbitals as the eigenvectors in the basis of the atomic orbitals, and their eigenvalues correspond to the occupation per molecular orbital \cite{illas_JCP_89_6376_1988}.
\section{MCSCF}
The equations presented here are based on one of the first MCSCF geminal works \cite{siegbahn_JCP_74_2384_1981}, its second order approximation \cite{chaban_TCA_97_88_1997}, the adaptation to ASCI \cite{levine_JCTC_16_2340_2020}, and also the multi-component extension for electron- nuclei in SHCI \cite{fajen_JCTC_17_965_2021}. and new \cite{malbon_JCTC_21_3968_2025}

general orbital opt for SCI \cite{yao_JCTC_17_4183_2021}

In MCSCF, the new set of optimized molecular orbitals $\mathbf{C}^{(n+1)}$ is obtained by means of a unitary  matrix $\mathbf{U}$ for each quantum species $\alpha$
\begin{equation}
\mathbf{C}^{\alpha,(n+1)} = \mathbf{C}^{\alpha,(n)} \mathbf{U}^{\alpha}
\end{equation}
represented as an exponential of an antisymmetric rotation parameters matrix $\mathbf{O}^{\alpha}$ 
\begin{equation}
\mathbf{U}^{\alpha} = \text{exp} ( \mathbf{O}^{\alpha}) = \mathbf{I}^{\alpha} + \mathbf{O}^{\alpha} + \frac{1}{2}( \mathbf{O}^{\alpha})^2 +\dots
\end{equation}
which is usually approximated in common applications by truncating the expansion up to first order
\begin{equation}
\text{exp} ( \mathbf{O}^{\alpha}) \approx \mathbf{I}^{\alpha} + \mathbf{O}^{\alpha}
\end{equation}
followed by a Schmidt orthogonalization. 
There $\mathbf{I}^{\alpha}$ is the identity matrix in $\alpha$ space, and the rotation parameters matrix is $2\times2$ skew-symmetric matrix 
\begin{equation}
 \mathbf{O}^{\alpha} = \left[ \begin{array}{cc}
      \mathbf{0}           & \mathbf{x}^{\alpha}   \\
      -\mathbf{x}^{\alpha} & \mathbf{0}
 \end{array}
 \right]
\end{equation}
where the elements $x_{pq}^{\alpha}$ corresponds to the rotation between an orbital $p$ and $q$. 
Considering that, the variation of an orbital $\phi$ due to the unitary transformation is then written as an expansion
\begin{equation}
\delta \phi_p = ( \mathbf{U} \phi_p) - \phi_p = \sum_q \left( x_{pq} + \frac{1}{2} \sum_r x_{pr}x_{pr} + \dots \right)\phi_p
\end{equation}
Similarly, one can express the energy of the system as a truncated second-order Taylor expansion over the rotation parameters $x_{pq}$ around an initial point $\mathbf{x}=0$,
\begin{equation}
E(\mathbf{x}) = E(0) + \sum_{pq} x_{pq} \left( \frac{\partial E}{\partial x_{pq}} \right)_0 + \frac{1}{2} \sum_{pqrs} x_{pq}x_{rs} \left( \frac{\partial^2 E}{\partial x_{pq} \partial x_{rs}} \right)_0 + \dots
\end{equation}

Then, the optimized orbital rotations are found when the total variational energy gradient is zero
\begin{equation}
\left.\frac{\partial E}{\partial x_{pq}^{\alpha}}\right\vert_{x^{\alpha}_{pq}=0} = g^{\alpha}_{pq} = 0
\end{equation}
This equation can be solved using the Newton-Raphson method starting from a zero rotated iteration $x_{pq}^{\alpha,(0)}=0$
\begin{equation}
 \mathbf{\delta}^{{\alpha},(n)} = \mathbf{x}^{{\alpha},(n+1)} -\mathbf{x}^{{\alpha},(n)} = -(\mathbf{B}^{{\alpha},(n)})^{-1} \cdot \mathbf{g}^{{\alpha},(n)}
\end{equation}
requiring the computation of the orbital rotation gradient $\mathbf{g}$ and Hessian $\mathbf{B}$ for each quantum species, which can be expressed in a compact and computationally practical form under second-quantization in terms of the 1- and 2- particle Reduced Density Matrices (RDMs) are needed.

%----------------------------------------------------------------------------------
\subsubsection{The 2-RDM intraspecies}
Similarly to the (1-RDM, the two-particle reduced density matrix (2-RDM) is
\begin{equation}
D_{pq,rs}^{\alpha\alpha} = \sum_{IJ} c_{I}  c_{J} \langle \Psi_I | a_{p}^{\dagger} a_{r}^{\dagger} a_q a_s | \Psi_J \rangle 
\quad \left\{ p,q,r,s \in \alpha \right\}
\end{equation}
which can be expressed as
\begin{align}
D_{pq,rs}^{\alpha\alpha} =& \sum_{I} c_{I} c_{I} \langle \Psi_I | a_{p}^{\dagger} a_q a_{r}^{\dagger} a_s - \delta_{qr} a_{p}^{\dagger} a_s | \Psi_I \rangle \notag \\
+& \sum_{I<J} (1 + \mathcal{P} ) c_{I} c_{J} \langle \Psi_I | a_{p}^{\dagger} a_q a_{r}^{\dagger} a_s - \delta_{qr} a_{p}^{\dagger} a_s | \Psi_J \rangle 
\quad \left\{ p,q,r,s \in \alpha \right\}
\label{eq:2rdm_aa}
\end{align}
due to the anti-commutation rules of fermionic annihilation and creation operators. 

Again, the contributions to each matrix element depend on the number of different orbitals between $I$ and $J$ due to their orthogonality conditions, allowing us to keep only the equal configurations on the bra and ket after applying the set of annihilation and creation operators.

%-------------------------------------------------------
{\bf (i) identical case $I=J$.}
There are two possibilities: first when two of the particles are annihilated and created in the same orbitals, leading to
\begin{align}
c_{I}  c_{J} \langle \Psi_I | a_{p}^{\dagger} a_q a_{r}^{\dagger} a_s | \Psi_J \rangle =
c_{I}  c_{J} \delta_{IJ} \delta_{pq} \delta_{rs} \\
\left\{ s \in |\Psi_{J}\rangle, q \in |\Psi_{J}\rangle \right\} \notag
\label{eq:2rdm_aa_0_1}
\end{align}
The second possibility appears when a particle is annihilated from $\Psi_{J}$ and then created in any unoccupied orbital of $\Psi_{J}$, then, to match the bra, that particle must be annihilated and created in the same starting orbital
for the first term of Eq. \ref{eq:2rdm_aa}, 
\begin{align}
c_{I}  c_{J} \langle \Psi_I | a_{p}^{\dagger} a_q a_{r}^{\dagger} a_s | \Psi_J \rangle =
c_{I}  c_{J} \delta_{IJ} \delta_{qr} \delta_{ps}
\\ 
\left\{ s \in |\Psi_{J}\rangle, r \notin |\Psi_{J}\rangle, r\ne s \right\} \notag
\end{align}
and for the second term Eq. \ref{eq:2rdm_aa} the only non-zero terms are
\begin{align}
c_{I}  c_{J} \langle \Psi_I | - \delta_{qr} a_{p}^{\dagger} a_s | \Psi_J \rangle  =
- c_{I}  c_{J} \delta_{IJ} \delta_{qr} \delta_{ps} 
\\ 
\left\{ s \in |\Psi_{J}\rangle, \forall\ rq \right\} \notag
\end{align}
then in that scenario the first and second terms of Eq. \ref{eq:2rdm_aa} can be grouped together, leaving only the contributions where $ rs \in |\Psi_I\rangle $, 
\begin{align}
c_{I}  c_{J} \langle \Psi_I | a_{p}^{\dagger} a_q a_{r}^{\dagger} a_s - \delta_{qr} a_{p}^{\dagger} a_s  | \Psi_J \rangle = - c_{I}  c_{J} \delta_{IJ} \delta_{qr} \delta_{ps}  
\left\{ rs \in |\Psi_{J}\rangle \right\} 
\label{eq:2rdm_aa_0_2}
\end{align}
Overall, the matrix elements derived from Eq. \ref{eq:2rdm_aa_0_1} and Eq. \ref{eq:2rdm_aa_0_2} can be expressed as
\begin{align}
{^{(0)}}D_{pq,rs}^{\alpha} =& \sum_{I} c_{I}  c_{J} 
\delta_{IJ} (\delta_{rs} \delta_{pq} - \delta_{qr} \delta_{ps} ) 
\to {^{(0)}}D_{pp,rr}^{\alpha} - {^{(0)}}D_{pr,rp}^{\alpha}
\label{eq:2rdm_aa_0} \\
&\left\{ s \in |\Psi_{J}\rangle, q \in |\Psi_{J}\rangle \right\} \notag
\end{align}
%-------------------------------------------------------
{\bf (ii) One-orbital difference ($j \to i$)} 
In this case, the excitation operators should restore the different orbital from one configuration $J$ to match the other one $I$, in order to retain a non-zero term. 
For that, there are two options from the first term in Eq. \ref{eq:2rdm_aa}.
The first is to excite the different orbital $j$ in $J$ to the position of the different orbital $i$ in $I$, and then the second excitation should occur between one occupied orbital of $J$ (or $I$) and the same orbital, as in 
\begin{align}
& c_{I}  c_{J} \langle \Psi_I | a_{p}^{\dagger} a_q a_{r}^{\dagger} a_s | \Psi_J \rangle 
= c_{I}  c_{J} \delta_{sj} \delta_{ri}\delta_{pq} \cdot \text{phase(I,J)} 
\label{eq:2rdm_aa_1_1} \\
&\left\{ j \notin |\Psi_{I}\rangle, j \in |\Psi_{J}\rangle, i \in |\Psi_{I}\rangle, i \notin |\Psi_{J}\rangle, q \in |\Psi_{I}\rangle \right\} \notag
\end{align}
The second option occurs when exciting the different orbital $j$ in $J$ to any unoccupied $r$ orbital in $J$, and then the second excitation brings back that orbital $r$ into the different orbital $i$ in $I$, matching the two configurations. 
\begin{align}
& c_{I}  c_{J} \langle \Psi_I | a_{p}^{\dagger} a_q a_{r}^{\dagger} a_s | \Psi_J \rangle = c_{I}  c_{J} \delta_{sj} \delta_{qr} \delta_{pi} \cdot \text{phase(I,J)} 
\label{eq:2rdm_aa_1_2} \\
&\left\{ j \notin |\Psi_{I}\rangle, j \in |\Psi_{J}\rangle, i \in |\Psi_{I}\rangle, i \notin |\Psi_{J}\rangle, r \notin a_s|\Psi_{J}\rangle \right\} \notag
\end{align}
However, for the second term in Eq.\ref{eq:2rdm_aa}, the only non-zero contribution is exciting the different orbital $j$ in $J$ to the different orbital $i$ in $I$ for any $rq$ pair 
\begin{align}
& c_{I}  c_{J} \langle \Psi_I | - \delta_{qr} a_{p}^{\dagger} a_s | \Psi_J \rangle =
- c_{I}  c_{J} \delta_{sj} \delta_{pi}\delta_{qr} \cdot \text{phase(I,J)} 
\label{eq:2rdm_aa_1_3} \\
&\left\{ j \notin |\Psi_{I}\rangle, j \in |\Psi_{J}\rangle, i \in |\Psi_{I}\rangle, i \notin |\Psi_{J}\rangle, \forall\ rq \right\} \notag
\end{align}
Thus, when adding up the two above equations, it will only survive the terms where $rq$ are in the configuration $J$
\begin{align}
&- c_{I} c_{J} \delta_{sj} \delta_{pi}\delta_{qr} \cdot \text{phase(I,J)}
\label{eq:2rdm_aa_1_23} \\
&\left\{ j \notin |\Psi_{I}\rangle, j \in |\Psi_{J}\rangle, i \in |\Psi_{I}\rangle, i \notin |\Psi_{J}\rangle, rq \in |\Psi_{I}\rangle \right\} \notag
\end{align}
Consequently, gathering the equations \ref{eq:2rdm_aa_1_1} and \ref{eq:2rdm_aa_1_23} and their equivalent counterpart when exciting orbitals in $I$ to match $J$ from the permuted term $\mathcal{P}_{IJ}$, one ends up with:
\begin{align}
^{(1)}D_{pq,rs}^{\alpha} =& \sum_{I<J} 
c_{I} c_{J} \cdot \text{phase(I,J)} \left[
\delta_{sj} ( \delta_{ri}\delta_{pq} - \delta_{pi}\delta_{qr} )
+ \delta_{si} ( \delta_{ri}\delta_{pq} - \delta_{pj}\delta_{qr} ) \right] \label{eq:2rdm_aa_1} \\
&\to {^{(1)}}D_{pp,ij}^{\alpha} - {^{(1)}}D_{iq,qj}^{\alpha}
+ {^{(1)}}D_{pp,ji}^{\alpha} - {^{(1)}}D_{jq,qi}^{\alpha} 
\notag \\
&\left\{ j \notin |\Psi_{I}\rangle, j \in |\Psi_{J}\rangle, i \in |\Psi_{I}\rangle, i \notin |\Psi_{J}\rangle, rq \in |\Psi_{I}\rangle \right\} \notag
\end{align}
{\bf (iii) Two-orbital difference ($l \to k$, $j \to i$ )} 
For this case, the only non-zero contribution is given by exciting the two different orbitals in $J$ to match the configuration $I$, or vice-versa, by exciting the two different orbitals, which can be done in four different ways.
First, exciting the first $l$ and second $j$ different orbital in $J$ to the occupied different orbital $k$ and $i$ in $I$
\begin{align}
&+c_{I}  c_{J} \cdot \text{phase(I,J)} 
\delta_{sl} \delta_{rk} \delta_{qj} \delta_{pi} \\
&\left\{ jl \notin |\Psi_{I}\rangle, jl \in |\Psi_{J}\rangle, ik \in |\Psi_{I}\rangle, ik \notin |\Psi_{J}\rangle \right\} \notag
\end{align}
Secondly, by exciting first the second different orbital $j$ in $J$ to the first occupied different orbital $k$ in $I$, and then $l$ to $i$
\begin{align}
&-c_{I}  c_{J} \cdot \text{phase(I,J)}
\delta_{sj} \delta_{rk} \delta_{ql} \delta_{pi} \\
&\left\{ jl \notin |\Psi_{I}\rangle, jl \in |\Psi_{J}\rangle, ik \in |\Psi_{I}\rangle, ik \notin |\Psi_{J}\rangle \right\} \notag
\end{align}
here a minus sign is introduced to match the order of configuration $I$ and $J$. 
Conversely, the third and fourth ways are those by exciting orbitals from configuration $I$ in the ket to match the occupied orbitals of $J$ in the bra 
$\langle \Psi_I | a_{p}^{\dagger} a_q a_{r}^{\dagger} a_s | \Psi_J \rangle = \langle \Psi_J | a_{p}^{\dagger} a_q a_{r}^{\dagger} a_s | \Psi_I \rangle $. 
Then operating in the ket to exciting the first $k$ and second $i$ different orbital in $I$ to the occupied different orbital $l$ and $j$ in $J$
\begin{align}
&+c_{J}  c_{I} \cdot \text{phase(I,J)}
\delta_{sk} \delta_{rl} \delta_{qi} \delta_{pj} \\
&\left\{ jl \notin |\Psi_{I}\rangle, jl \in |\Psi_{J}\rangle, ik \in |\Psi_{I}\rangle, ik \notin |\Psi_{J}\rangle \right\} \notag
\end{align}
and analogously to the second term by exciting first the second different orbital in $I$
\begin{align}
&-c_{J}  c_{I} \cdot \text{phase(J,I)}
\delta_{sk} \delta_{rj} \delta_{qi} \delta_{pl} \\
&\left\{ jl \notin |\Psi_{I}\rangle, jl \in |\Psi_{J}\rangle, ik \in |\Psi_{I}\rangle, ik \notin |\Psi_{J}\rangle \right\} \notag
\end{align}
Combining the last four terms leads to
\begin{align}
^{(2)}D_{pq,rs}^{\alpha} =& \sum_{I < J} 
c_{I}  c_{J} \cdot \text{phase(I,J)}
(\delta_{sl} \delta_{rk} \delta_{qj} \delta_{pi}
-\delta_{sj} \delta_{rk} \delta_{ql} \delta_{pi}
+\delta_{sk} \delta_{rl} \delta_{qi} \delta_{pj}
-\delta_{sk} \delta_{rj} \delta_{qi} \delta_{pl}
) \label{eq:2rdm_aa_2} \\
&\to {^{(2)}}D_{ij,kl}^{\alpha} - {^{(2)}}D_{il,kj}^{\alpha}
 + {^{(2)}}D_{ji,lk}^{\alpha} - {^{(2)}}D_{li,jk}^{\alpha} \notag \\
&\left\{ jl \notin |\Psi_{I}\rangle, jl \in |\Psi_{J}\rangle, ik \in |\Psi_{I}\rangle, ik \notin |\Psi_{J}\rangle \right\} \notag
\end{align}
In the end, the 2-RDM for the same quantum species is computed as
\begin{align}
D_{pq,rs}^{\alpha} = {^{(0)}}D_{pq,rs}^{\alpha} + {^{(1)}}D_{pq,rs}^{\alpha} + {^{(2)}}D_{pq,rs}^{\alpha} 
\end{align}
\subsubsection{The 2-RDM inter-species}
On the other hand, for different quantum species, another 2-RDM is defined
\begin{equation}
D_{pq,rs}^{\alpha\beta} = \sum_{IJ} c_{I}  c_{J} \langle \Psi_I |  a_{p}^{\dagger} a_q a_{r}^{\dagger} a_s  | \Psi_J \rangle \quad p,q \in \alpha, \quad r,s \in \beta
\label{eq:2rdm_ab}
\end{equation}
Again, their matrix elements are computed on cases according to the number of different orbitals between configurations

{\bf (i) identical case $I=J$.}
Now, since the two creation and annihilation operators act on different spaces for $\alpha$ and $\beta$ species, the only non-zero contribution appears when two distinct types of particles are annihilated and created in the same orbitals
\begin{align}
^{(0)}D_{pq,rs}^{\alpha\beta} = \sum_{IJ} c_{I}  c_{J} \delta_{IJ} \delta_{pq} \delta_{rs} 
\to {^{(0)}}D_{pp,rr}^{\alpha\beta} 
\label{eq:2rdm_ab_0} \\
\left\{ s^{\beta} \in |\Psi_{J}\rangle, q^{\alpha} \in |\Psi_{J}\rangle \right\} \notag
\end{align}
{\bf (ii) One-orbital difference ($j^{\beta} \to i^{\beta}$)} 
Similarly, since both excitation operators act on different spaces, the only possibility is to excite the different orbitals in $J$ to match $I$, and the other species should excite into the same orbital.
\begin{align}
&^{(1)}D_{pq,rs}^{\alpha\beta} = \sum_{IJ} 
c_{I}  c_{J} \cdot \text{phase(I,J)} \delta_{sj} \delta_{ri} \delta_{pq}
\to {^{(1)}}D_{pp,ij}^{\alpha\beta}
\label{eq:2rdm_ab_1} \\
&\left\{ j^{\beta} \notin |\Psi_{I}\rangle, j^{\beta} \in |\Psi_{J}\rangle, i^{\beta} \in |\Psi_{I}\rangle, i^{\beta} \notin |\Psi_{J}\rangle, p^{\alpha}q^{\alpha} \in |\Psi_{I}\rangle \right\} \notag
\end{align}
{\bf (ii) Two-orbital difference ($l^{\beta} \to k^{\beta}$, $j^{\alpha} \to i^{\alpha}$) } 
Here also the surviving terms are only those that restore the configuration $I$ from $J$, or vice-versa, by exciting the different orbitals for each species, which can be done in two different ways.
First to excite $J$ orbitals in the ket to match $I$ in the bra
\begin{align}
&+c_{I}  c_{J} \cdot \text{phase(I,J)} \delta_{sl} \delta_{rk} \delta_{qj} \delta_{pi}
\\
&\left\{ l^{\beta} j^{\alpha} \notin |\Psi_{I}\rangle, l^{\beta} j^{\alpha} \in |\Psi_{J}\rangle, k^{\beta} i^{\alpha} \in |\Psi_{I}\rangle, k^{\beta} i^{\alpha} \notin |\Psi_{J}\rangle \right\} \notag
\end{align}
and secondly, to excite $I$ orbitals in the ket to match $J$ in the bra
\begin{align}
&+c_{J}  c_{I} \cdot \text{phase(J,I)} \delta_{sk} \delta_{rl} \delta_{qi} \delta_{pj}
\\
&\left\{ l^{\beta} j^{\alpha} \notin |\Psi_{I}\rangle, l^{\beta} j^{\alpha} \in |\Psi_{J}\rangle, k^{\beta} i^{\alpha} \in |\Psi_{I}\rangle, k^{\beta} i^{\alpha} \notin |\Psi_{J}\rangle \right\} \notag
\end{align}
gathering both to obtain
\begin{align}
^{(2)}D_{pq,rs}^{\alpha\beta} =& \sum_{I\le J} 
c_{I}  c_{J} \cdot \text{phase(I,J)} (
\delta_{sl} \delta_{rk} \delta_{qj} \delta_{pi}
+\delta_{sk} \delta_{rl} \delta_{qi} \delta_{pj} )
\to {^{(2)}}D_{ij,kl}^{\alpha\beta} + {^{(2)}}D_{ji,lk}^{\alpha\beta}
\label{eq:2rdm_ab_2} \\
&\left\{ l^{\beta} j^{\alpha} \notin |\Psi_{I}\rangle, l^{\beta} j^{\alpha} \in |\Psi_{J}\rangle, k^{\beta} i^{\alpha} \in |\Psi_{I}\rangle, k^{\beta} i^{\alpha} \notin |\Psi_{J}\rangle \right\} \notag
\end{align}
then the 2-RDM inter-species is computed as
\begin{align}
D_{pq,rs}^{\alpha\beta} = {^{(0)}}D_{pq,rs}^{\alpha\beta} + {^{(1)}}D_{pq,rs}^{\alpha\beta} + {^{(2)}}D_{pq,rs}^{\alpha\beta} 
\end{align}
Finally, due to the anti-commutation relations of creation and annihilation operators of the same quantum species, and the commutation relations between different species.
The RDMs have the following symmetry rules:
\begin{equation}
D^{\alpha}_{pq} = D^{\alpha}_{qp}
\end{equation}
\begin{equation} 
D^{\alpha}_{pq,rs} = D^{\alpha}_{rs,pq} = D^{\alpha}_{sr,qp} = D^{\alpha}_{qp,sr} \quad p,q,r,s \in \alpha
\end{equation}
\begin{equation} 
D^{\alpha\beta}_{pq,rs} = D^{\alpha}_{qp,sr} \quad p,q \in \alpha \quad r,s \in\beta
\end{equation}

\subsubsection{Gradient and Hessian}
From this point, we will employ the following notation for orbital indexes: general orbital indexes $\{p,q,r,s,t\}$; inactive (or core) $\{i,j,k,l\}$; active $\{u,v,w,x,y\}$; and virtual virtual: $\{a,b,c,d\}$.
Given that, the generalized orbital gradient elements are
\begin{equation}
g_{pq} = \frac{\partial E}{\partial x_{pq}^{\alpha}} = 2(F^{\alpha}_{pq}-F^{\alpha}_{qp})
\end{equation}
introducing the generalized Fock matrix $\mathbf{F}$
\begin{equation} % fixed 12  JCTC-ASCI
F_{pq}^{\alpha} =  \sum_r D^{\alpha}_{pr}h_{qr}^{\alpha} + \sum_{rst} D^{\alpha\alpha}_{pr,st} \left[ (qr|st) \right]
+ \sum_{r\in\alpha}\sum_{\beta \ne \alpha} q_{\beta}\sum_{st\in\beta} D^{\alpha\beta}_{pr,st} \left[ (qr|st) \right]
\end{equation}
%\begin{equation} % fixed 12  JCTC-ASCI
%F_{pq}^{\alpha} =  \sum_r h_{qr}^{\alpha}D^{\alpha}_{pr} + \sum_{rst} D^{\alpha\alpha}_{pr,st} \left[ (qr|st) \right]
%+ \sum_{r\in\alpha}\sum_{\beta \ne \alpha} q_{\beta}\sum_{st\in\beta} D^{\alpha\beta}_{pr,st} \left[ (qr|st) \right]
%\end{equation}
While the second-order derivatives for the Hessian matrix $\mathbf{B}$ are
\begin{equation}
B_{pq,rs} = \left(\frac{\partial^2 E}{\partial x_{pq} \partial x_{rs}} \right)_0 = 
w_{pq,rs} - w_{qp,rs} - w_{pq,sr} + w_{qp,sr}   
\end{equation}
where 
\begin{align}
w_{pq,rs} &= 2D_{pr}h_{qs} 
+ 2\sum_{mn} [ D_{rp,mn} ( qs|mn) + (D_{rn,mp} + D_{rn,pm} ) ( qm |sn) ] \\
&+ 2\sum_{\beta \ne \alpha} q_{\beta} \sum_{mn\in\beta} D_{rp,mn} ( qs|mn)
+ \delta_{qr} ( F_{ps} + F_{sp} )
\end{align} 
However, for practical calculations, the Hessian is usually approximated to its diagonal elements only 
\begin{equation}
B_{pq,pq} = \left(\frac{\partial^2 E}{\partial x_{pq} \partial x_{pq}} \right)_0 = 
w_{pq,pq} - w_{qp,pq} - w_{pq,qp} + w_{qp,qp}   
\end{equation}
which, for the diagonal case, reduces to two unique two dimensional matrices
\begin{align}
w_{x,y}^{(xyxy)}&= 2D_{xx}h_{yy} 
+ 2\sum_{mn} [ D_{xx,mn} ( yy|mn) + (D_{xn,mx} + D_{xn,xm} ) ( ym |yn) ] \\
&+ 2\sum_{\beta \ne \alpha} q_{\beta} \sum_{mn\in\beta} D_{xx,mn} ( yy|mn)
+ \delta_{yx} ( F_{xy} + F_{yx} )
\end{align} 
and
\begin{align}
w_{x,y}^{(yxxy)} &= 2D_{yx}h_{xy} 
+ 2\sum_{mn} [ D_{xy,mn} ( xy|mn) + (D_{xn,my} + D_{xn,ym} ) ( xm |yn) ] \\
&+ 2\sum_{\beta \ne \alpha} q_{\beta} \sum_{mn\in\beta} D_{xy,mn} ( xy|mn)
+ \delta_{xx} ( F_{yy} + F_{yy} )
\end{align} 
then all elements can be connected through symmetries
\begin{equation}
B_{pq,pq} = \left(\frac{\partial^2 E}{\partial x_{pq} \partial x_{pq}} \right)_0 = 
w^{(xyxy)}_{p,q} - w^{(yxxy)}_{p,q} - w^{(yxxy)}_{q,p} + w^{(xyxyx)}_{q,p}   
\end{equation}

The previous gradients and Hessian expressions are valid for any general MCSCF expansion. 
Although for computational efficiency, those expressions can be simplified for a CASSCF calculation by splitting the terms into inactive, active, and virtual orbital rotations. 

\section{ASCI algorithms}

\begin{algorithm}
\caption{Compute ASCI amplitudes }
\label{al:asci_amplitudes}
\begin{minted}[escapeinside=||]{fortran}
subroutine Compute_amplitudes ( |$\mathbf{x}$|, |$\mathbf{x}_{conf} $| )
integer :: n ! OMP thread ID
integer :: |$b_{pos}$(omp\_threads)| ! Position in buffer array
integer :: |$\mathcal{N}$| ! Size of input configuration vector
real :: |$\mathbf{x}(\mathcal{N})$| ! Coefficients configuration vector
real :: |$\mathbf{b}(\mathcal{N}_{\mathcal{B}})$| ! Amplitudes in buffer space
type(configuration) :: |$\mathbf{x}_{conf}(\mathcal{N})$| ! Orbital representation of input space
type(configuration) :: |$\mathbf{b}_{conf}(\mathcal{N}_{\mathcal{B}})$|  ! Orbital representation of buffer space

|$b_{pos}$(:) = 1| ! Initial position in buffer array
!|\$|omp parallel  
n = omp_get_thread_num() ! Getting OMP thread ID
!|\$|omp do 
do |$J$| =  1, |$\mathcal{N}$|
    |${J}_{conf}$| = |$\mathbf{x}_{conf}$|(J)
    ! Generate all connected excitations from the J-th configuration 
    call Single_Connected( n, |$b_{pos}(n)$|, |$\mathbf{x}$|(J), |$\mathbf{b}$|, |$J_{conf}$|, |$\mathbf{b}_{conf}$| )  ! Algorithm |\ref{al:asci_single}|
    call DoubleIntra_Connected( n, |$b_{pos}(n)$|, |$\mathbf{x}$|(J), |$\mathbf{b}$|, |$J_{conf}$|, |$\mathbf{b}_{conf}$| ) ! Algorithm |\ref{al:asci_double_intra}|
    call DoubleInter_Connected( n, |$b_{pos}(n)$|, |$\mathbf{x}$|(J), |$\mathbf{b}$|, |$J_{conf}$|, |$\mathbf{b}_{conf}$| )  ! Algorithm |\ref{al:asci_double_inter}|
enddo
!|\$|omp end do
!|\$|omp end parallel 
\end{minted}
\end{algorithm}

\begin{algorithm}
\caption{ Sort buffer arrays: configurations and amplitudes}
\label{al:asci_sort_buffer}
\begin{minted}[escapeinside=||]{fortran}
Subroutine Sort_Buffer ( |$\mathbf{b}$|, |$\mathbf{B}_{conf} $| )
integer, optional :: n ! OMP thread ID
integer :: |$b_{pos}$(omp\_threads)| ! Position in buffer array
real :: |$\mathbf{B}(\mathcal{N}_{\mathcal{B}})$| ! Amplitudes in buffer space 
type(configuration) :: |$\mathbf{B}_{conf}(\mathcal{N}_{\mathcal{B}})$| ! Orbital representation of buffer space
integer :: |$\mathbf{B}_{indx}(\mathcal{N}_{\mathcal{B}})$| ! Tracking original positions
integer :: b_interval_per_omp(2,n) ! Buffer interval per OMP thread

b1 = 1, b2 = |$N_{\mathcal{B}}$| ! Sorting the whole buffer space
if ( present (n) ) then
    b1 = b_interval_per_omp(1,n) 
    b2 = b_interval_per_omp(2,n)
endif 
! Sort by configuration occupation
call Quick_Sort_Array (|$\mathbf{B}_{conf}(b1:b2)$|, (|$\mathbf{B}_{indx}(b1:b2)$| )
! Find duplicates and sum their amplitudes 
call Merge_Duplicated_Configurations (|$\mathbf{B}_{conf}(b1:b2)$|, (|$\mathbf{B}_{index}(b1:b2)$| )
! Sort amplitudes by absolute value 
call Quick_Sort_Vector (|$\mathbf{B}(b1:b2)$|, (|$\mathbf{B}_{indx}(b1:b2)$| )
! Mark the new position for overwrite new configurations
|$b_{pos}$| = 0.5*b2
\end{minted}
\end{algorithm}

\begin{algorithm}
\caption{ PT2 computation per batch }
\label{al:asci_pt2_batches}
\begin{minted}[escapeinside=||]{fortran}
Subroutine Compute_PT2 ( |$E_{SCI}$|, |$E_{SCI+PT2}$|, |$\mathbf{t}$|, |$\mathbf{b}$|, |$\mathbf{t}_{conf}$|, |$\mathbf{b}_{conf}$| )
integer :: n ! OMP thread ID
integer :: |$b_{pos}$(omp\_threads)| ! Position in buffer array
integer :: |$\mathbf{X}_{\text{batch}}(2,\mathcal{N}_{\chi} )$| ! Lower and upper batch index
real :: |$E_{SCI}$|, |$E_{SCI+PT2}$| 
real :: |$\mathbf{t}(\mathcal{N}_{\mathcal{T}})$| ! Coefficients configuration vector
real :: |$\mathbf{b}(\mathcal{N}_{\mathcal{B}})$| ! Amplitudes in buffer space
type(configuration) :: |${I}_{conf}$| 
type(configuration) :: |${M}_{conf}$| 
type(configuration) :: |$\mathbf{t}_{conf}(\mathcal{N})$| ! Orbital representation of input space
type(configuration) :: |$\mathbf{b}_{conf}(\mathcal{N}_{\mathcal{B}})$|  ! Orbital representation of buffer space

|$E_{SCI+PT2}$| = 0.0_dp

|Copy Target to Buffer| (|$\mathbf{t}_{conf}$|, |$\mathbf{b}_{conf}$| ) 
|$\mathbf{b}(1:\mathcal{N}_{\mathcal{T}} ) $| = 1.0E6 
Compute Batch distribution interval ( |$\mathbf{X}_{\text{batch}}$|, |$\mathbf{b}_{conf}$| )

do |$\chi$| = 1, |$\mathcal{N}_{\chi}$|
    !|\$|omp parallel  
    n = omp_get_thread_num() ! Getting OMP thread ID
    !|\$|omp do 
    do |$I$| = 1, |$\mathcal{N}$|
        |${I}_{conf}$| = |$\mathbf{t}_{conf}$|(I)
        ! Generate all connected excitations from the I-th configuration 
        call Single_Connected( n, |$b_{pos}(n)$|, |$\mathbf{t}$|(I), |$\mathbf{b}$|, |$I_{conf}$|, |$\mathbf{b}_{conf}$|, |$\mathbf{X}_{\text{batch}}( \chi )$| )  ! Algorithm |\ref{al:asci_single}|
        call DoubleIntra_Connected( n, |$b_{pos}(n)$|, |$\mathbf{t}$|(I), |$\mathbf{b}$|, |$I_{conf}$|, |$\mathbf{b}_{conf}$|, |$\mathbf{X}_{\text{batch}}( \chi )$| ) ! Algorithm |\ref{al:asci_double_intra}|
        call DoubleInter_Connected( n, |$b_{pos}(n)$|, |$\mathbf{t}$|(I), |$\mathbf{b}$|, |$I_{conf}$|, |$\mathbf{b}_{conf}$|, |$\mathbf{X}_{\text{batch}}( \chi )$| )  ! Algorithm |\ref{al:asci_double_inter}|
    enddo
    !|\$|omp end do
    !|\$|omp end parallel 
    !|\$|omp parallel do reduction (+:|$E_{SCI+PT2}$|)
    do M = |$\mathcal{N}_{\mathcal{T}}$| + 1, |$\mathcal{N}_{\mathcal{B}}$| ! loop over connected configurations
        |$M_{conf}$| = |$\mathbf{B}_{conf}(M)$|
        call Energy_zero ( |${M}_{conf}$|, |$H_{MM}$| ) ! Eq. |\ref{eq:sc_rules_0}|
        A = |$\mathbf{b}$(M)|
        |$E_{SCI+PT2}$| = |$E_{SCI+PT2}$| + |A|**2 / ( |$E_{SCI}$| - |$H_{MM}$| ) ! Eq. |\ref{eq:pt2}|
    enddo
    !|\$|omp end parallel do
    |Reset Buffer| ( |$\mathbf{b}$|, |$\mathbf{b}_{conf}$| ) 
enddo


\end{minted}
\end{algorithm}

\begin{algorithm}
\caption{ sBk computation from pre-stored buffer space }
\label{al:asci_sBk}
\begin{minted}[escapeinside=||]{fortran}
Subroutine Compute_sBk ( |$E_{SCI}$|, |$\mathbf{t}$|, |$\mathbf{b}$|, |$\mathbf{t}_{conf}$|, |$\mathbf{b}_{conf}$| )
real :: |$E_{SCI}$|
type(Configurations) :: |$\mathbf{T}_{conf}$|, |$\mathbf{B}_{conf}$| ! Target and buffer configurations

while do ( |$\Delta E$| <= 1E-5 ) ! Convergence criteria
    k = k + 1
    ! Solve |$\textbf{H}^{\text{eff}}\textbf{t}^{(k)} = E_{\text{SCI}}^{(k)}\textbf{t}^{(k)}$|, Eq. |\ref{eq:effective_hamiltonian}|
    call Eigen_Solver( |$\mathbf{t}$|, |$\mathbf{t}_{conf}$|, |$E_{SCI}$|, Matrix_Vector_Function) 
enddo while

Function Matrix_Vector_Product ( | $\mathbf{v}$ |, |$\mathbf{\sigma}$| )
real :: | $\mathbf{v} (\mathcal{N}_{\mathcal{T}})$| , |$\mathbf{\sigma} (\mathcal{N}_{\mathcal{T}})$|, | $\mathbf{h} (\mathcal{N}_{\mathcal{T}})$| , |$\hat{v}$| 
! Build |$ \mathbf{\sigma} = \mathbf{H}^{\text{(00)}} \mathbf{v} $|
! Build |$ \mathbf{\sigma} = \mathbf{H}^{\text{(00)}} \mathbf{v} + \mathbf{\Delta} \mathbf{v} $|
!|\$|omp parallel do reduction (+:|$\mathbf{\sigma}$|)
do M = |$\mathcal{N}_{\mathcal{T}}$| + 1, |$\mathcal{N}_{\mathcal{B}}$| ! loop over connected configurations
    |${M}_{conf}$| = |$\mathbf{B}_{conf}(M)$|
    call Energy_zero ( |${M}_{conf}$|, |$H_{MM}$| ) ! Eq. |\ref{eq:sc_rules_0}|
    |$\mathbf{h}$| = 0.0_dp 
    do J = 1, |$\mathcal{N}_{\mathcal{T}}$| ! loop over target space
        |${J}_{conf}$| = |$\mathbf{T}_{conf}(J)$|
        ! Find number of different orbitals between I and J configurations 
        coupling = Coupling_Configurations( M , J )
        ! Compute CI matrix element |$H_{M}$| 
        select case (coupling)
          case ("single") call Energy_One( |${M}_{conf}$|,  |${J}_{conf}$|, |$H_{MJ}$| ) ! Eq. |\ref{eq:sc_rules_1}| 
          case ("double-intra") call Energy_TwoIntra( |${M}_{conf}$|, |${J}_{conf}$|, |$H_{MJ}$| ) ! Eq. |\ref{eq:sc_rules_2aa}|
          case ("double-inter") call Energy_TwoInter( |${M}_{conf}$|, |${J}_{conf}$|, |$H_{MJ}$| ) ! Eq. |\ref{eq:sc_rules_2ab}|
        end case
        | $\mathbf{h}$|(J) = |$H_{MJ}$| ; | $\hat{v}$| = |$\hat{v}$| + |$H_{MJ}$| * |$\mathbf{v}(J)$|
    enddo
    |$\hat{v}$| = |$\hat{v}$| / ( |$E_{SCI}$| - |$H_{MM}$| ) ! Eq. |\ref{eq:sbk_davidson}|
    do I = 1, |$\mathcal{N}_{\mathcal{T}}$| ! loop over target space
        |$ \mathbf{\sigma}(I) $| = |$ \mathbf{\sigma}(I) $| + | $\mathbf{h}$|(I) * |$\hat{v}$|
    enddo 
enddo
!|\$|omp end parallel do
enddo Function
\end{minted}
\end{algorithm}

\begin{algorithm}
\caption{Generate single connected configurations }
\label{al:asci_single}
\begin{minted}[escapeinside=||]{fortran}
Subroutine Single_Connected ( n, |$b_{pos}$|, |$C_J$|, |$\mathbf{b}$|, |$I_{conf}$|, |$\mathbf{b}_{conf}$|, |$\mathbf{x}_{\text{batch}}$| )
integer :: n ! OMP thread ID
integer :: |$b_{pos}$| ! Position in buffer array
real :: |${C}_J$| ! linear combination coefficient
real :: |$H_{IJ}$|, |$H_{II}$| ! Hamiltonian matrix elements
real :: |$A$| ! Amplitude value
real :: |$\mathbf{b}(\mathcal{N}_{\mathcal{B}})$| ! Amplitudes in buffer space 
type(configuration) :: |${J}_{conf}$|, |${I}_{conf}$| ! I-th and J-th configurations
type(configuration) :: |$\mathbf{b}_{conf}(\mathcal{N}_{\mathcal{B}})$| ! Orbital representation of buffer space
integer, optional :: |$\mathbf{x}_{\text{batch}}(2)$| ! Lower and upper batch index

|${I}_{conf} = {J}_{conf}$|
do |$\alpha$| = 1, |$N_{sp}$| ! loop for all quantum species
    do p = |$N_{\alpha}^{core}$| + 1, |$N_{\alpha}^{occ}$| ! occupied orbitals
        ! excite from o1 orbital 
        o1 = |${J}_{conf}\%occ(\alpha,p)$| 
        |${I}_{conf}\%orb(\alpha,o1)$| = 0
        do q = 1, |$N_{\alpha}^{basis} - N_{\alpha}^{occ}$| ! unoccupied orbitals
            ! excite from o1 to v1 orbital 
            v1 = |${J}_{conf}\%vir(\alpha,q)$|
            |${I}_{conf}\%orb(\alpha,v1)$| = 1

            if ( sum(|${I}_{conf}\%orb$|) not in |$\mathbf{x}_{\text{batch}}$| ) continue
            
            call Energy_zero( |$H_{II}$|, |${I}_{conf}$| ) ! Eq. |\ref{eq:sc_rules_0}|
            call Energy_one( |$H_{IJ}$|, |${I}_{conf}$|, |${J}_{conf}$| ) ! Eq. |\ref{eq:sc_rules_1}|
            ! Compute amplitude Eq. |\ref{eq:asci_amplitude}|
            A = |$H_{IJ} * C_J / ( H_{II} - E_{SCI})$|
            ! Add amplitude and configuration to buffer space 
            call Append_Amplitude( n, |$b_{pos}$|, A, |$\mathbf{b}$|, |${I}_{conf}$|, |$\mathbf{b}_{conf}(:)$| )
            |${I}_{conf}\%orb(\alpha,v1)$| = 0 ! restore I configuration 
       enddo
       |${I}_{conf}\%orb(\alpha,o1)$| = 1 ! restore I configuration
   enddo
enddo
\end{minted}  
\end{algorithm}

\begin{algorithm}
\caption{Generate double intra-species connected configurations}
\label{al:asci_double_intra}
\begin{minted}[escapeinside=||]{fortran}
Subroutine DoubleIntra_Connected ( n, |$b_{pos}$|, |$C_J$|, |$\mathbf{b}$|, |$I_{conf}$|, |$\mathbf{b}_{conf}$|, |$\mathbf{x}_{\text{batch}}$| )
integer :: n ! OMP thread ID
integer :: |$b_{pos}$| ! Position in buffer array
real :: |${C}_J$| ! linear combination coefficient
real :: |$H_{IJ}$|, |$H_{II}$| ! Hamiltonian matrix elements
real :: |$A$| ! Amplitude value
real :: |$\mathbf{b}(\mathcal{N}_{\mathcal{B}})$| ! Amplitudes in buffer space 
type(configuration) :: |${J}_{conf}$|, |${I}_{conf}$| ! I-th and J-th configurations
type(configuration) :: |$\mathbf{b}_{conf}(\mathcal{N}_{\mathcal{B}})$| ! Orbital representation of buffer space
integer, optional :: |$\mathbf{x}_{\text{batch}}(2)$| ! Lower and upper batch index

|${I}_{conf}$| = |${J}_{conf}$|
do |$\alpha$| = 1, |$N_{sp}$| ! loop for all quantum species 
    do p = |$N_{\alpha}^{core}$| + 1, |$N_{\alpha}^{occ}$| ! occupied orbitals
        ! excite from o1 orbital 
        o1 = |${J}_{conf}\%occ(\alpha,p)$|
        |${I}_{conf}\%orb(\alpha,o1)$| = 0
        do q = 1, |$N_{\alpha}^{basis} - N_{\alpha}^{occ}$| ! unoccupied orbitals
            ! excite from o1 to v1 orbital 
            v1 = |${J}_{conf}\%vir(\alpha,q)$|
            |${I}_{conf}\%orb(\alpha,v1)$| = 1
            do r = |$N_{\alpha}^{core}$| + 1, |$N_{\alpha}^{occ}$| ! occupied orbitals
                ! excite from o2 orbital 
                o2 = |${J}_{conf}\%occ(\alpha,p)$|
                |${I}_{conf}\%orb(\alpha,o2)$| = 0
                do s 1 = |$N_{\alpha}^{basis} - N_{\alpha}^{occ}$| ! unoccupied orbitals 
                    ! excite from o2 to v2 orbital  
                    v2 = |${J}_{conf}\%vir(\alpha,q)$|
                    |${I}_{conf}\%orb(\alpha,v2)$| = 1

                    if ( sum(|${I}_{conf}\%orb$|) not in |$\mathbf{x}_{\text{batch}}$| ) continue
                    
                    call Energy_zero( |$H_{II}$|, |${I}_{conf}$|, |${I}_{conf}$| ) ! Eq. |\ref{eq:sc_rules_0}|
                    call Energy_TwoIntra( |$H_{IJ}$|, |${I}_{conf}$|, |${J}_{conf}$| ) ! Eq. |\ref{eq:sc_rules_2aa}|
                    ! Compute amplitude Eq. |\ref{eq:asci_amplitude}|
                    A = |$H_{IJ} * C_J / ( H_{II} - E_{SCI})$|
                    ! Add amplitude and configuration to buffer space 
                    call Append_Amplitude( n, |$b_{pos}$|, A, |$\mathbf{B}$|, |${I}_{conf}$|, |$\mathbf{B}_{conf}$| )
                    |${I}_{conf}\%orb(\alpha,v2)$| = 0 ! restore I configuration
                enddo 
                |${I}_{conf}\%orb(\alpha,o2)$| = 1 ! restore I configuration
            enddo 
            |${I}_{conf}\%orb(\alpha,v1)$| = 0 ! restore I configuration
        enddo 
        |${I}_{conf}\%orb(\alpha,o1)$| = 1 ! restore I configuration
    enddo 
enddo 
\end{minted}
\end{algorithm}

\begin{algorithm}
\caption{Generate double inter-species connected configuration }
\label{al:asci_double_inter}
\begin{minted}[escapeinside=||]{fortran}
Subroutine DoubleInter_Connected ( n, |$b_{pos}$|, |$C_J$|, |$\mathbf{b}$|, |$I_{conf}$|, |$\mathbf{b}_{conf}$|, |$\mathbf{x}_{\text{batch}}$| )
integer :: n ! OMP thread ID
integer :: |$b_{pos}$| ! Position in buffer array
real :: |${C}_J$| ! linear combination coefficient
real :: |$H_{IJ}$|, |$H_{II}$| ! Hamiltonian matrix elements
real :: |$A$| ! Amplitude value
real :: |$\mathbf{b}(\mathcal{N}_{\mathcal{B}})$| ! Amplitudes in buffer space 
type(configuration) :: |${J}_{conf}$|, |${I}_{conf}$| ! I-th and J-th configurations
type(configuration) :: |$\mathbf{b}_{conf}(\mathcal{N}_{\mathcal{B}})$| ! Orbital representation of buffer space
integer, optional :: |$\mathbf{x}_{\text{batch}}(2)$| ! Lower and upper batch index

|${I}_{conf}$ = ${J}_{conf}$|
do |$\alpha$| = 1, |$N_{sp}$| ! loop for all quantum species
    do p = |$N_{\alpha}^{core}$ + 1, $N_{\alpha}^{occ}$| ! occupied orbitals 
        ! excite from o1 orbital 
        o1 = |${J}_{conf}\%occ(\alpha,p)$| 
        |${I}_{conf}\%orb(\alpha,o1)$| = 0
        do q = 1, |$N_{\alpha}^{basis} - N_{\alpha}^{occ}$| ! unoccupied orbitals
            ! excite from o1 to v1 orbital 
            v1 = |${J}_{conf}\%vir(\alpha,q)$|
            |${I}_{conf}\%orb(\alpha,o1)$| = 1
            do |$\beta$| = |$\alpha$ + 1, $N_{sp}$| ! loop for all other quantum species
                do r = |$N_{\beta}^{core}$ + 1|, |$N_{\beta}^{occ}$| ! occupied orbitals
                    ! excite from o2 orbital 
                    o2 = |${I}_{conf}\%occ(\beta,p)$|
                    |${I}_{conf}\%orb(\beta,o2)$| = 0
                    do s = 1, |$N_{\beta}^{basis}$ - $N_{\beta}^{occ}$| ! unoccupied orbitals
                        ! excite from o2 to v2 orbital 
                        v2 = |${I}_{conf}\%vir(\alpha,q)$|
                        |${I}_{conf}\%orb(\beta,v2)$| = 1

                        if ( sum(|${I}_{conf}\%orb$|) not in |$\mathbf{x}_{\text{batch}}$| ) continue
                        
                        call Energy_zero( |$H_{II}$|, |${I}_{conf}$, ${I}_{conf}$| ) ! |Eq. \ref{eq:sc_rules_0}|
                        call Energy_TwoInter( |$H_{IJ}$|, |${I}_{conf}$|, |${I}_{conf}$| ) ! |Eq. \ref{eq:sc_rules_2ab}|
                        ! Compute amplitude Eq. |\ref{eq:asci_amplitude}|
                        A = |$H_{IJ} * C_J / ( H_{II} - E_{SCI})$|
                        ! Add amplitude and configuration to buffer space 
                        call Append_Amplitude ( n, |$b_{pos}$|, A, |$\mathbf{B}$|, |${I}_{conf}$|, |$\mathbf{B}_{conf}$| )
                        |${I}_{conf}\%orb(\beta,v2)$| = 0 ! restore I configuration
                    enddo
                    |${I}_{conf}\%orb(\beta,o2)$| = 1 ! restore I configuration
                enddo
            enddo
            |${I}_{conf}\%orb(\alpha,v1)$| = 0 ! restore I configuration
        enddo
        |${I}_{conf}\%orb(\alpha,01)$| = 1 ! restore I configuration
    enddo
enddo
\end{minted}
\end{algorithm}

\section{Memory scaling}
\begin{figure}
    \centering
    \includegraphics[width=0.5\linewidth]{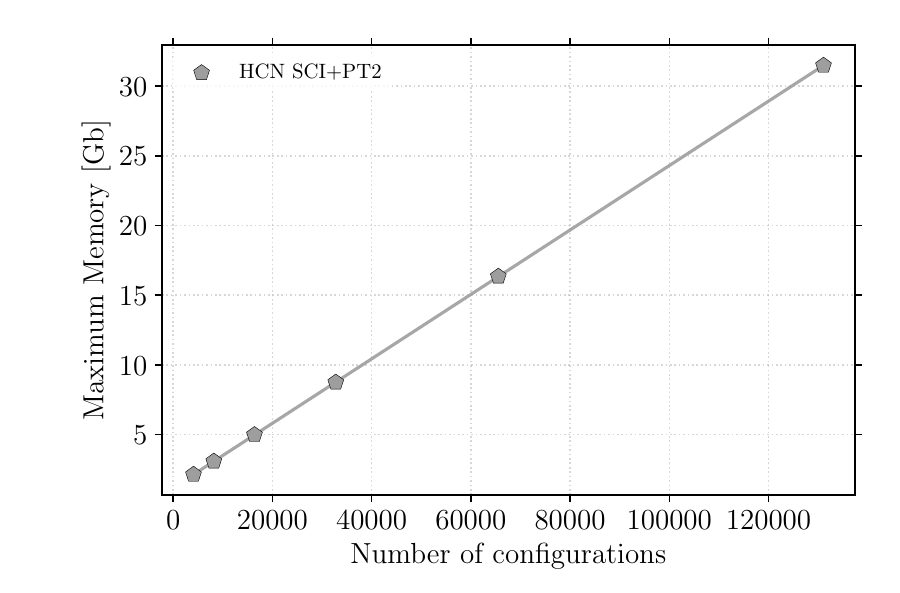}
    \caption{
    Computational scaling for the maximum  Random Access Memory (RAM) used for a single SCI+PT2 calculation as a function of the variational space size.
    }
    \label{sm:fig:memory}
\end{figure}
\section{HCN}

\begin{figure}[h!]
    \centering
    \includegraphics[width=0.495\linewidth]{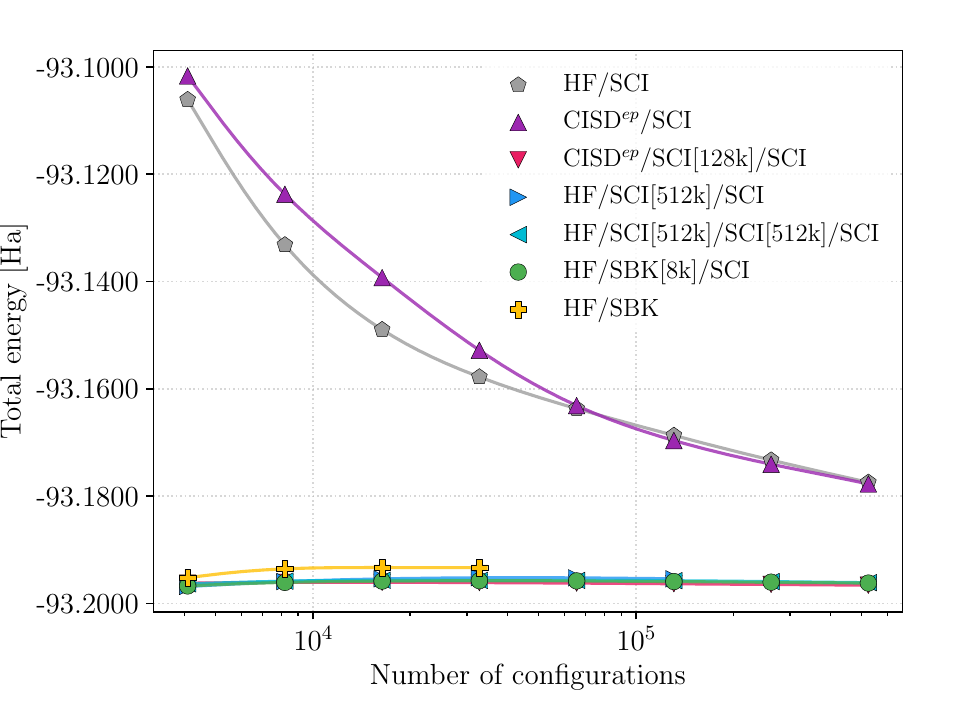}
    \caption{Convergence in \ce{HCN} with quantum hydrogen nuclei for the SCI+PT2 energies as a function of the size of the variational space.
    The slash symbol ($/$) is used to indicate the natural orbital passing between two calculations. 
    For restarts from a previous SCI calculation, the number inside square brackets indicates the size of the variational space.
    }
    \label{si:fig:results:hcn}
\end{figure}

\section{Excitations populations}

\begin{figure}[h]
    \centering
    \includegraphics[width=0.495\linewidth]{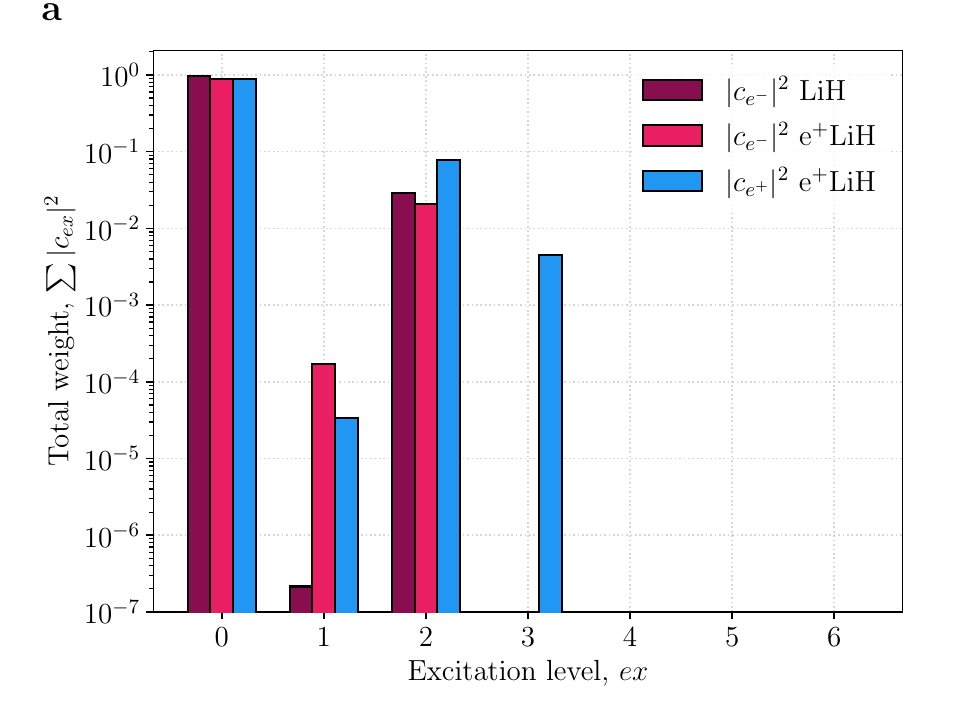}
    \includegraphics[width=0.495\linewidth]{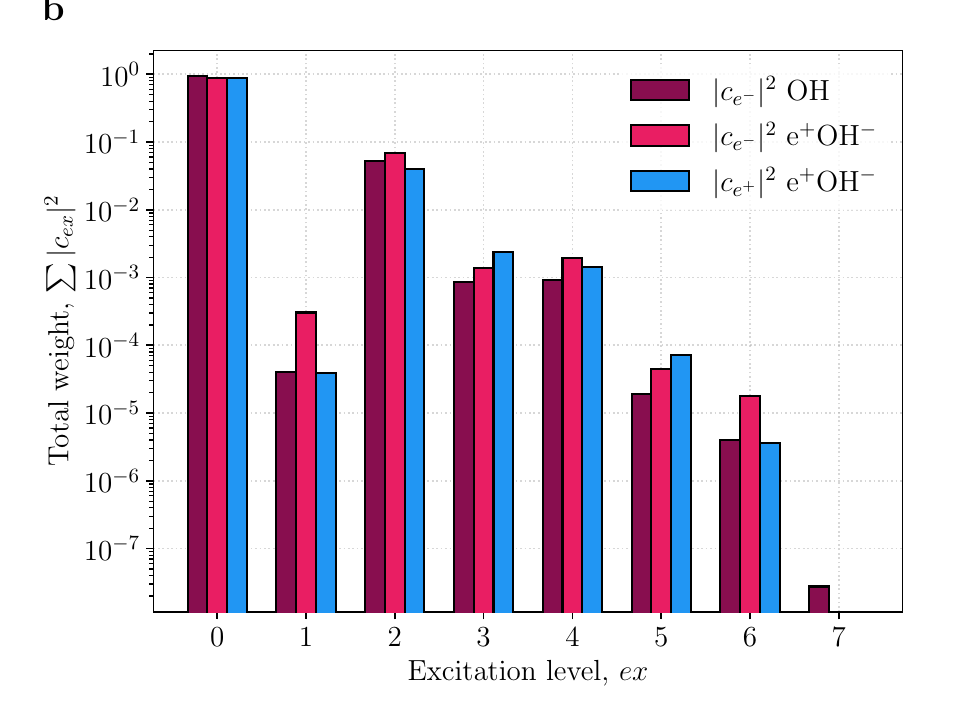} \\
    \includegraphics[width=0.495\linewidth]{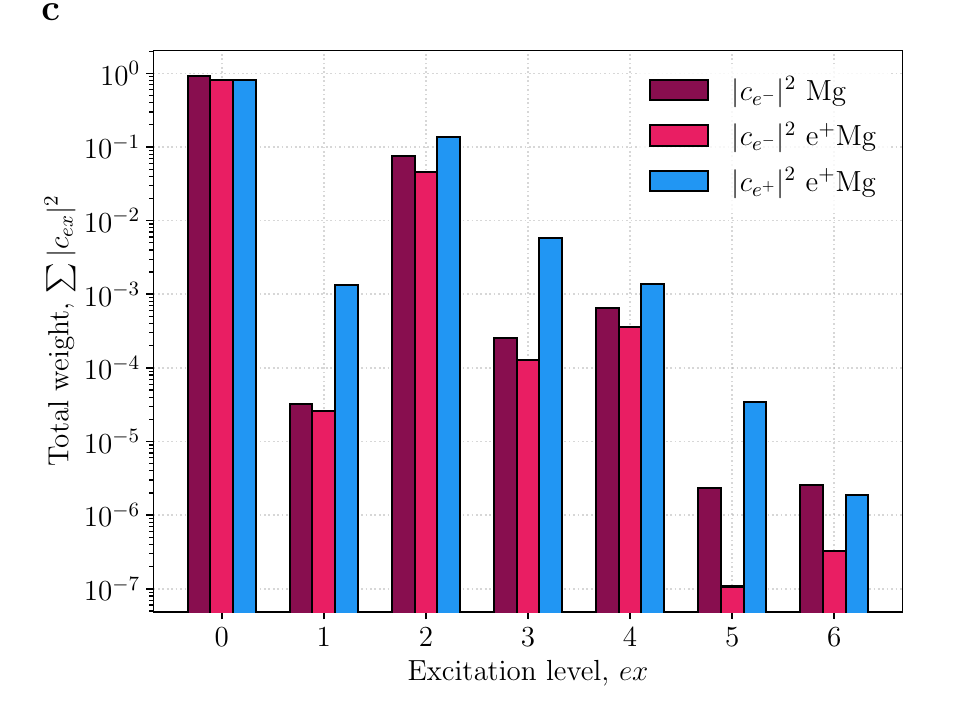}
    \includegraphics[width=0.495\linewidth]{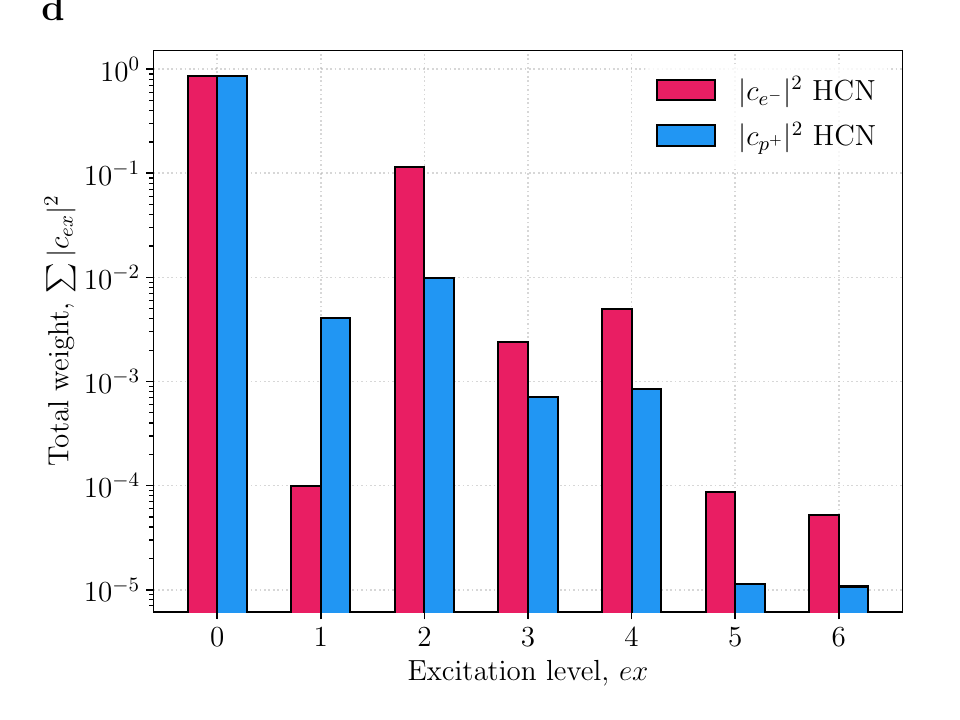}
    \caption{
    Total weight as the sum of squared linear combination CI coefficients grouped by type of excitation level (0: ground state, 1: single, 2: double, and so on) either with or without positron or proton excitations for \ce{e^+LiH} (\textbf{a}), \ce{e^+OH^-} (\textbf{b}), \ce{e^+Mg} (\textbf{c}), \ce{HCN} (\textbf{d}).
    }
    \label{sm:fig:populations}
\end{figure}

\section{Configurations distributions}

\begin{figure}[h]
    \centering
    \includegraphics[width=0.495\linewidth]{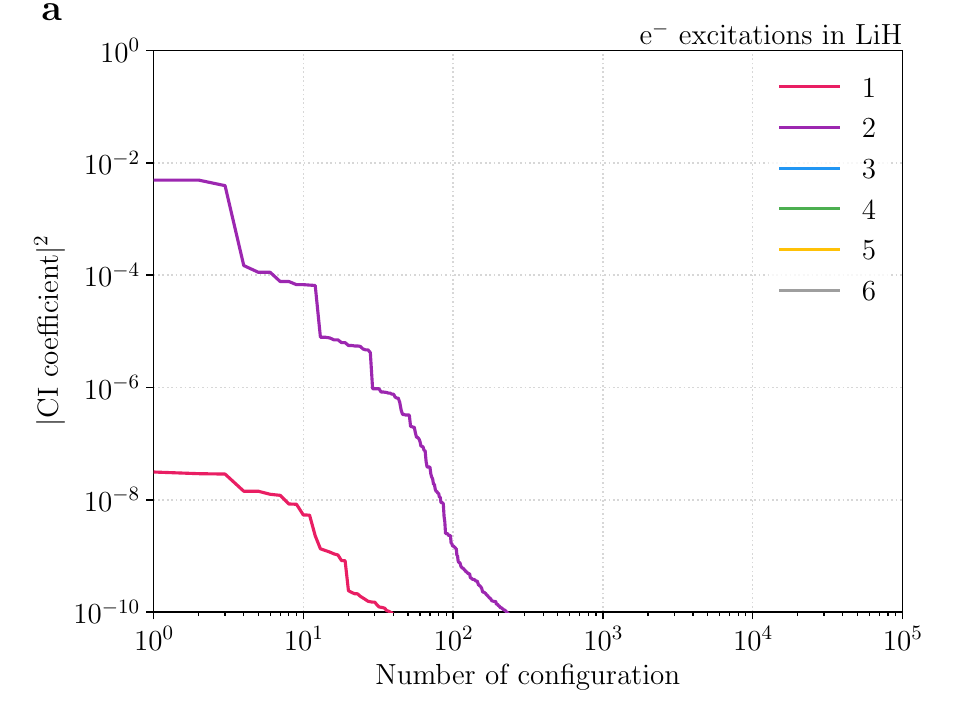} 
    \includegraphics[width=0.495\linewidth]{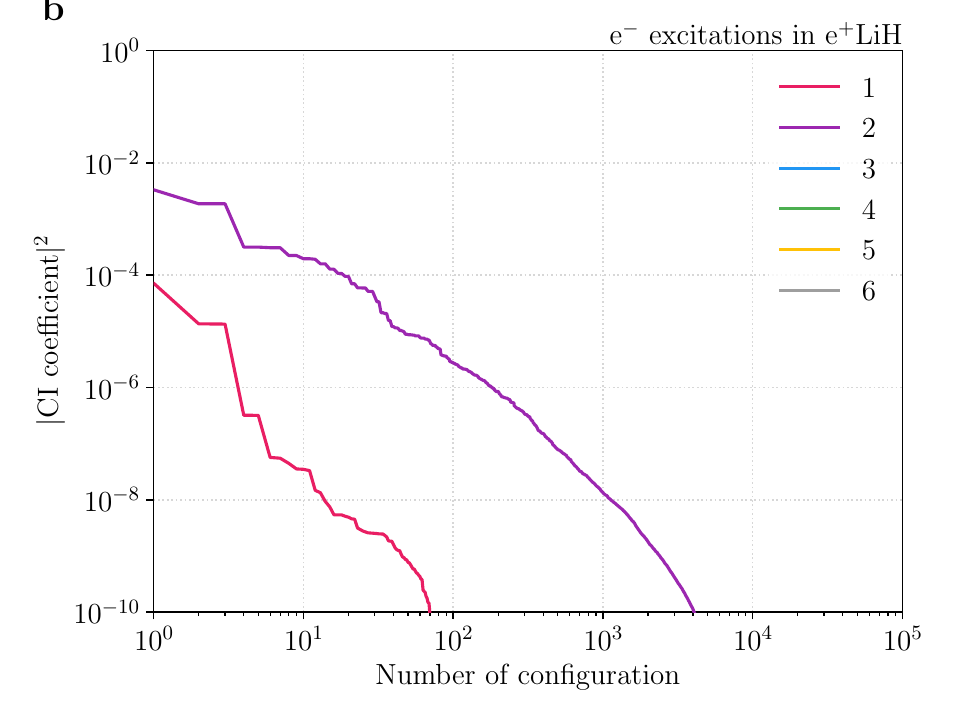}
    \includegraphics[width=0.495\linewidth]{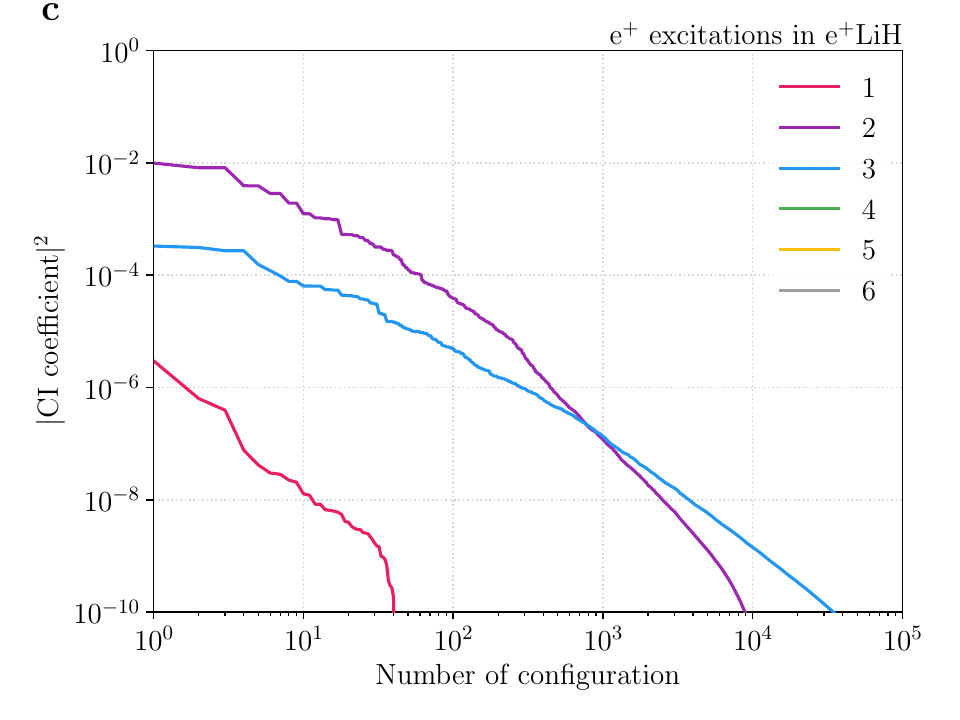}
    \caption{
    CI linear combination coefficients distribution for all 131072 SCI configurations grouped by excitation type as electronic excitations in \ce{LiH} (\textbf{a}), electronic excitations in \ce{e^+LiH} (\textbf{b}), and electronic-positron excitations in \ce{e^+LiH} (\textbf{c}).
    }
    \label{sm:fig:e+lih_distribution}
\end{figure}

\begin{figure}[h]
    \centering
    \includegraphics[width=0.495\linewidth]{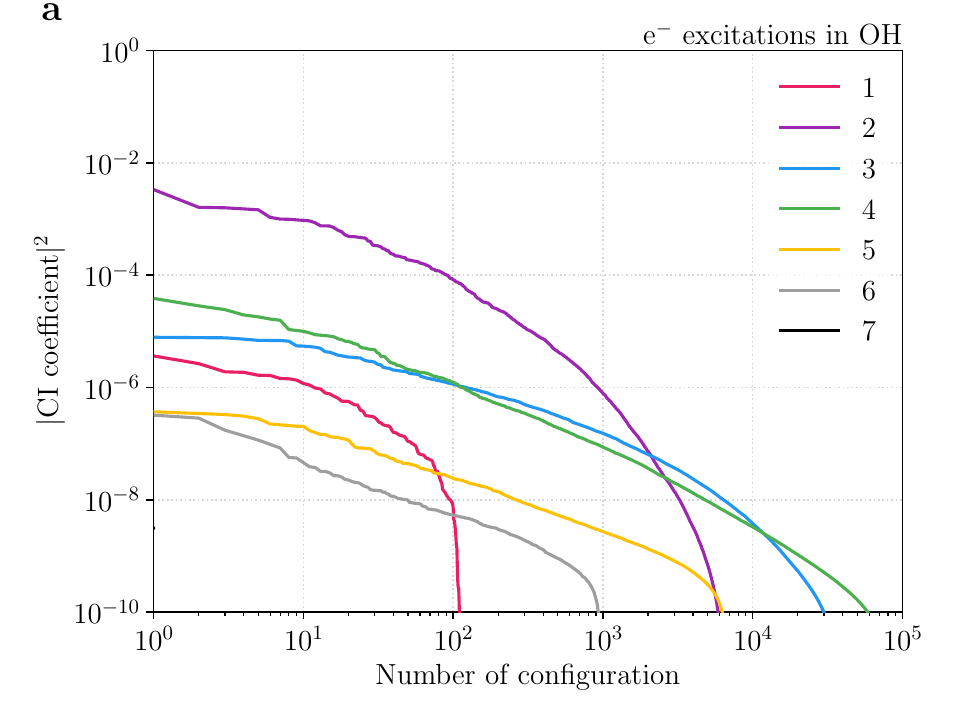}
    \includegraphics[width=0.495\linewidth]{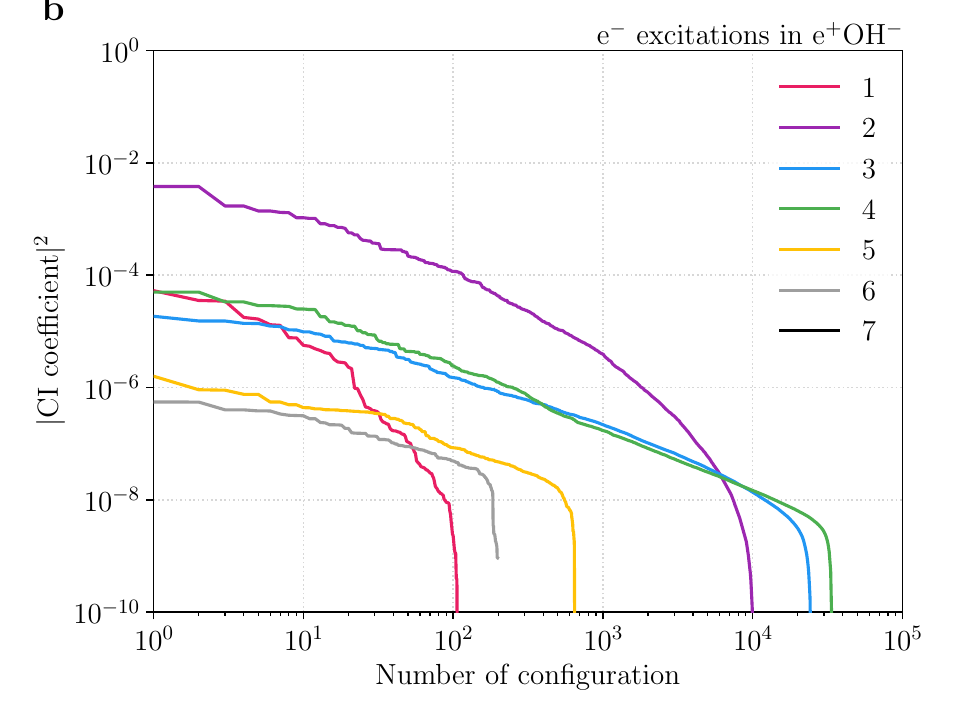}
    \includegraphics[width=0.495\linewidth]{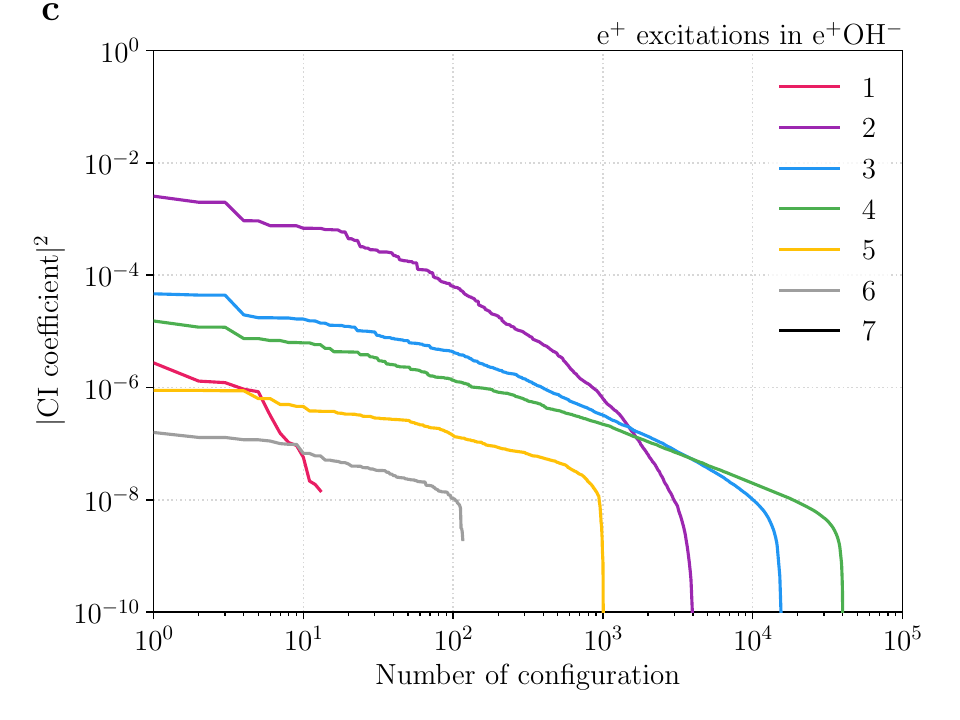}
    \caption{
    CI linear combination coefficients distribution for all 131072 SCI configurations grouped by excitation type as electronic excitations in \ce{OH} (\textbf{a}), electronic excitations in \ce{e^+OH^-} (\textbf{b}), and electronic-positron excitations in \ce{e^+OH^-} (\textbf{c}).
    }
    \label{sm:fig:e+oh-_distribution}
\end{figure}

\begin{figure}[h]
    \centering
    \includegraphics[width=0.495\linewidth]{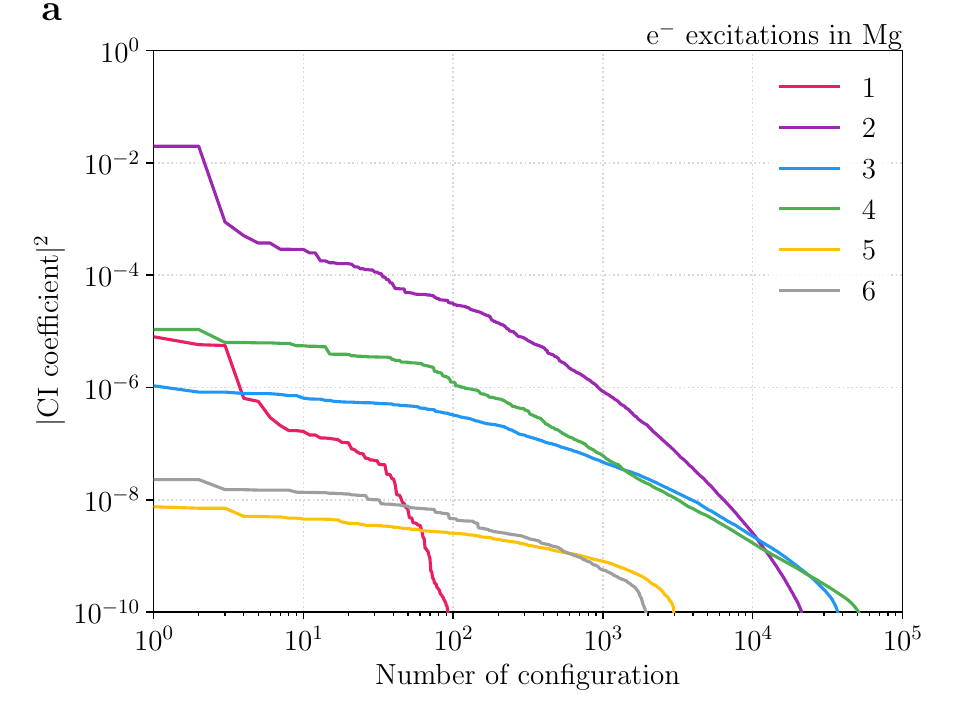}
    \includegraphics[width=0.495\linewidth]{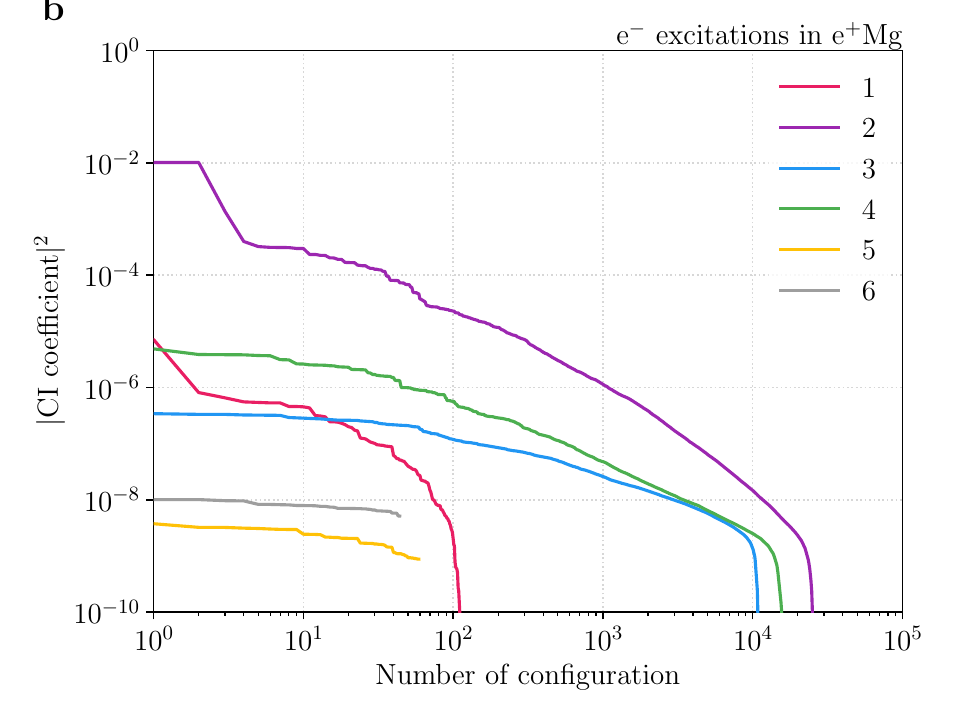}
    \includegraphics[width=0.495\linewidth]{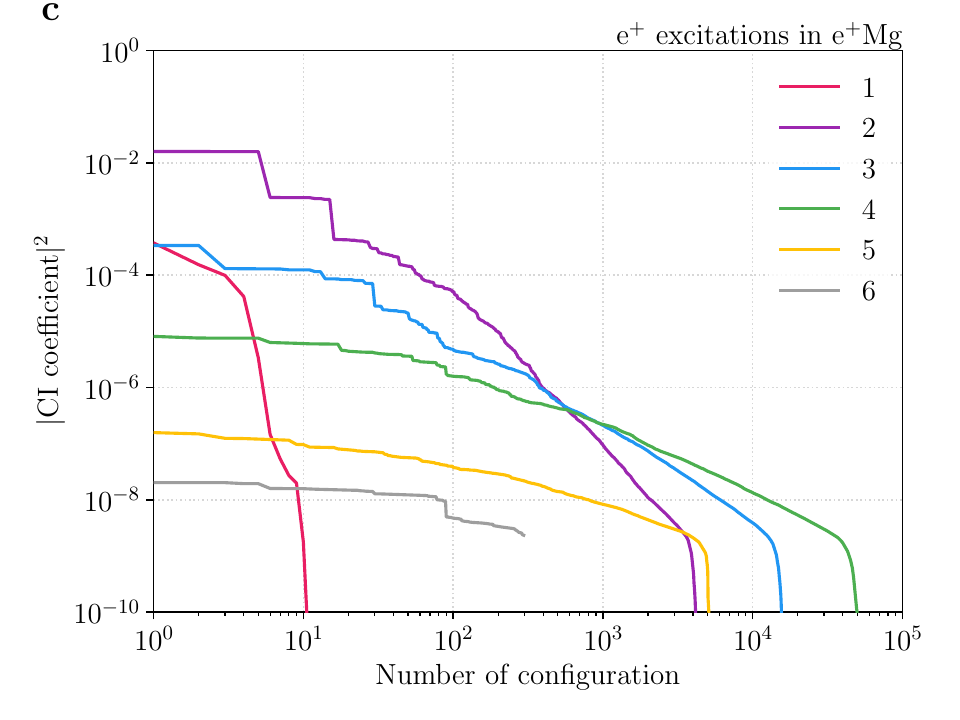}
    \caption{
    CI linear combination coefficients distribution for all 131072 SCI configurations grouped by excitation type as electronic excitations in \ce{Mg} (\textbf{a}), electronic excitations in \ce{e^+Mg} (\textbf{b}), and electronic-positron excitations in \ce{e^+Mh} (\textbf{c}).
    }
    \label{sm:fig:e+mg_distribution}
\end{figure}

\begin{figure}[h]
    \centering
    \includegraphics[width=0.495\linewidth]{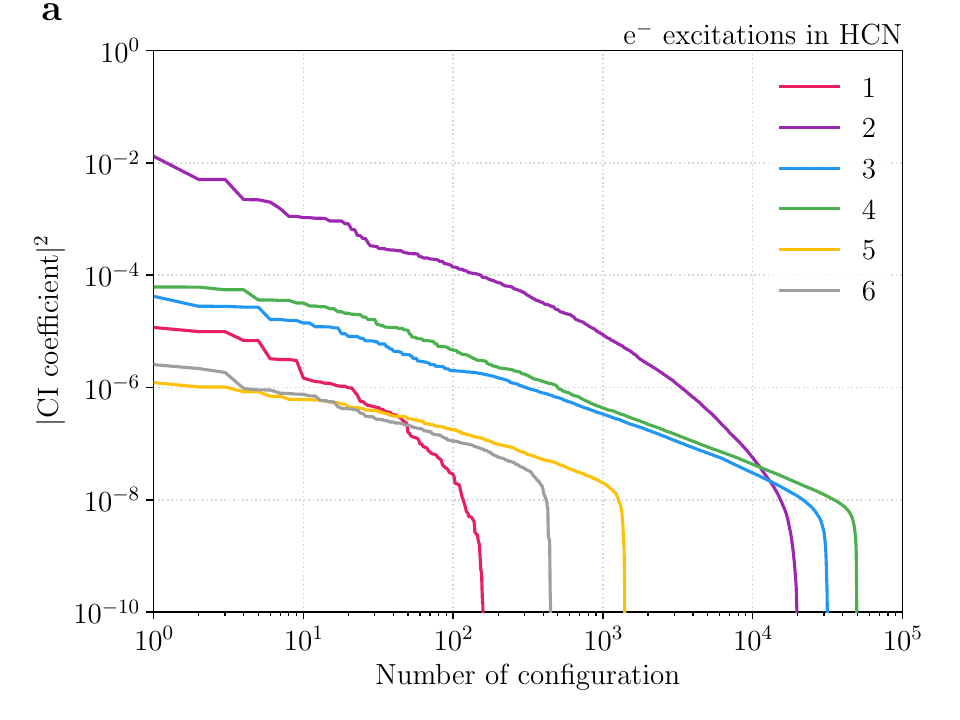}
    \includegraphics[width=0.495\linewidth]{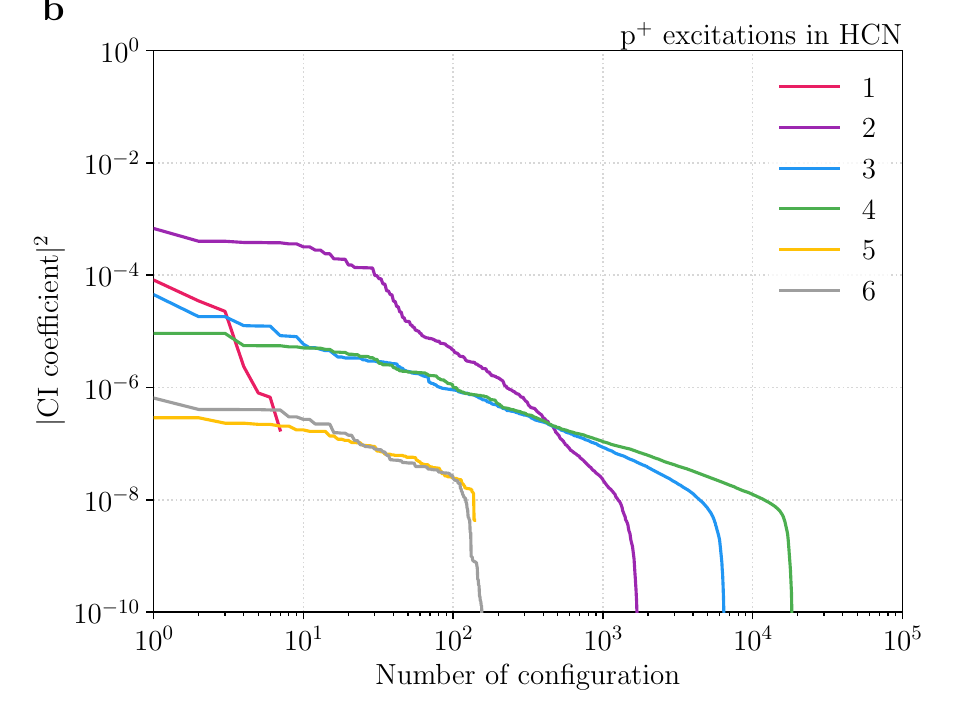}
    \caption{
    CI linear combination coefficients distribution for all 131072 SCI configurations grouped by excitation type as electronic excitations in \ce{HCN} (\textbf{a}) and electronic-proton excitations in \ce{HCN} (\textbf{b}).
    }
    \label{sm:fig:hcn_distribution}
\end{figure}

\end{document}